\documentclass[
 aip,
 amsmath,amssymb,
 reprint,
]{revtex4-1}

\usepackage{graphicx}
\usepackage{dcolumn}
\usepackage{bm}

\usepackage[utf8]{inputenc}
\usepackage[T1]{fontenc}
\usepackage{mathptmx}
\usepackage{etoolbox}

\usepackage{float}
\usepackage[caption = false]{subfig}

\usepackage{xcolor}

\makeatletter
\def\@email#1#2{%
 \endgroup
 \patchcmd{\titleblock@produce}
  {\frontmatter@RRAPformat}
  {\frontmatter@RRAPformat{\produce@RRAP{*#1\href{mailto:#2}{#2}}}\frontmatter@RRAPformat}
  {}{}
}%
\makeatother
\begin{document}

\preprint{AIP/123-QED}

\title[Elliptical instability and convection]{The interactions of the elliptical instability and convection}

\author{Nils B. de Vries}
 \email{N.B.deVries@leeds.ac.uk.}
 
\author{Adrian J. Barker}

\author{Rainer Hollerbach}
\altaffiliation[Also at ]{Isaac Newton Institute for Mathematical Sciences, 20 Clarkson Road, Cambridge CB3 0EH}
\affiliation{ 
Department of Applied Mathematics, School of Mathematics, University of Leeds, Leeds LS2 9JT, UK}
\date{\today}

\begin{abstract}
The elliptical instability is an instability of elliptical streamlines, which can be excited by large-scale tidal flows in rotating fluid bodies, and excites inertial waves if the dimensionless tidal amplitude ($\epsilon$) is sufficiently large. It operates in convection zones but its interactions with turbulent convection have not been studied in this context. We perform an extensive suite of Cartesian hydrodynamical simulations in wide boxes to explore the interactions of the elliptical instability and Rayleigh-B\'{e}nard convection. We find that geostrophic vortices generated by the elliptical instability dominate the flow, with energies far exceeding those of the inertial waves. Furthermore, we find that the elliptical instability can operate with convection, but it is suppressed for sufficiently strong convection, primarily by convectively-driven large-scale vortices. We examine the flow in Fourier space, allowing us to determine the energetically dominant frequencies and wavenumbers. We find that power primarily concentrates in geostrophic vortices, in wavenumbers that are convectively unstable, and along the inertial wave dispersion relation, even in non-elliptically deformed convective flows. Examining linear growth rates on a convective background, we find that convective large-scale vortices suppress the elliptical instability in the same way as the geostrophic vortices created by the elliptical instability itself. Finally, convective motions act as an effective viscosity on large-scale tidal flows, providing a sustained energy transfer (scaling as $\epsilon^2$). Furthermore, we find that the energy transfer resulting from bursts of elliptical instability, when it operates, is consistent with the $\epsilon^3$ scaling found in prior work.
\end{abstract} 

\pacs{47.20.-k; 47.20.Bp; 47.27.-i; 47.32.-y; 47.32.Ef; 47.35.+i}

\maketitle

\section{Introduction}

The elliptical instability arises when elliptically deformed streamlines 
excite pairs of inertial waves through parametric resonances \citep[e.g.][]{Waleffeellipinstab,Ellipticalinstability,BBO2016}. As long as retarding processes such as viscous damping can be overcome, an arbitrarily small elliptical deformation can yield instability. The resulting inertial waves couple with the deformation \citep[][]{Waleffeellipinstab}, leading to exponential growth of their amplitudes. This mechanism is in essence a triadic resonance interaction in which waves extract energy from the elliptical flow.

Its nonlinear evolution has been studied extensively. As the linear instability saturates, the inertial waves appear to collapse to rotating turbulence, which typically dissipates over time, leading to the flow becoming unstable again\cite[e.g.][]{Malkus1989,Barker2013,Barker2014,Favier2015,Barker2016,LeReun2017,le_reun2019_intwavebreakdownexp}. This collapse to turbulence either occurs via weak inertial wave ``turbulence"\cite{Malkus1989, Barker2014, LeReun2017, le_reun2019_intwavebreakdownexp}, or rotating turbulence involving large-scale geostrophic vortices or zonal flows\cite{Barker2013,Favier2015, Barker2016,LeReun2017,Grannan2017,le_reun2019_intwavebreakdownexp}. The inertial wave ``turbulence" (involving a sea of weakly interacting inertial waves) may occur when the forcing amplitude is weak \cite{le_reun2019_intwavebreakdownexp,Astoul2022}, or when geostrophic modes are suppressed, either by artificial frictional damping\cite{LeReun2017} or via an external process such as the imposition of a magnetic field \cite{Barker2014}.

In recent years, the elliptical instability also finds application as a tidal dissipation mechanism in stars and planets\cite{Rieutord2004,LE_BARS_elliptical,Cebron2010,Cebron_terrestrial,CebronellipHJ,Barker2013,Ogilviereview,Barker2016}, extracting energy from a tidal flow. Tidal flows in stars or planets are usually split up into a large-scale equilibrium or non-wave-like tide, and a dynamical or wave-like tide \cite{Zahn1977tidesplit,OgilvieIW}. The equilibrium tide is the quasi-hydrostatic fluid bulge rotating around the body \cite{Zahn1977tidesplit}, while the dynamical tide consists of waves generated by resonant tidal forcing. The equilibrium tide is thought to be dissipated through its interaction with turbulence, usually of a convective nature \cite{Zahn1989Turbvisceqtide,GoldreichNicholson1977}, or by its own fluid instabilities
\cite{Rieutord2004,LE_BARS_elliptical,Cebron2010,librationellip,CebronellipHJ,Barker2013,Barker2016}, among which is the elliptical instability. This however requires careful consideration of the dynamics of the elliptical instability, particularly the properties of the turbulence which is expected, as well as its interaction with other processes in the system, such as magnetic fields or (stable or unstable) stratification \cite{LeReun2018}.

The equilibrium tide deforms a body (body 1) into an ellipsoidal shape that follows an orbiting companion (body 2), and its deformation is represented by the ellipticity, or tidal amplitude parameter
\begin{equation}
    \epsilon=\left(\frac{m_2}{m_1}\right)\left(\frac{R_1}{a}\right)^3,
\end{equation} 
with $m_1$ and $m_2$ the masses of bodies 1 and 2 (e.g.\ a planet and its host star), $R_1$ is the radius of body 1, and $a$ is the orbital separation (semi-major axis). In an asynchronously rotating planet or star, the equilibrium tide in the frame rotating with the tidal bulge is an elliptical flow inside the planet. The rotation rate of this flow is the difference of the spin of the planet $\Omega$ and the orbital rotation rate $n$ and is denoted by $\gamma\equiv\Omega-n$. In this work we model a small patch of an equilibrium tidal flow, which is treated as a background flow $\tilde{\textbf{U}}_0$ in the bulge frame (rotating at the rate $n$ about the axis of rotation of the planet) following \cite{Barker2013}, such that
\begin{equation} 
    \tilde{\textbf{U}}_0=  \gamma\begin{pmatrix}
0 & -(1+\epsilon) & 0\\
1-\epsilon & 0 & 0\\
0 & 0 & 0
\end{pmatrix}\textbf{x},
\label{eq:backgroundflow non rotating}
\end{equation}
where $\textbf{x}$ represents the position vector from the centre of the planet. In the frame rotating with the planet at the rate $\Omega$, it can be written alternatively as the flow 
\begin{equation}
    \textbf{U}_0=\textbf{Ax}=-\gamma \epsilon\begin{pmatrix}
\sin(2\gamma t) & \cos(2\gamma t) & 0\\
\cos(2\gamma t) & -\sin(2\gamma t) & 0\\
0 & 0 & 0
\end{pmatrix}\textbf{x},
\label{eq:backgroundflow}
\end{equation}
where $\textbf{x}$ now represents the position vector from the centre of the planet in the frame rotating with the planet. This description represents the exact equilibrium tide of a uniformly rotating incompressible fluid body perturbed by an orbiting companion \citep{Chandrasekhar_ellipsoidal,Barker2016}, but approximates the main features of the equilibrium tide in more realistic models \cite{OgilvieIW,Barker2020}. We choose to work in the frame rotating with the planet at the rate $\Omega$ in this study. Larger deformations $\epsilon$ result in faster growth of the waves which means that they can overcome stronger viscous damping by either the viscosity of the fluid or by a turbulent viscosity.

The elliptical instability has been studied previously in simulations using a local Cartesian box model both with \citep[][]{Barker2014} and without \citep[][]{Barker2013} weak magnetic fields. The earlier study found that the elliptical instability leads to bursty behaviour, involving the interaction of instability-generated inertial waves with geostrophic columnar vortical flows produced by their nonlinear interactions. Irregular cyclic `predator-prey behaviour' was obtained in which the elliptical instability first excited inertial waves, these interacted nonlinearly to produce vortices that inhibited further growth of waves until these vortices were damped sufficiently by viscosity, thereby enabling further growth of the waves. Similar behaviour features in global hydrodynamical simulations of the elliptical instability \citep[][]{Barker2016}, where zonal flows take the place of columnar vortices in the predator-prey dynamics. Upon taking magnetic fields into account in the local model, the behaviour changed from bursts to a sustained energy input into the flow, as magnetic fields served to break up or prevent formation of strong vortices \cite{Barker2014}. Similar sustained behaviour is observed if the vortices are damped by an artificial frictional force mimicking Ekman friction on no-slip boundaries \cite{LeReun2017}. 

These prior studies analysed the elliptical instability in the convective regions of planetary (or stellar) interiors, but did not incorporate convection explicitly (except perhaps by motivating a choice of viscosity if this is due to turbulence). However, convection can potentially interact with the elliptical instability in a number of ways. Firstly, we might imagine that smaller-scale turbulent convective eddies could act like an effective viscosity in damping larger-scale inertial waves, thereby inhibiting or reducing the growth of the elliptical instability. Secondly, convection under the influence of rotation is known to generate mean flows (zonal flows or vortices), and these flows could also interact with those generated by the elliptical instability. In this study we wish to address the following questions: Can the elliptical instability operate in a turbulent convective background? How do convectively-driven flows interact with the elliptical instability and modify (tidal) energy transfer rates?

The interaction of the elliptical instability with convection has been studied within linear theory  \citep[e.g.][]{LeBars2006_linellipcylinder,Ellipticalinstability}, experimentally in cylindrical containers \cite{Lavorelexperimentalellip} and using idealised laminar global simulations in a triaxial ellipsoid \cite{Cebron2010}. These studies illustrate that the elliptical instability can modify heat transport, though they did not focus on the dissipation of tidal flows which is our primary focus here. The dimensionless heat transport is usually represented by the Nusselt number (Nu), which is a measure of the ratio of the total heat flux to the conductive flux (i.e. with no transport by fluid motions), as a function of the Rayleigh number (Ra), the dimensionless ratio of buoyancy driving to viscous and thermal damping, which measures the strength of convective driving. The Nusselt number was observed to be increased by the elliptical instability for Rayleigh numbers close to and below the value required for onset of convection. It was also observed to be larger than one even with stable stratification (Ra $<0$), indicating that the elliptical instability can contribute to heat transport in this regime also.

\begin{figure}
    \centering
    \includegraphics[width=0.8\linewidth]{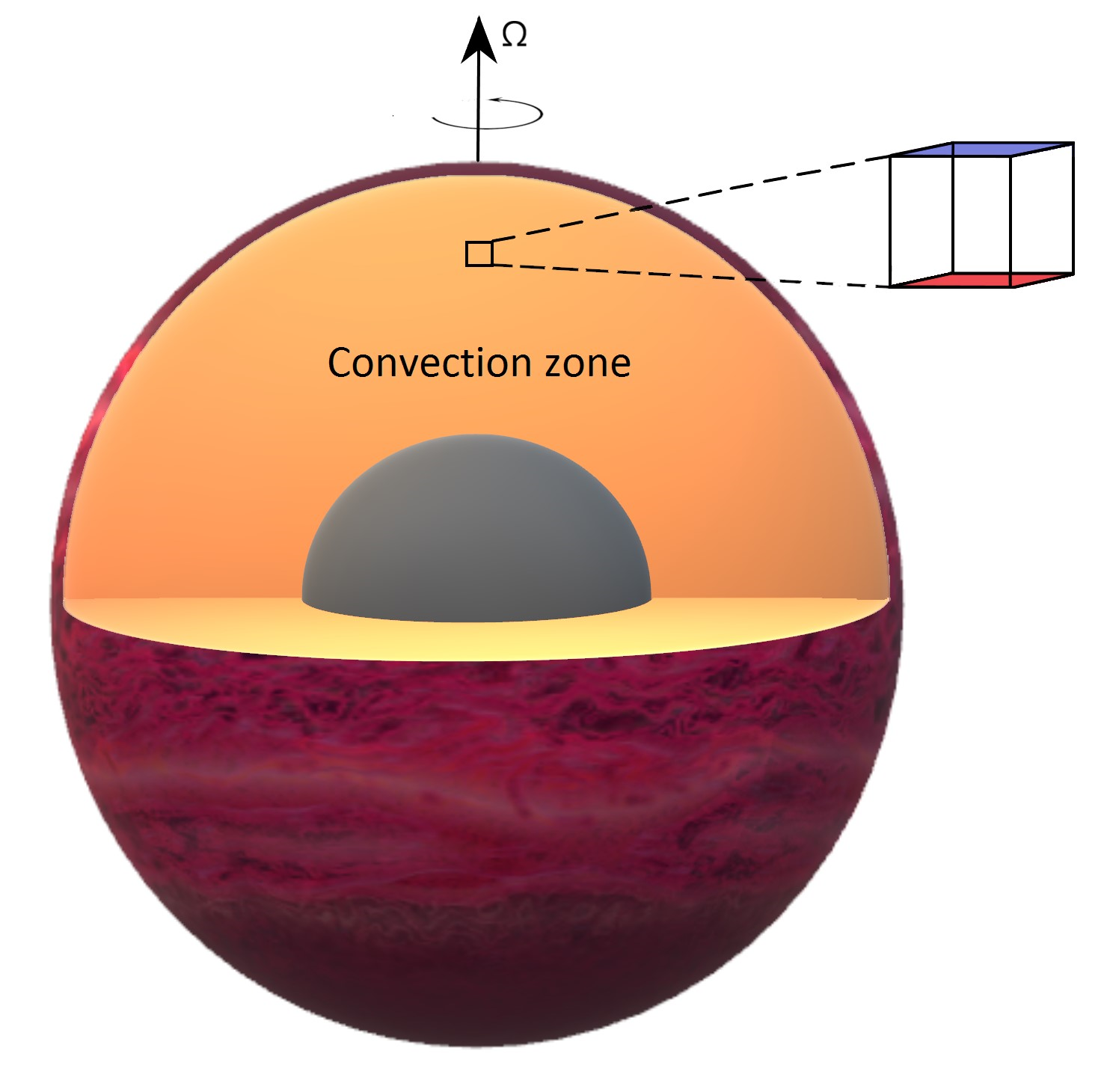}
    \caption{Location of the local box in the convection zone of a Hot Jupiter, indicating the rotation axis, and the temperature gradient represented by the red (hot) and blue (cold) sides of the box.}
    \label{fig:HJlocalbox}
\end{figure}

In this paper, we build upon the local box model -- which represents a small patch of a planet or star (see Fig.~\ref{fig:HJlocalbox}) -- in \citet[][]{Barker2013,Barker2014,LeReun2017} to study the interaction of the elliptical instability with convection with a focus on the resulting tidal dissipation. Our local model allows for higher resolution studies, which in turn allows us to reach more turbulent regimes than e.g. \citet[]{Lavorelexperimentalellip,Cebron2010}. One of our aims is to study the behaviour of the elliptical instability, and to see if introducing convection could also lead to sustained energy injection in a similar fashion to a magnetic field (or frictional damping). As a secondary goal, we are interested in studying the modifications to heat transport by the elliptical instability, both in the weakly-driven regime of convection and in stably-stratified layers like those that may exist in giant planet interiors.

In Section \ref{sec:modelsetup} we will discuss the linear properties of the elliptical instability and describe the model used, and in Section \ref{sec:paramresults} we describe the results obtained from our parameter survey in a qualitative manner. We investigate the sustained energy injection and analyse frequency and wavenumber spectra for the flow in Section \ref{sec:sustained}. Then, we briefly discuss the scaling of the energy transfer from the background flow with $\epsilon$ in Section \ref{sec:scalinglaws}. We discuss and conclude in Section \ref{sec:discussion}.

\section{Model Setup}~
\label{sec:modelsetup}

\subsection{The elliptical instability}
\label{sec:ellip}

The linear properties of the elliptical instability have been reviewed by \citet{Ellipticalinstability}. This instability operates when two inertial waves have frequencies that approximately add up to the tidal frequency $2\gamma$ (see introduction).
In the short wavelength limit, this occurs for two waves with frequencies $\omega=\pm\gamma$, which must each satisfy the dispersion relation for inertial waves, $\omega=\pm2\Omega k_z/k$, where $k_z/k=\cos\theta$. The elliptical instability grows at a rate proportional to $\epsilon\gamma$.

Since we are investigating a small patch of a planet, we assume the tidal flow can be modelled locally as an unbounded strained vortex in the bulge frame \citep[as in Eq.~5, where its stability is reviewed in][]{Ellipticalinstability}. This approach yields a growth rate which depends on the angle the inertial waves make with respect to the rotation axis and the strain direction in the horizontal plane. For illustration, when the tidal bulge is stationary ($n=0$), $\gamma=\Omega$, and thus $k_z/k=\pm1/2$ for the most unstable modes \citep[as reviewed in][]{Ellipticalinstability}. 
Furthermore, the fastest growing waves have a phase aligned with the strain direction, by an angle of $\pm\pi/4$ with respect to the elliptical deformation in the equatorial plane (i.e. plane containing the vortical flow).

The effects of viscosity, detuning, convection and rotation of the elliptical bulge are also reviewed by \citet{Ellipticalinstability}. Viscosity reduces the growth rate according to: $\sigma-\nu|\textbf{k}|^2$, where $\sigma$ is the inviscid growth rate, $\nu$ is the viscosity and $\textbf{k}$ is the wavevector of the fastest growing mode. Detuning is a reduction of the growth rate as a result of the wave not satisfying the resonance conditions exactly, which also reduces the maximum growth rate. 
The rotation around the companion, and thus the rotation of the elliptical bulge, modifies the growth rate depending on the rotation speed ($n$). The growth rate is decreased for most values of\citep[e.g. see Fig.1 in ][]{Barker2013} $n$, 
and it cannot operate in the interval $n=[-3/2\gamma, -1/2\gamma]$. In this interval no inertial waves exist that satisfy the dispersion relation defined by: 
\begin{equation}
    \omega=\pm 2\Omega\cos \theta.
    \label{eq:dispersionrelation}
\end{equation}

In the interval $n=[-1/2\gamma, 0]$ the growth rate is increased, though everywhere else it is decreased, over the case with $n=0$. The linear growth rate of an inviscid fluid at small $\epsilon$ without convection is given by \citep{MF1992,Ellipticalinstability}:
\begin{equation}
    \sigma=\frac{9}{16}\gamma\epsilon\frac{(3\gamma+2n)^2}{9(\gamma+n)^2}.
    \label{eq:ellipgrowthrate}
\end{equation}
If an (un)stable stratification (aligned with the rotation axis) is present the dispersion relation is modified, as well as the above equation. The stratification introduced into the equation is represented by the Brunt-V\"ais\"al\"a (or buoyancy) frequency $N$. The modified dispersion relation is:
\begin{equation}
    \omega^2=4\Omega^2\cos^2(\theta)+N^2\sin^2(\theta).
    \label{ref:dispersionrelation_Nsqr}
\end{equation}

The modified version of Eq.~\ref{eq:ellipgrowthrate} for small $\epsilon$ is then \cite{Ellipticalinstability}:
\begin{equation}
    \sigma=\frac{9}{16}\gamma\epsilon\frac{4(3\gamma+2n)^2(\gamma^2-N^2)}{9\gamma^2(4(\gamma+n)^2-N^2)}.
    \label{eq:kerswellgrowthconvrotat}
\end{equation}
Both the effects of unstable stratification (negative $N^2$) and stable stratification (positive $N^2$) can be computed using this equation.
We observe that stable stratification typically inhibits elliptical instability, but that convection typically enhances growth.

For clarity of presentation $\gamma=\Omega$ is chosen, resulting in $n=0$, i.e. strictly representing the unphysical case where there is no rotation of the bulge. The body in question is not rotating around its companion which causes the tidal effects. However, it turns out that for simulations the only linear effects of choosing a different value of $\Omega$, and therefore a non-zero value of $n$, would be to modify the growth rate of the elliptical instability \cite{Barker2013} as well as the wavenumber of the most unstable mode \citep[][]{Barker2014}.\\

\subsection{Governing equations and setup of the simulations}
\label{sec:governeq}
\label{sec:simsetup}

We model the convective instability using Rotating Rayleigh-B\'enard Convection (RRBC) and the Boussinesq approximation. RRBC is chosen as it is the simplest model of rotating convection \cite{Chandrsekharbook} which allows us to study its interaction with elliptical instability (an even simpler model is ``homogeneous convection" with periodic boundaries in the vertical, but this has unphysical nonlinear behaviour). The Boussinesq approximation is justified when studying small-scale convective (and wavelike) flows, which satisfies the required conditions that flows are much slower than the sound speed, $u\ll c_s$, and the vertical size of the domain $d$ is much smaller than a pressure or density scale height, $d\ll H_p$ \cite{bousapprox}. However, this neglects variations of the properties of a planet, i.e. density and temperature, and of course, any large-scale circulations cannot be modelled using this approximation. 

The rotation axis is aligned with the $z$-direction, as is the temperature gradient, as indicated in Fig.~\ref{fig:HJlocalbox}. The box in the current setup thus represents a polar region, because the local rotation and gravity vectors are either aligned or anti-aligned (depending on the sign of $\Omega$). The conduction state temperature profile $T(z)$ between the hot plate at the bottom and the cold plate at the top, about which we will perturb, is
\begin{equation}
    \alpha g (T-T_0)=\frac{z N^2}{d},
\end{equation}
where $g$ is the local gravitational acceleration (assumed constant) and $\alpha$ is the (constant) thermal expansion coefficient.
Without loss of generality $T_0$ is set to zero, so the temperature at the bottom (hot plate, typically) is $T(z=0)=0$, while the temperature at the top is $T(z=d)=N^2/(\alpha g)$, such that the temperature drop is $\Delta T =-N^2/\alpha g$. 

We non-dimensionalise by scaling lengths with the vertical domain size $d$ (distance between the plates), scaling time using the corresponding thermal timescale $d^2/\kappa$, thus scaling velocities with $\kappa/d$, pressures with $\rho_0 \kappa^2/d^2$, and finally scaling temperature with $T=\Delta T\theta$ (i.e. by the temperature difference between the plates).
The governing equations for the dimensionless velocity and temperature perturbations $\textbf{u}$ and $\theta$ to the background flow $\textbf{U}_0$ and conduction state temperature $T(z)$ in the Boussinesq approximation, in a frame rotating at the rate $\Omega$ about $z$, are then:

\begin{equation}
    \frac{D\textbf{u}}{Dt}+\textbf{u}\cdot\nabla \textbf{U}_0+\frac{ \textrm{Pr}}{ \textrm{Ek}}\hat{\textbf{z}}\times \textbf{u} = -\nabla p +  \textrm{PrRa}\theta \hat{\textbf{z}} + \text{Pr}\nabla^2\textbf{u},
    \label{eq:maingovern}
\end{equation}
\begin{equation}
    \nabla \cdot \textbf{u}=0,
    \label{eq:govern2}
\end{equation}
\begin{equation}
    \frac{D\theta}{Dt}-u_z=\nabla^2\theta,
    \label{eq:govern3}
\end{equation}
where
\begin{equation}
    \frac{D}{Dt}\equiv\frac{\partial}{\partial t}+\textbf{U}_0\cdot \nabla+ \textbf{u}\cdot \nabla,
\end{equation}
with $\textbf{u}=(u_x,u_y,u_z)$ and $p$ being the perturbation to the pressure.
The non-dimensional parameters describing the convection are the Rayleigh number
\begin{equation}
\mathrm{Ra} = \frac{\alpha g (-N^2) d^4}{\nu\kappa},
\end{equation}
where $\nu$ and $\kappa$ are the constant kinematic viscosity and thermal diffusivity, the Ekman number (ratio of viscous to Coriolis terms)
\begin{equation}
\mathrm{Ek} = \frac{\nu}{2\Omega d^2},
\end{equation}
and the Prandtl number $\mathrm{Pr}=\nu/\kappa$. Note that we can relate the dimensional squared buoyancy frequency $N^2 = -\text{Ra\,Pr}\,\kappa^2/(\alpha g d^4)$, so that when $\text{Pr}=1$, the dimensionless value (in thermal time units) is $N^2=-\text{Ra}$. The tidal background flow also introduces the dimensionless ellipticity $\epsilon$ and the frequency $\gamma$ in our chosen units.

Our simulations are executed in a small Cartesian box of dimensionless size $[L_x,L_y,1]$ with $L_x=L_y=L$. The boundary conditions in the horizontal directions are periodic, while in the vertical direction they are impermeable, $u_z(z=0)=u_z(z=1)=0$, and stress-free, $\partial_z u_x(z=0)=\partial_z u_x(z=1)=\partial_z u_y(z=0)=\partial_z u_y(z=1)=0$. Stress-free boundary conditions are chosen both for numerical convenience and because they are probably more physically relevant than no-slip boundary conditions for modelling convection in the deep interior of a planet, far from boundaries. The convection in our box thus represents a single convection cell in the vertical. Boundary conditions in the vertical for the temperature perturbation are assumed to be perfectly conducting, $\theta(z=0)=\theta(z=1)=0$.

In the derivation of the elliptical instability the choice is often made to work with so-called shearing
waves \cite{Waleffeellipinstab,Ellipticalinstability}. Shearing waves have time-dependent wavevectors, which allows us to account for the stretching and rotation of waves due to a background flow, such as the equilibrium tide in our simulation. A single shearing wave (sometimes also referred to as a Kelvin wave\cite{Ellipticalinstability}, although strictly different to a coastal Kelvin wave) is represented as:
\begin{equation}
    \begin{aligned}
    \textbf{u}&=\text{Re}[(\hat{u}_x(t)\cos(k_zz),\hat{u}_y(t)\cos(k_zz),\hat{u}_z(t)\sin(k_zz))\,\exp^{i\textbf{k}_\perp (t)\cdot \textbf{x}}],\\ p&=\text{Re}[\hat{p}(t)\cos(k_zz)\,\exp^{i\textbf{k}_\perp(t)\cdot \textbf{x}}],
    \label{basis}
    \end{aligned}
\end{equation}
where $\dot{\textbf{k}}_\perp=-A^T\textbf{k}_\perp$ and $\textbf{k}_\perp=(k_x,k_y,0)$, and we use a basis of these waves in our simulations following \citet{Barker2013}, except that we use a sine-cosine decomposition in $z$ similar to \citet[][]{Craig2019effvisc}.

The simulations are executed using the Snoopy code \citep{snoopycode}. The Snoopy code implements a Fourier pseudo-spectral method using FFTW3 in a Cartesian box. We use a sine-cosine decomposition in $z$, as in Eq.~\ref{basis}, and shearing waves (i.e.\ Fourier modes) in $x$ and $y$. A 3rd-order Runge-Kutta scheme is used for the time-stepping, together with a CFL safety factor to ensure the timesteps are small enough to accurately capture non-linear effects, usually set to 1.5 (which is smaller than the stability limit of $\sqrt{3}$). The anti-aliasing in the code uses the standard 2/3 rule \citep{boyd2001chebyshev}. A variety of different Rayleigh numbers were analysed using the simulations. In addition, some simulations were performed with Rayleigh numbers in the stably stratified regime, i.e.~with $\text{Ra}<0$. The values of the Rayleigh number are typically reported as $\text{Ra}/\text{Ra}_c$ for clarity, where $\text{Ra}_c$ is the onset Rayleigh number (determined numerically), and the range of this ratio studied at $\textrm{Ek}=5\cdot10^{-5.5}$ is from $2$ to $20$ and from $-10$ to $0.8$, in the convectively unstable and stable regimes, respectively. We vary $\epsilon$ from $0.01$ to $0.20$.

In simulations of RRBC in a local box model large-scale vortex (LSV) structures emerge in the flow when rotation dominates \citep{CelineLSV,FavierLSV,RubioLSV2014}. This LSV emerges with our chosen boundary conditions (more details below) and grows to the size of the box in the horizontal. One of the effects of the LSV is to reduce heat transport, as the vertical motions are suppressed by such a vortex \citep[][]{CelineLSV}. An additional reason to study convective LSVs is that the elliptical instability can be suppressed by the presence of strong vortices \cite[][]{Barker2013}. The convective LSV might suppress the elliptical instability as well, potentially preventing it from operating efficiently. Since our flow is likely to be rotationally dominated due to our choice of Ekman number and computationally feasible Rayleigh numbers, a horizontal box size was chosen that would capture this LSV. The question remains however what effect changing the aspect ratio of the box would have on the effects presented in this report, as the aspect ratio $L/d$ (the ratio of horizontal length of the box to its vertical length) influences the ratio of the vertical to total kinetic energy \citep{CelineLSV}.

\subsection{Energetic analysis of simulations}
\label{diagnostics}

To analyse the energy of the flow we derive a kinetic energy equation by taking the scalar product with $\textbf{u}$ of Eq.~\ref{eq:maingovern} and then averaging over the box. We define our averaging operation on a quantity $X$ as $\langle X\rangle=\frac{1}{L^2d}\int_V X\text{ }dV$. We obtain:
\begin{equation}
    \frac{d}{dt} K= I + \langle \text{PrRa}\theta u_z\rangle - D_{\nu},
\end{equation}
where we have defined the mean kinetic energy:
\begin{equation}
    K\equiv\frac{1}{2}\langle|\textbf{u}|^2\rangle,
\end{equation}
the mean viscous dissipation rate
\begin{equation}
    D_{\nu}\equiv-\text{Pr} \langle \textbf{u} \cdot \nabla^2 \textbf{u} \rangle,
\end{equation}
and the energy injection rate (more generally, energy transfer rate) from the tidal to convective flows (or vice versa)
\begin{equation}
    I\equiv-\langle\textbf{u}\text{A}\textbf{u}\rangle.
\end{equation}
To obtain an equation for the thermal (potential) energy, we multiply Eq. \ref{eq:govern3} by $-$RaPr$\theta$ and average over the box to obtain:
\begin{equation}
    \frac{d}{dt}P=-\langle\text{PrRa}\theta u_z\rangle-D_{\kappa},
\end{equation}
where we have defined the mean thermal energy as
\begin{equation}
    P\equiv-\textrm{PrRa}\frac{1}{2}\langle\theta^2\rangle,
\end{equation}
and the thermal dissipation rate as
\begin{equation}
    D_\kappa\equiv\textrm{PrRa}\langle\theta\nabla^2\theta\rangle.
\end{equation}
The total energy is $E=K+P$, which thus obeys:
\begin{equation}
    \frac{d}{dt}E=I-D_\nu-D_\kappa.
\end{equation}

In a statistically steady state it is expected that the (time-averaged value of the) energy injected balances the total dissipation, i.e. $I\approx D\equiv D_\nu+D_\kappa$ (on average). This sum of the two dissipation rates then represents the tidal energy dissipation rate resulting from the tidal energy injected. Therefore, to interpret the tidal energy dissipation rate we examine the tidal energy injection rate $I$. 

Arguments to describe scaling laws for the dissipation due to the elliptical instability were first described by picturing the instability saturation as involving a single most unstable mode whose amplitude saturates when its growth rate ($\sigma$) balances its nonlinear cascade rate\cite{Barker2013}. Thus, if the most important mode of the elliptical instability satisfies $\sigma\sim k u$, where $k$ is its wavenumber magnitude and $u$ is its velocity amplitude, then we find $u\sim\epsilon\gamma/k$. The total dissipation rate $D$ therefore scales as $D\sim u^2\sigma\sim \epsilon^3\gamma^3/k^2$. Thus, in such a statistically steady state the dissipation and energy injection rate are expected to scale as
\begin{equation}
    D=I\propto \epsilon^3,
    \label{eq:epsiloncubed}
\end{equation}
and this is consistent with some local and global simulations \citep{Barker2013,Barker2016} as well as the scaling found for related instabilities like the precessional instability \cite{Barker2016precession,Pizzi2022}. We are interested in exploring whether convection could lead to a different result, and potentially reduce this steep $\epsilon$ scaling.

Since we know both the elliptical instability \citep[][]{Barker2013} and convection \cite{CelineLSV} in isolation can produce geostrophic flows such as vortices, we introduce further diagnostics to analyse these flows and the roles they play. To do this, we decompose the total energy injection from the background flow into:
\begin{equation}
    I = I_{2D}+I_{3D},
\end{equation}
where we have defined $I_{2D}=-\langle\textbf{u}_{2D}\text{A}\textbf{u}_{2D}\rangle$
and $I_{3D}=-\langle\textbf{u}_{3D}\text{A}\textbf{u}_{3D}\rangle$.
$I_{2D}$ and $\textbf{u}_{2D}$ are defined to include all (geostrophic) modes where the wavevector has only non-vanishing $x$ and $y$ components, with $k_z=0$, and $I_{3D}$ and $\textbf{u}_{3D}$ includes all the modes with $k_z\neq0$. Thus we have decomposed the total energy injection rate into energy injection into the barotropic ($k_z=0$) and baroclinic ($k_z\neq0$) flow. A pure inertial wave with $k_z=0$ would have zero frequency, while in convectively unstable simulations, which is the main focus of this work, no gravity waves exist which could have $\omega\neq0$ even when $k_z=0$, and therefore one can crudely think of this decomposition as one into geostrophic vortex modes ($I_{2D}$) and waves ($I_{3D}$). We have found that the time-averaged energy input into the vortical motions $I_{2D}$ is approximately zero \citep[or small, see also][]{Barker2013}, but that the input into the waves $I_{3D}$ is on average non-zero (which it must be when the elliptical instability operates) and clearly demonstrates any bursty behaviour observed. Based on this observation, only results derived from $I_{3D}$ will be plotted in this paper. The total kinetic energy $K$ is also split up into a 2D and 3D component in a similar manner by defining $K_{2D}$ and $K_{3D}$, so as to allow us to determine which components contribute the most and dominate the flow. 

To further analyse the energy transfer rates $I$ and $I_{3D}$, we also convert these to an effective viscosity $\nu_{eff}$ and $\nu_{eff,3D}$ respectively. The effective viscosity represents the energy dissipation that would result from a constant kinematic viscosity with the value $\nu=\nu_{eff}$, and this quantity allows us to interpret the value of $I$. In particular, this is a useful comparison to quantify the rate at which turbulent convection could damp our tidal flow, if this interaction behaves like a turbulent viscosity. To define the effective viscosity, we equate the work done by the tidal flow on the convective flow with the viscous dissipation rate of the tidal flow, assuming this is due to a constant kinematic viscosity $\nu_{eff}$, following \citet{Goodmaneffvisc,Ogilvieeffvisc,Craig2019effvisc,VB2020}. We note that
\begin{equation}
    I=-\frac{1}{V}\int_V \textbf{u}\cdot(\textbf{u}\cdot\nabla)\textbf{U}_0 dV.
\end{equation}
We define the strain rate tensor for the tidal flow as $e_{ij}^0\equiv\frac{1}{2}(\partial_iU_{0,j}+\partial_jU_{0,i})$, such that the rate at which energy is dissipated is given by
\begin{equation}
    \frac{2\nu_{eff}}{V}\int_V e_{ij}^0e_{ij}^0 dV=4\nu_{eff}\gamma^2\epsilon^2.
\end{equation}
The effective viscosity is then defined by
\begin{equation}
    \nu_{eff}=I/(4\gamma^2\epsilon^2).
    \label{eq: I_and_nueff}
\end{equation}
The injection terms represent energy being transferred from the background flow to the perturbations or vice versa. This, by definition, impacts the energy in these flows. The evolution of the tidal flow $\textbf{U}_0$, however, isn't explicitly accounted for in our model, which we treat as a fixed (but time-dependent) background flow. The idea is that it has much larger energy than the perturbations, so it is treated as an infinite reservoir in our simulations. These results therefore give us only a snapshot at a certain point in time of the evolution of the system, which is reasonable because tidal evolutionary processes usually occur slowly relative to convective or rotational timescales.

Finally, we compute the vertical heat transport in our simulations, represented by the Nusselt number, which we define as:
\begin{equation}
    \text{Nu}=1+\text{RaPr}\langle\theta u_z\rangle.
    \label{eq:Nusseltdefinition}
\end{equation}
This gives the ratio of the total heat flux to the conductive flux, and would take the value one in the absence of flows (i.e.~heat transported purely by conduction).

\begin{figure}
     \centering
     \begin{minipage}[b]{0.45\textwidth}
         \centering
         \includegraphics[height=50mm,width=\textwidth]{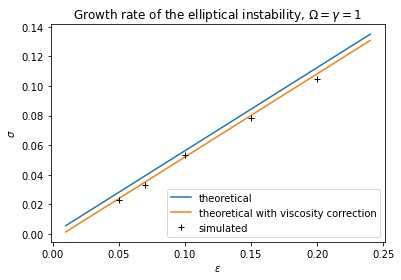}
         \label{fig:growth ratetestplotepsilon}
     \end{minipage}
     \hfill
     \begin{minipage}[b]{0.45\textwidth}
         \centering
         \includegraphics[height=50mm,width=\textwidth]{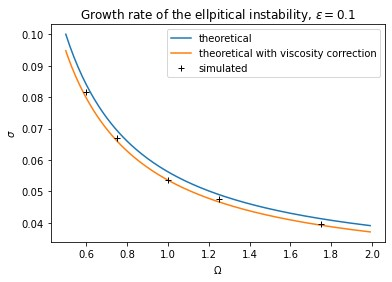}
         \label{fig:growth ratetestplotomega}
     \end{minipage}
     \hfill
     \begin{minipage}[b]{0.45\textwidth}
         \centering
         \includegraphics[height=50mm,width=\textwidth]{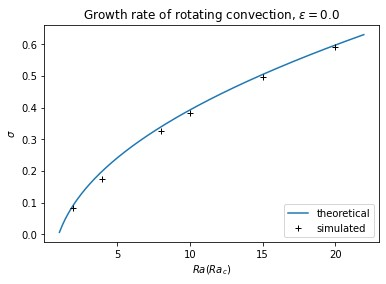}
         \label{fig:growth ratetestplotconv}
     \end{minipage}
     \caption{Growth rates of the elliptical instability and convection studied in isolation. Top, middle: growth rate of the elliptical instability ($\sigma$ with time units of $\gamma^{-1}$) showing simulations compared with the theoretical prediction based on Eq.~\ref{eq:ellipgrowthrate}, as a function of $\epsilon$ ($n=0$, $\gamma=\Omega=1$) in the top panel, and Eq.~\ref{eq:ellipgrowthrate} as a function of $\Omega$ (keeping $\gamma=1$) in the middle panel. The simulations are in excellent agreement, with a slight reduction due to viscosity and detuning. Bottom: growth rate of rotating convection, compared with the theoretical prediction of RRBC for the fastest growing mode ($\sigma$ in thermal time units) with $\textrm{Ek}=5\cdot10^{-5.5}$.}
    \label{fig:growth ratetestplotellip}
\end{figure}

\subsection{Linear growth rates and numerical validation}
\label{sec:linearstab}

 The predicted growth rate of the elliptical instability is given in Eq.~\ref{eq:kerswellgrowthconvrotat}, however this was derived for an unbounded flow in $z$. Since we have adopted the RRBC setup with impermeable walls in $z$, we must determine how this affects the growth rate of instability, although we expect it remains unchanged (and we have also demonstrated this analytically, though we omit this derivation). Hence, we performed multiple test simulations analysing the linear growth rates of both the elliptical and convective instabilities. We initialised them with random noise and the non-linearities were switched off, thus leading to a continuous exponential growth, allowing for easy extraction of the growth rate. Fits were performed to the mean kinetic energy on a log-scale to determine $\sigma$, by noting that if velocity components grow as $\exp(\sigma t)$, then $K\propto \exp(2\sigma t)$. 

The results, along with the theoretical growth rate predictions for both instabilities, are plotted in Fig.~\ref{fig:growth ratetestplotellip}. The top panel shows the growth rate of the elliptical instability as a function of $\epsilon$ (when $n=0$), where we have adopted a time unit $\gamma^{-1}$, equivalent to using $\gamma=1$, for the purposes of this figure. It can be concluded that the modelling of the inviscid growth rate is approximately correct. We also accounted for viscous damping by including the viscous decay rate so that the total growth rate is $\sigma-\nu k^2$ (where $k$ is the wavevector magnitude of the mode), which is in excellent agreement with our simulations. We obtain values that are very slightly smaller than the theoretical prediction though, even when taking into account viscosity, which is likely due to the detuning effect discussed previously (i.e.~that the mode does not precisely satisfy $k_z/k=1/2$). In numerical simulations this detuning arises because of the finite number of grid points, which, in addition to a chosen aspect ratio, prohibits the waves from precisely satisfying the aforementioned condition. However, because this condition does not stipulate the size of the wavenumbers, instead stipulating their ratio, and as such direction, the fastest growing mode that dominates the volume-averaged energy will be as large-scale as possible while still adequately satisfying the resonance condition in order to reduce the viscosity correction. This means that it should be unaffected by resolution, instead being controlled by the aspect ratio of the box. The effect on the growth rate of this detuning, which here corresponds to the difference between the markers and the viscosity corrected growth rate, at this aspect ratio and $\gamma=\Omega=1$, corresponding to the top panel of Fig.~\ref{fig:growth ratetestplotellip} and the other simulations in this work, is determined numerically to be $\approx0.002$.

The growth rate of the elliptical instability as a function of $\Omega$, keeping $\gamma=1$, is shown in the middle panel of Fig.~\ref{fig:growth ratetestplotellip}, and also follows the theoretical prediction well, but is again slightly lower for the same reasons. The growth rate of the convective instability for $\textrm{Ek}=5\cdot10^{-5.5}$ is shown in the bottom panel of Fig.~\ref{fig:growth ratetestplotellip}, now using thermal time units, and this is also in very good agreement with the linear convective growth rate for the fastest growing mode  (which scales as $\textrm{Ek}^{-1/3}$ as obtained from solving the relevant cubic dispersion relation numerically) as expected. We can therefore be confident that both instabilities have been captured correctly.

\section{Numerical results}
\label{sec:paramresults}

\subsection{Qualitative analysis of illustrative simulations}
\label{sec:qualitative}

\begin{figure*}
    \centering\begin{minipage}[b]{0.45\textwidth}
         \centering
         \includegraphics[width=0.9\textwidth]{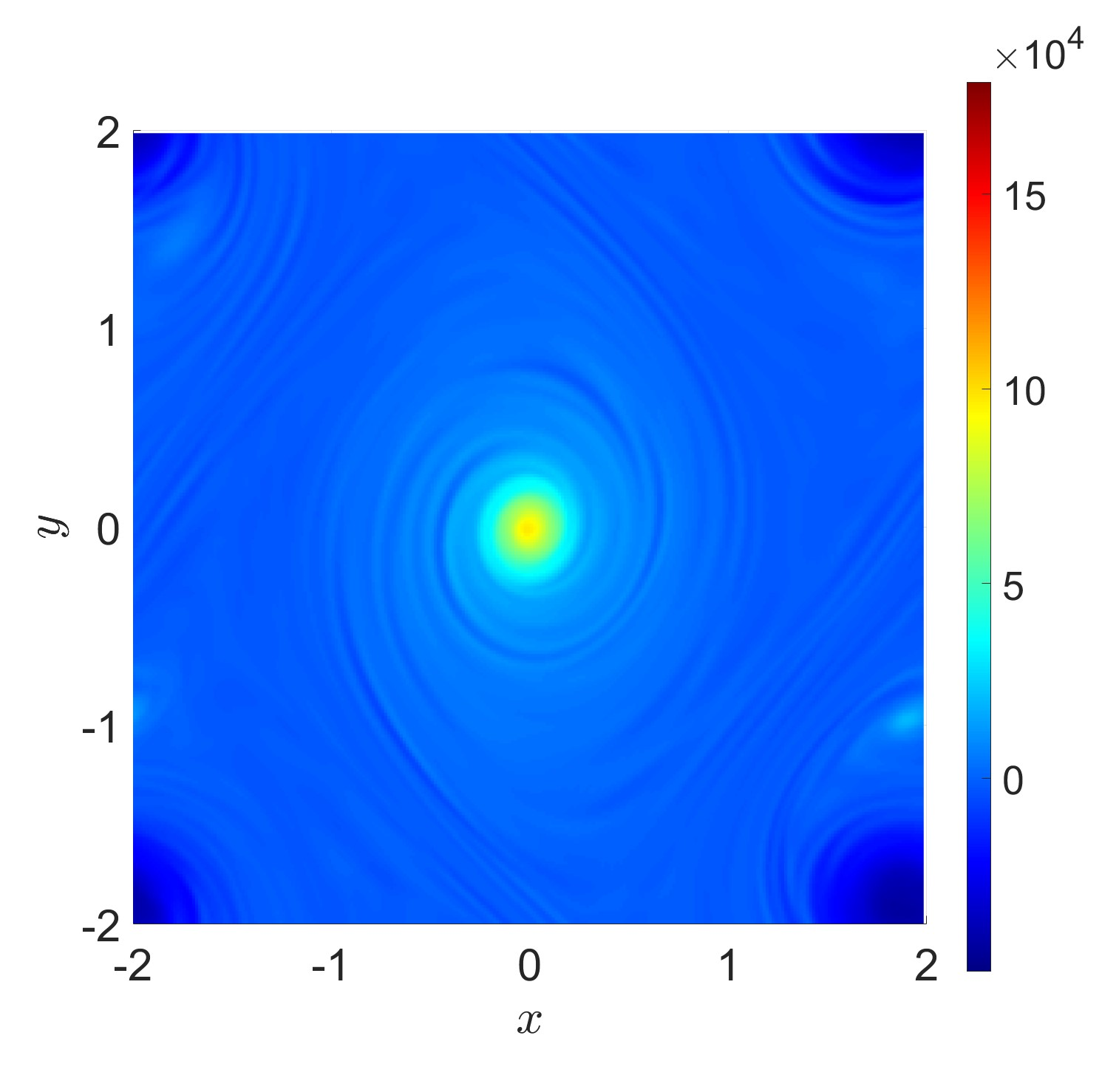}
         \label{fig:vorticityellip}
     \end{minipage}
     \hfill
     \begin{minipage}[b]{0.45\textwidth}
         \centering
         \includegraphics[width=0.9\textwidth]{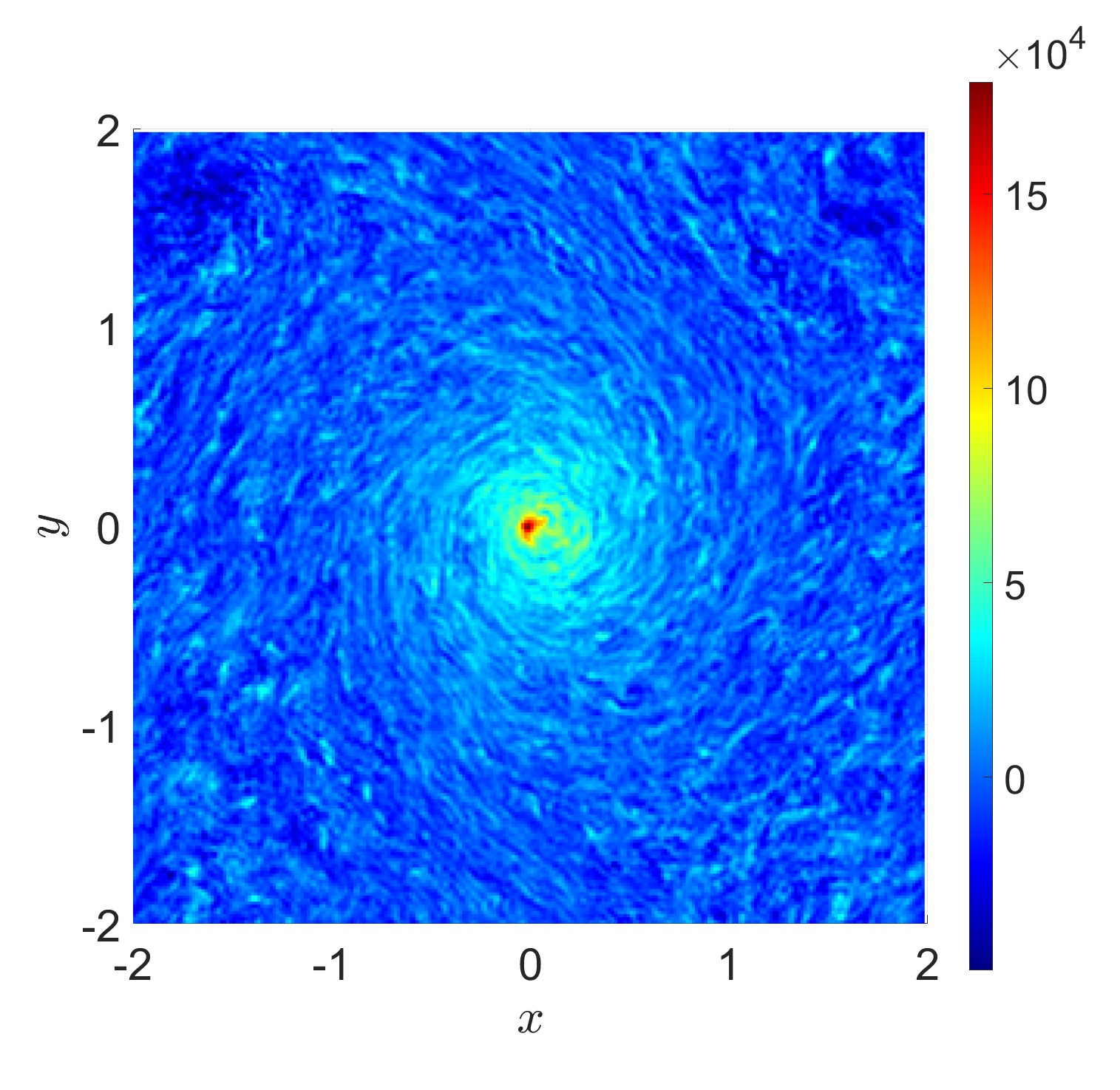}
         \label{fig:vorticityellip+conv}
     \end{minipage}
     \caption{The vertical vorticity averaged over $z$ ($\langle\omega_z\rangle_z$) of the flow at $t=0.08$. The cyclonic vortex is centred for clarity in both images. Left: elliptical instability with $\text{Ra}=0$, $\epsilon=0.1$, $\textrm{Ek}=5\cdot10^{-5.5}$. Right: elliptical instability and convection with $\text{Ra}=6\text{Ra}_c$, $\epsilon=0.1$, $\textrm{Ek}=5\cdot10^{-5.5}$.}
     \label{fig:vorticity_flowpictures}
\end{figure*}

We begin our discussion of simulation results by presenting the $z$-averaged vertical vorticity $\langle\omega_z\rangle_z$ of the flow in Fig.~\ref{fig:vorticity_flowpictures}, taken from a snapshot at $t=0.08$ in a simulation in which only the elliptical instability and its associated nonlinear dynamics are present with $\epsilon=0.1$, $\text{Ra}=0$ (left) and a simulation in which both the elliptical instability and the convective instability, as well as both their associated nonlinear dynamics, are present with $\epsilon=0.1$, $\text{Ra}=6\text{Ra}_c$ (right). This time is after the initial saturation of instability, in which an LSV has formed in the flow in both cases. These LSVs are an important feature produced by nonlinear evolution of both elliptical instability and convection. Note that with our chosen aspect ratio and Ekman number the convective LSV emerges when $\textrm{Ra}\gtrsim3\textrm{Ra}_c$. Thus they emerge at similar values of the Rayleigh number as the convective LSV in \citet[][]{CelineLSV}, which can be readily seen upon taking into account the factor of $\approx8.7$ included in the definition of $\textrm{Ra}_c$, as a result of which the parameter $\tilde{\textrm{Ra}}=\textrm{Ra}\textrm{Ek}^{4/3}\gtrsim20$ from their work\citep[][]{CelineLSV} translates to: $\textrm{Ra}\gtrsim20\textrm{Ek}^{-4/3}\approx2.5\textrm{Ra}_c$.
The vortices are centred in these images for clarity. The nonlinear evolution of the elliptical instability in the left panel creates a cyclonic vortex and a smaller and weaker anticylonic vortex (at the corners, noting the periodic boundaries). In the right panel, the convection dominates the flow, and as a result the convective LSV dominates and results in a single cyclonic vortex, with a primarily anticyclonic background. The vorticity in the centre of this vortex is larger than the vorticity in the centre of the one in the left panel. In the right panel, small-scale convective eddies are present throughout the box, making the flow appear much noisier compared to the elliptical instability in isolation.

\citet{Barker2013} found that these vortices are produced by the nonlinear saturation of the elliptical instability and play a key role in the predator-prey behaviour of the inertial waves and vortices that they observed. They simulated cubic boxes ($L_x=d=1$) and tall thin boxes $(L_x/d<1)$, but the dynamics in wider boxes $(L_x/d>1)$, such as those that are typically used to study convection, were not examined there.

\begin{figure}
     \centering
     \begin{minipage}[b]{0.45\textwidth}
         \centering
         \includegraphics[width=\textwidth, height=50mm]{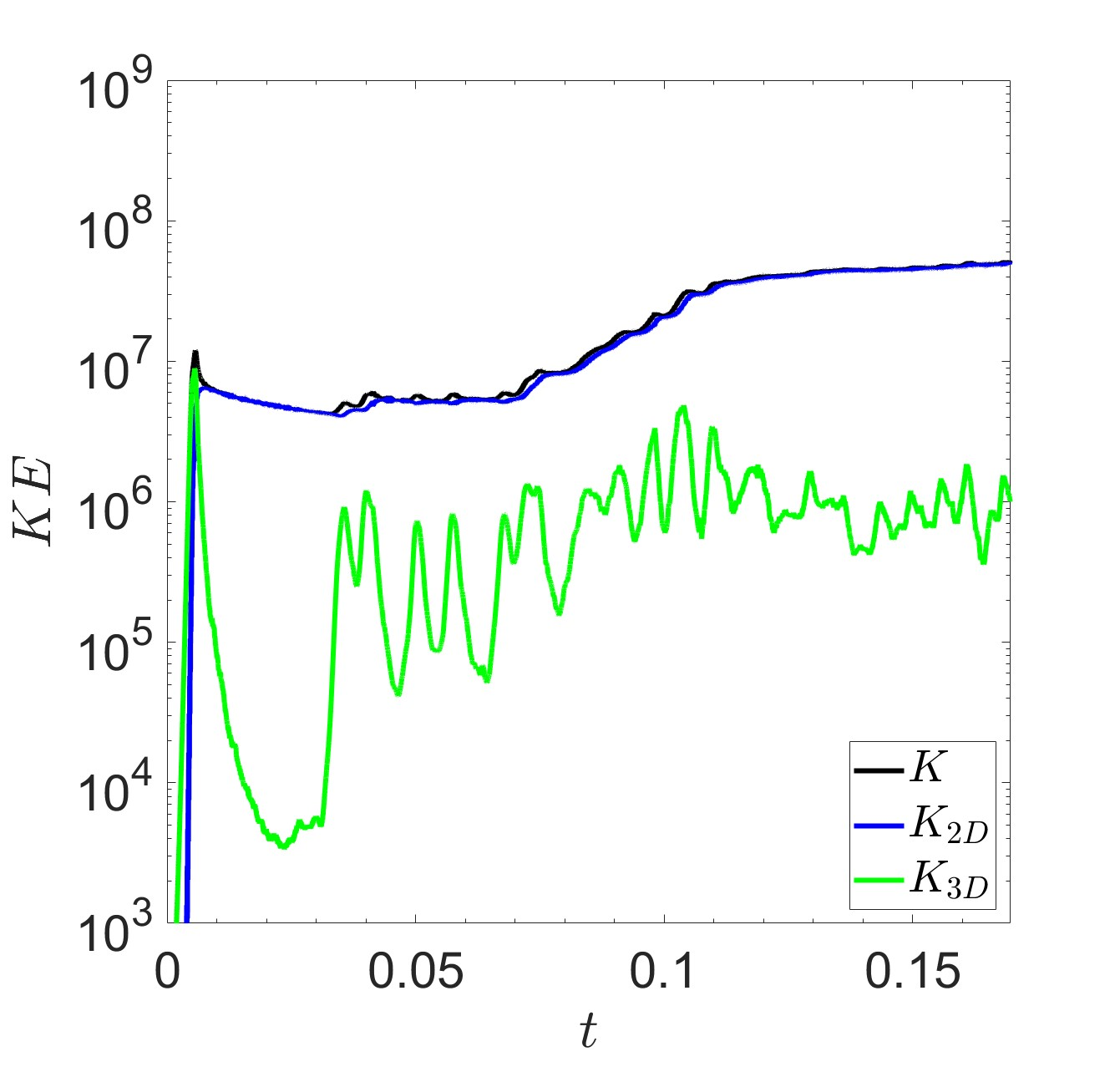}
         \label{fig:KEellipinstab}
     \end{minipage}
     \hfill
     \begin{minipage}[b]{0.45\textwidth}
         \centering
         \includegraphics[width=\textwidth, height=50mm]{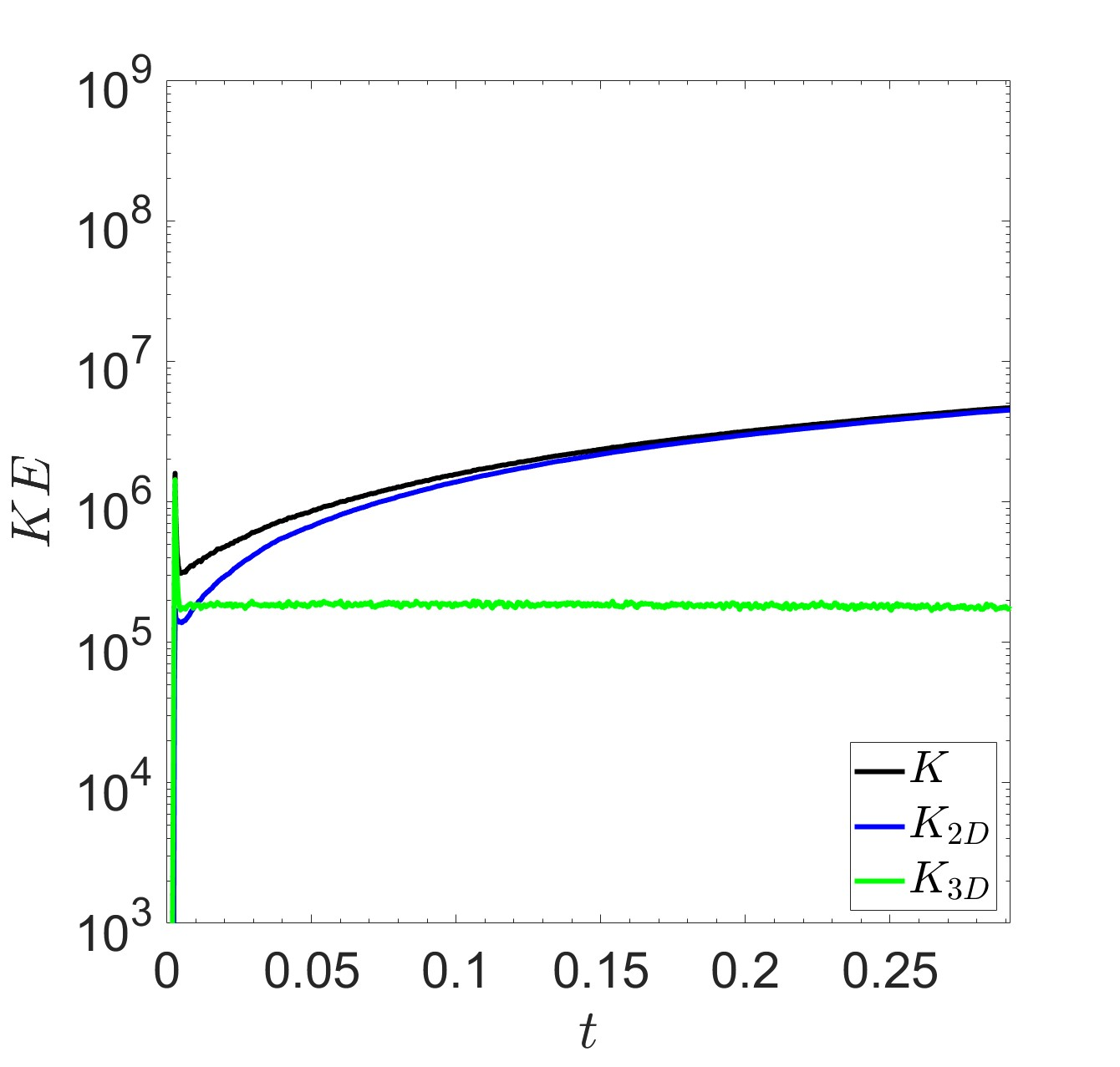}
         \label{fig:KErotconv}
     \end{minipage}
     \caption{Kinetic energy of simulations of the elliptical instability in isolation (top) with $\text{Ra}=0$, $\epsilon=0.1$ and convection in isolation (bottom) with $\text{Ra}=4\text{Ra}_c$, $\epsilon=0$. The 2D (blue) and 3D (green) components of the energy represent the energy in the vortical motions, and the waves and convective eddies, respectively. The total kinetic energy is plotted as the black line.}
    \label{fig:instabisolation}
\end{figure}
\begin{figure}
     \centering
     \begin{minipage}[b]{0.45\textwidth}
         \centering\includegraphics[width=\textwidth,height=50mm]{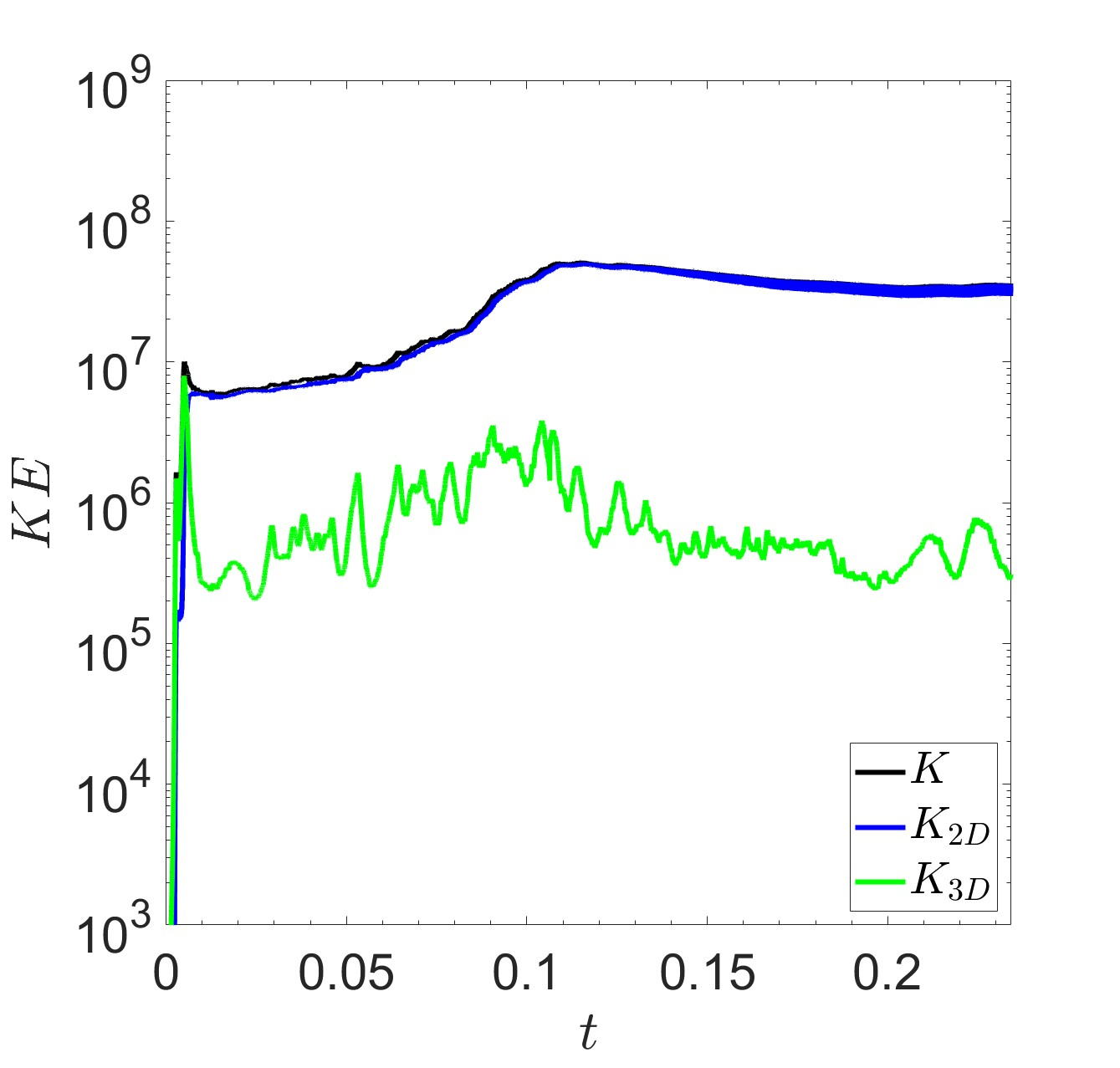}
        \label{fig:KERa4eps0.1}
     \end{minipage}
     \hfill
     \begin{minipage}[b]{0.45\textwidth}
         \centering
         \includegraphics[width=\textwidth, height=50mm]{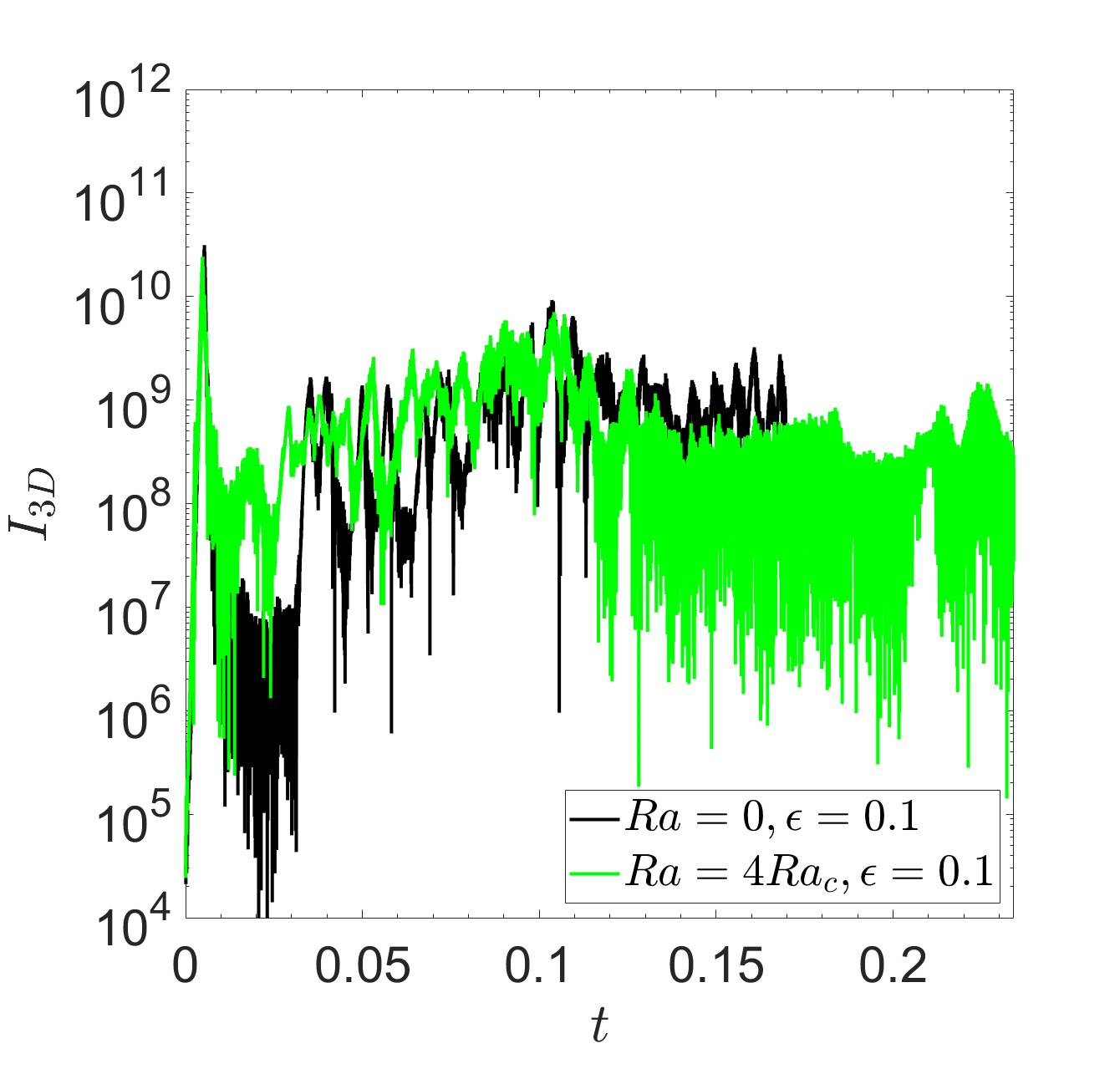}
         \label{fig:I3DRa4eps0.1}
     \end{minipage}
     \caption{Kinetic energy (top) of the elliptical instability and convection with $\text{Ra}=4\text{Ra}_c$, $\epsilon=0.1$, $\textrm{Ek}=5\cdot10^{-5.5}$. The 2D (blue) and 3D (green) components are plotted in addition to the total kinetic energy. The energy injection (bottom) of both simulations with $\text{Ra}=0$, $\epsilon=0.1$ (black) and $\text{Ra}=4\text{Ra}_c$, $\epsilon=0.1$ (green), both with $\textrm{Ek}=5\cdot10^{-5.5}$.}
    \label{fig:I3DKERa4eps0.1}
\end{figure}

The top panel of Fig.~\ref{fig:instabisolation} shows the volume-averaged kinetic energy as a function of time in a simulation of the elliptical instability in a wide box with $\epsilon=0.1$, $\text{Ra}=0$. The vortex observed previously in a 1-by-1-by-1 box by \citet{Barker2013} dominates the flow to an even greater extent in the 4-by-4-by-1 box, as we can see by the dominance of the energy in the 2D component of the flow ($K_{2D}$) at all times after the initial saturation. The 2D, or geostrophic, modes have energies ($K_{2D}$) much larger than the inertial waves (quantified by the energy in the 3D modes, $K_{3D}$), but the inertial waves undergo transient bursts temporarily increasing their energy, though $K_{2D}$ remains dominant unlike in the 1-by-1-by-1 case in \citet{Barker2013}. Each burst in the 3D energy later results in an increase in the 2D energy, indicating energy transfers from inertial waves to vortices. The vortex slowly decays viscously, however, the bursts of inertial wave energy are sufficient to compensate this lost energy, enhancing it further until a quasi-steady state is reached after $t\sim 0.1$. The corresponding energy injection ($I_{3D}$) in the bottom panel of Fig.~\ref{fig:I3DKERa4eps0.1} in black shows that the 3D energy increase is a result of a direct energy injection into those modes. The 2D injection ($I_{2D}$, not shown) is oscillatory in sign and has a small value consistent with 0. Meanwhile, the bottom panel of Fig.~\ref{fig:instabisolation} shows the clear dominance of the LSV in a purely convective simulation with $\epsilon=0$, $\text{Ra}=4\text{Ra}_c$, as the 2D energy of the LSV, and by extension the LSV itself, continuously grows for all times plotted. The LSV will continue to grow until it reaches either the horizontal box-scale or its growth is balanced by viscous dissipation. A steady level of 3D energy is present in this simulation after the initial saturation, representing the energy in the convective eddies.

The interaction of convection and the elliptical instability varies according to the parameters chosen. First, we present a simulation with weak convection but strong ellipticity in Fig.~\ref{fig:I3DKERa4eps0.1}, with $\text{Ra}=4\text{Ra}_c$ and $\epsilon=0.1$. The convection in this simulation leads to an LSV which results in continuous growth of the 2D modes. This however doesn't inhibit the elliptical instability, and a multitude of bursts is observed. The elliptical instability in fact enhances the energy in the 2D modes by at least one order of magnitude compared to the purely convective simulation in the bottom panel of  Fig.~\ref{fig:instabisolation}, as the bursts input more energy into the LSV. There is a continuous decrease of the 2D energy, from $t=0.1$ to $t=0.17$. In this period of time neither the bursts, weakened by the strong vortex, nor the convective eddies provide enough energy to compensate the viscous dissipation. The corresponding energy injection in the bottom panel of Fig.~\ref{fig:I3DKERa4eps0.1} in green roughly matches the energy injection of the purely elliptical simulation, although it is initially maintained at a higher value. From $t=0.11$ onwards, when the LSV has reached its strongest value, the behaviour gradually changes from bursts to an almost continuous energy injection. From $t=0.21$, once the LSV has been sufficiently weakened, bursty behaviour is again observed.

 \begin{figure*}
\subfloat[$K$ of the simulation with $\text{Ra}=1.99\text{Ra}_c$, $\epsilon=0.2$. \label{fig:KE1.99}]{\includegraphics[height=47mm,width= 0.45\textwidth]{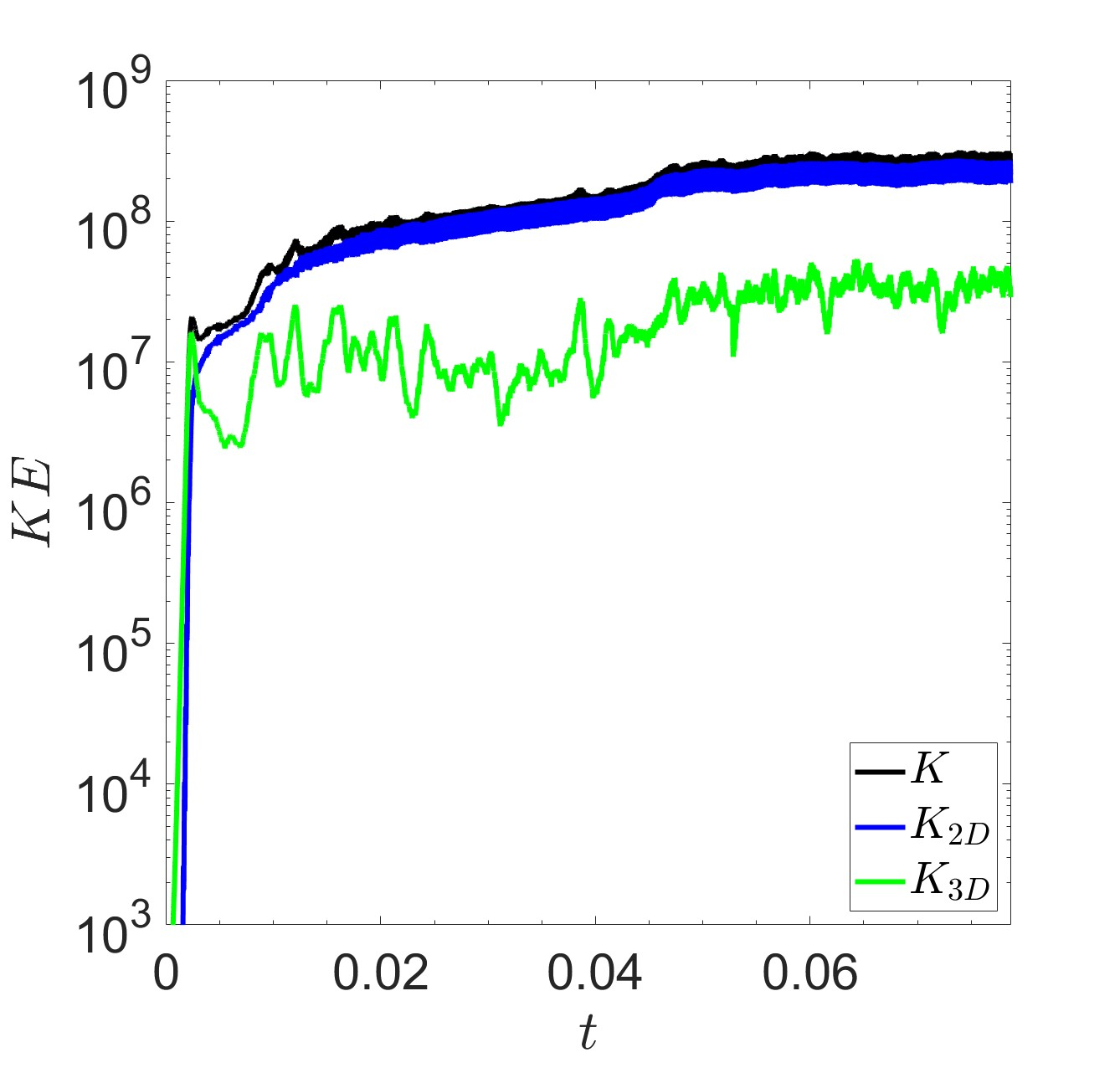}}
\hfill
\subfloat[$I_{3D}$ of the simulation with $\text{Ra}=1.99\text{Ra}_c$, $\epsilon=0.2$. \label{fig:injection1.99}]{\includegraphics[height=47mm,width= 0.45\textwidth]{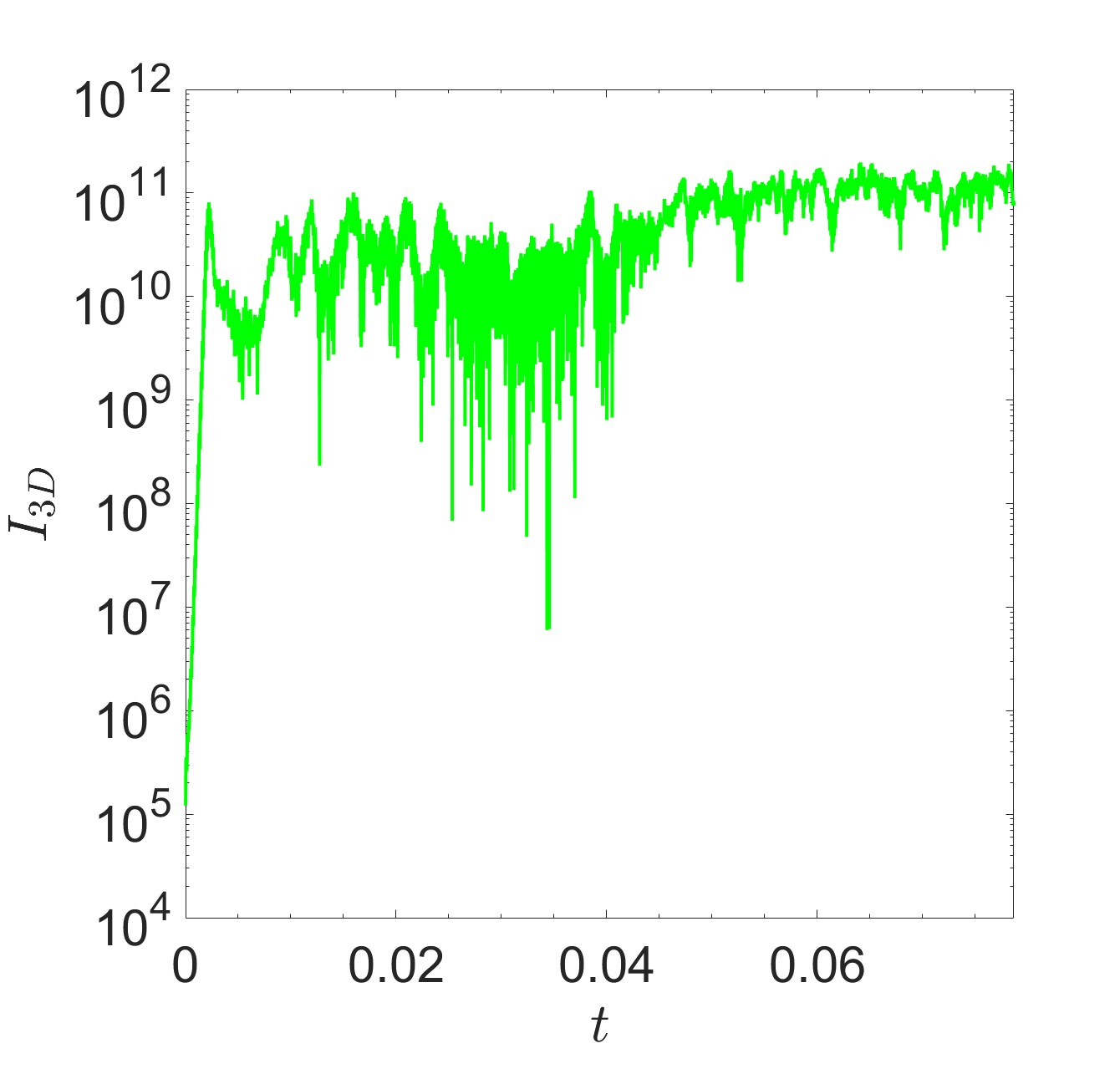}}\\

\subfloat[$K$ of the simulation with $\text{Ra}=6\text{Ra}_c$, $\epsilon=0.1$.\label{fig:KE6,eps0.1}]{\includegraphics[height=47mm,width= 0.45\textwidth]{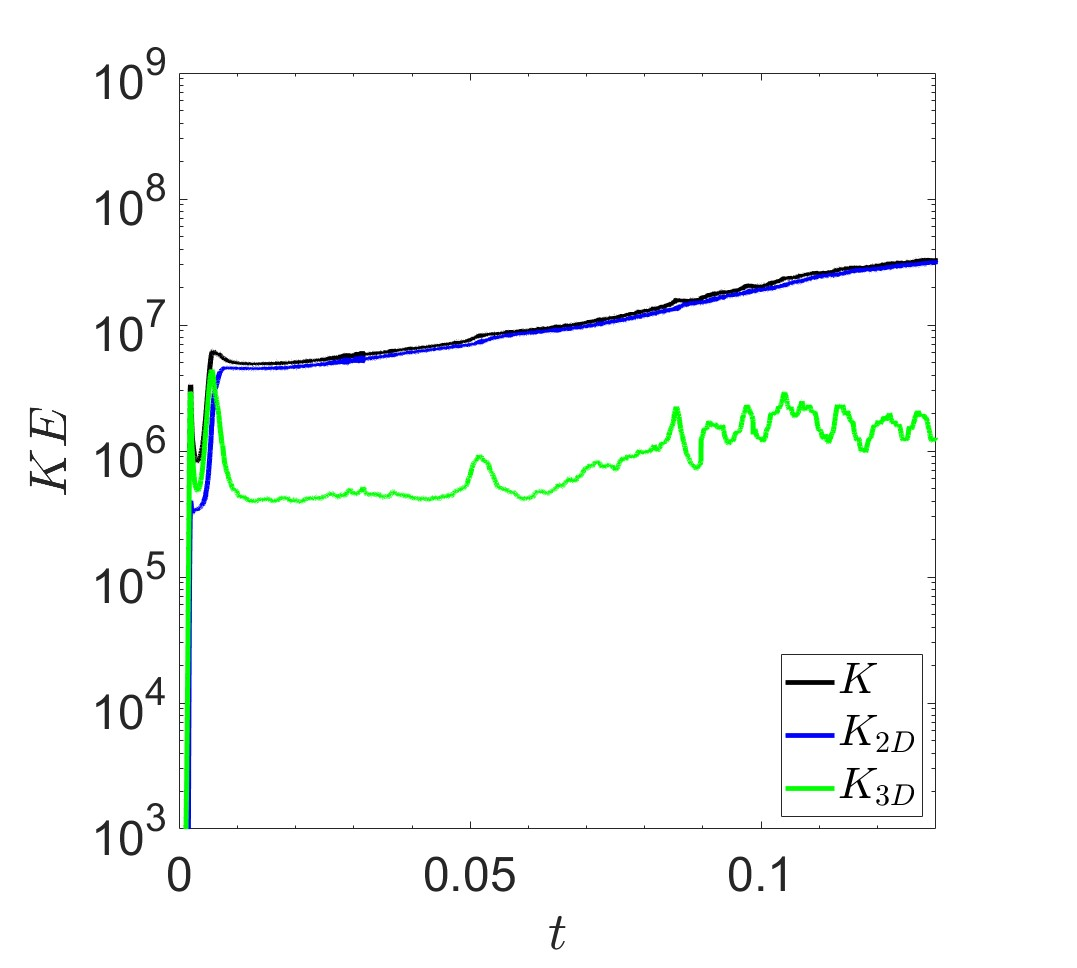}}
\hfill
\subfloat[$I_{3D}$ of the simulation with $\text{Ra}=6\text{Ra}_c$, $\epsilon=0.1$. \label{fig:injection6,eps0.1}]{\includegraphics[height=47mm,width= 0.45\textwidth]{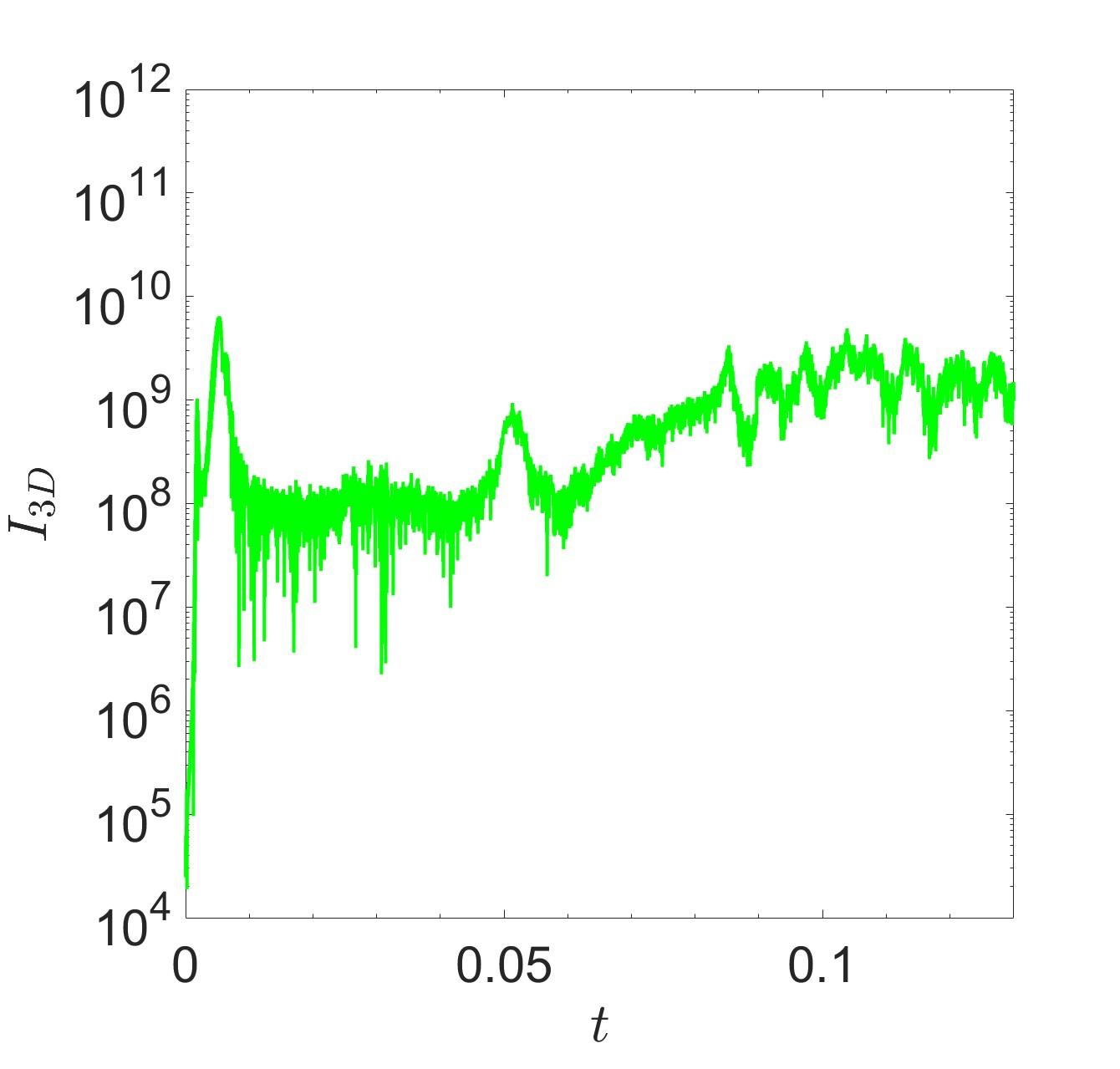}}\\

\subfloat[$K$ of the simulation with $\text{Ra}=6\text{Ra}_c$, $\epsilon=0.05$. \label{fig:KE6,eps0.05}]{\includegraphics[height=47mm,width= 0.45\textwidth]{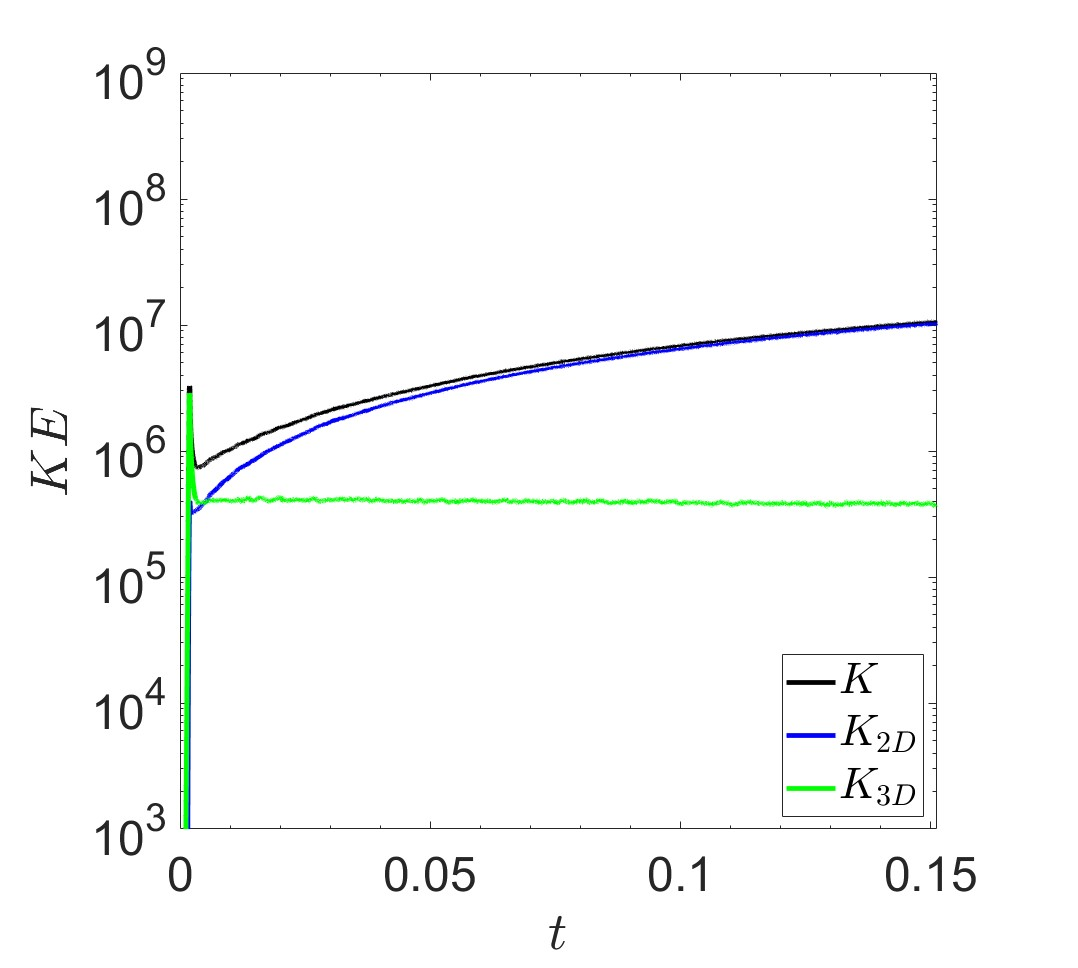}}
\hfill
\subfloat[$I_{3D}$ of the simulation with $\text{Ra}=6\text{Ra}_c$, $\epsilon=0.05$.\label{fig:injection6,eps0.05}]{\includegraphics[height=47mm,width= 0.45\textwidth]{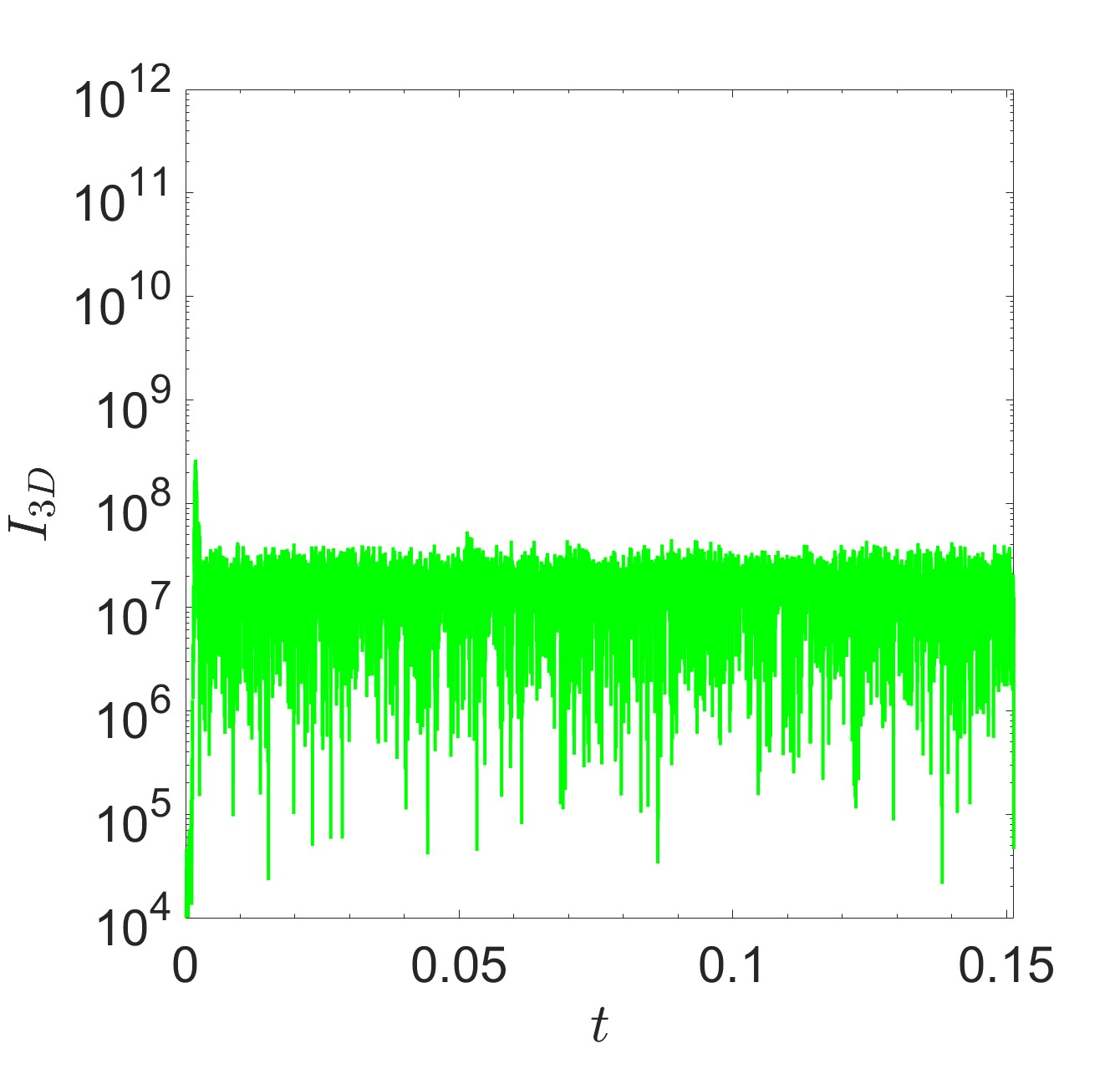}}\\

\subfloat[$K$ of the simulation with $\text{Ra}=20\text{Ra}_c$, $\epsilon=0.1$. \label{fig:KE20}]{\includegraphics[height=47mm,width= 0.45\textwidth]{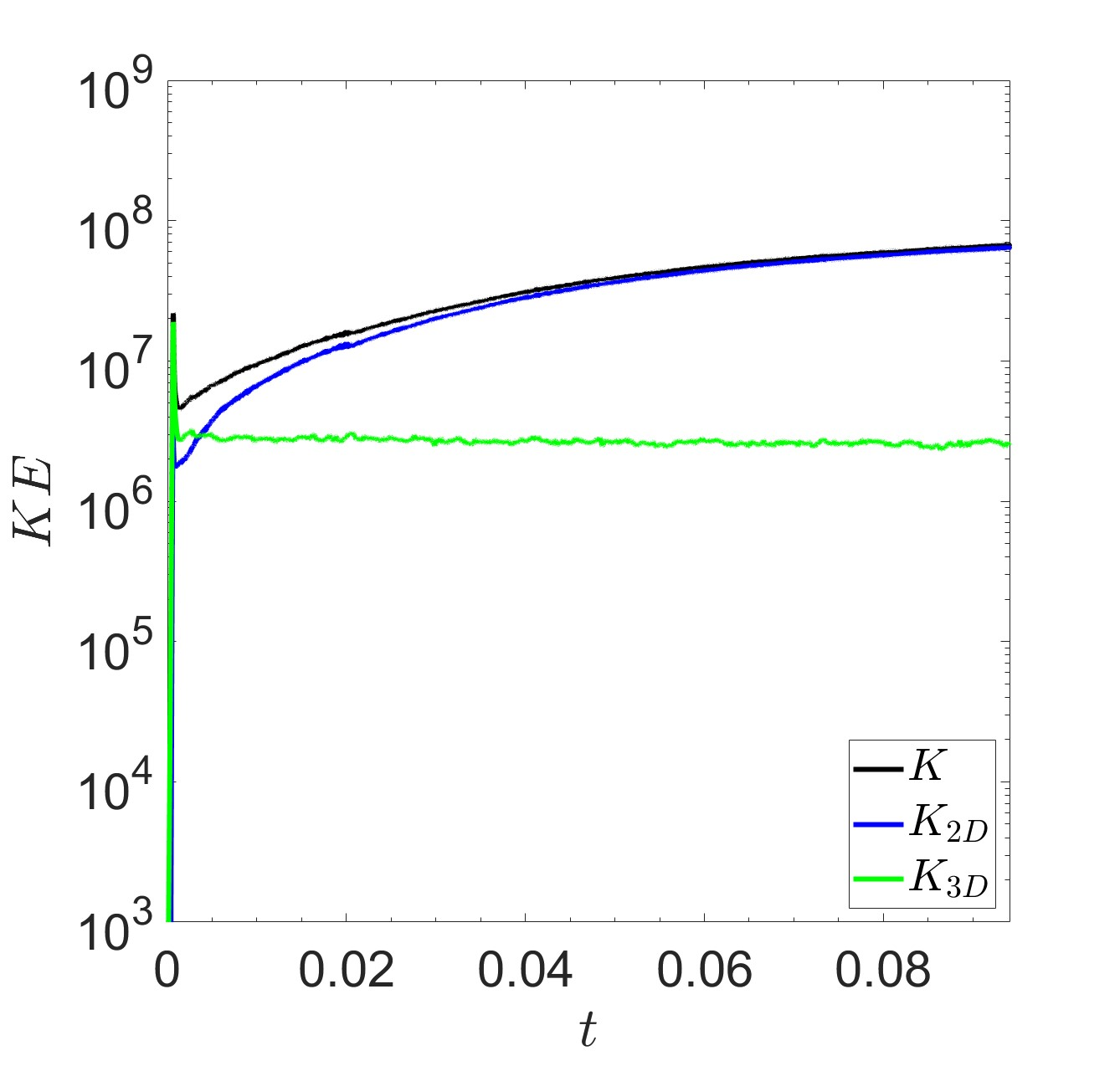}}
\hfill
\subfloat[$I_{3D}$ of the simulation with $\text{Ra}=20\text{Ra}_c$, $\epsilon=0.1$. \label{fig:injection20}]{\includegraphics[height=47mm,width= 0.45\textwidth]{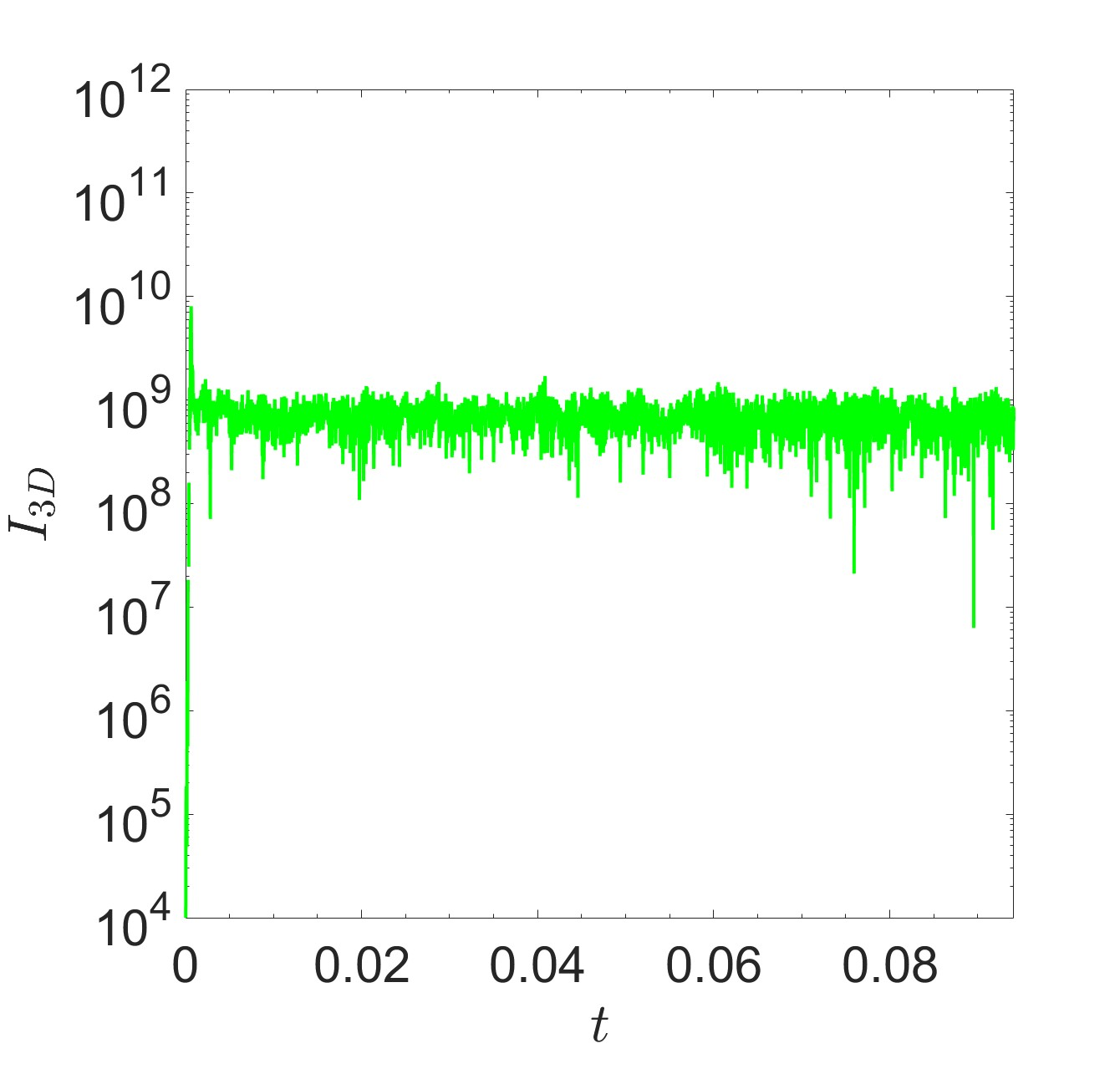}}
     \caption{Kinetic energy (left) and the energy injection contribution $I_{3D}$ (right) for a range of simulations. Convection results in a sustained energy input into the flow from the tidal flow, allowing for sustained tidal dissipation. If the convection is sufficiently strong compared to the elliptical instability, it can suppress the bursts leaving only this sustained energy injection.}
    \label{fig:allthefiguresKEI}
\end{figure*}

\subsection{Varying strength of convective driving and ellipticity}
\label{vatiation}

In Fig.~\ref{fig:allthefiguresKEI} we present results for a range of values of $\text{Ra}/\text{Ra}_c$ and $\epsilon$ with $\textrm{Ek}=5\cdot10^{-5.5}$. The figures on the left show the time evolution of the kinetic energy components, and those on the right the energy injection term $I_{3D}$. Fig.~\ref{fig:KE1.99} and Fig.~\ref{fig:injection1.99} show simulations with $\text{Ra}=6\text{Ra}_c$ and $\epsilon=0.2$. \citet[][]{Barker2013} observed a change in behaviour at $\epsilon\gtrsim 0.15$, seeing a sharp increase in the frequency and strength of the elliptical instability bursts. The 3D component of the kinetic energy is maintained at a higher level in this case, but is still lower than the energy in the 2D component. The energy injection features many bursts in a short time frame, with multiple bursts injecting energy at the same rate as the initial burst in linear growth phase. The increased burst frequency also leads to a sustained energy injection throughout the simulation. There appears to be a secondary transition around $t=0.045$ where the energy injection increases steeply and maintains a significant non-zero energy injection, much larger than the initial burst. We observe a correspondingly higher minimum level of the 3D component of the energy during this simulation.

The kinetic energy in Fig.~\ref{fig:KE6,eps0.1} shows that increasing the Rayleigh number, i.e. making the convection stronger compared to the elliptical instability, results in fewer visible bursts into the 3D component in the first half of the simulation compared with the top panel of Fig.~\ref{fig:I3DKERa4eps0.1}, and the total kinetic energy is further dominated by the 2D component. The increased convection strength, and therefore (for our parameters) stronger LSV compared with Fig.~\ref{fig:I3DKERa4eps0.1}, drowns out most of the bursts from the elliptical instability. The reduced presence of the elliptical instability is also clearly visible from the $I_{3D}$ term in Fig.~\ref{fig:injection6,eps0.1}, showing considerably fewer bursts in the first half of the simulation, decreasing the tidal dissipation. On the other hand, a ``floor value" corresponding to a non-zero continuous energy injection arises, which is most clearly visible in between bursts. This sustained energy injection arises from the interaction between the convection and the equilibrium tidal flow.

Fig.~\ref{fig:KE6,eps0.05} and especially Fig.~\ref{fig:injection6,eps0.05} confirm this sustained injection occurs as the convection is strengthened relative to the elliptical instability. The ellipticity has been reduced to $\epsilon=0.05$, thus weakening the elliptical instability (whose growth rate is proportional to $\epsilon)$. As a result, the bursts from the elliptical instability have vanished, with only a short initial burst remaining, after which a continuous injection arises. These simulations suggest there is a point at which the convection, with both its LSVs and its resulting effective viscosity acting to damp the inertial modes, overpowers the elliptical instability such that the bursts are completely suppressed.

Increasing the Rayleigh number to $\text{Ra}=20\text{Ra}_c$ and maintaining $\epsilon=0.1$ instead of decreasing $\epsilon$ in Fig.~\ref{fig:KE20} and Fig.~\ref{fig:injection20} leads to similar behaviour, with no bursty behaviour for the elliptical instability and instead a sustained energy injection. Thus, increasing Ra inhibits bursts even if they were present at lower values of the Rayleigh number. Additionally, the sustained injection term has increased by a factor of about 20 compared to Fig.~\ref{fig:injection6,eps0.05}. The sustained energy injection then increases with Ra and $\epsilon$. Thus, introducing convection has two effects: 1) the bursts of elliptical instability are suppressed to a greater extent as the strength of convection increases relative to the strength of the elliptical instability, and 2) a sustained energy injection arises from the interaction between convection and the background flow. 

\subsection{Heat transport modification by elliptical instability}
\label{heat}

\begin{figure}
     \centering
     \begin{minipage}[b]{0.45\textwidth}
         \centering
    \includegraphics[width=0.9\linewidth]{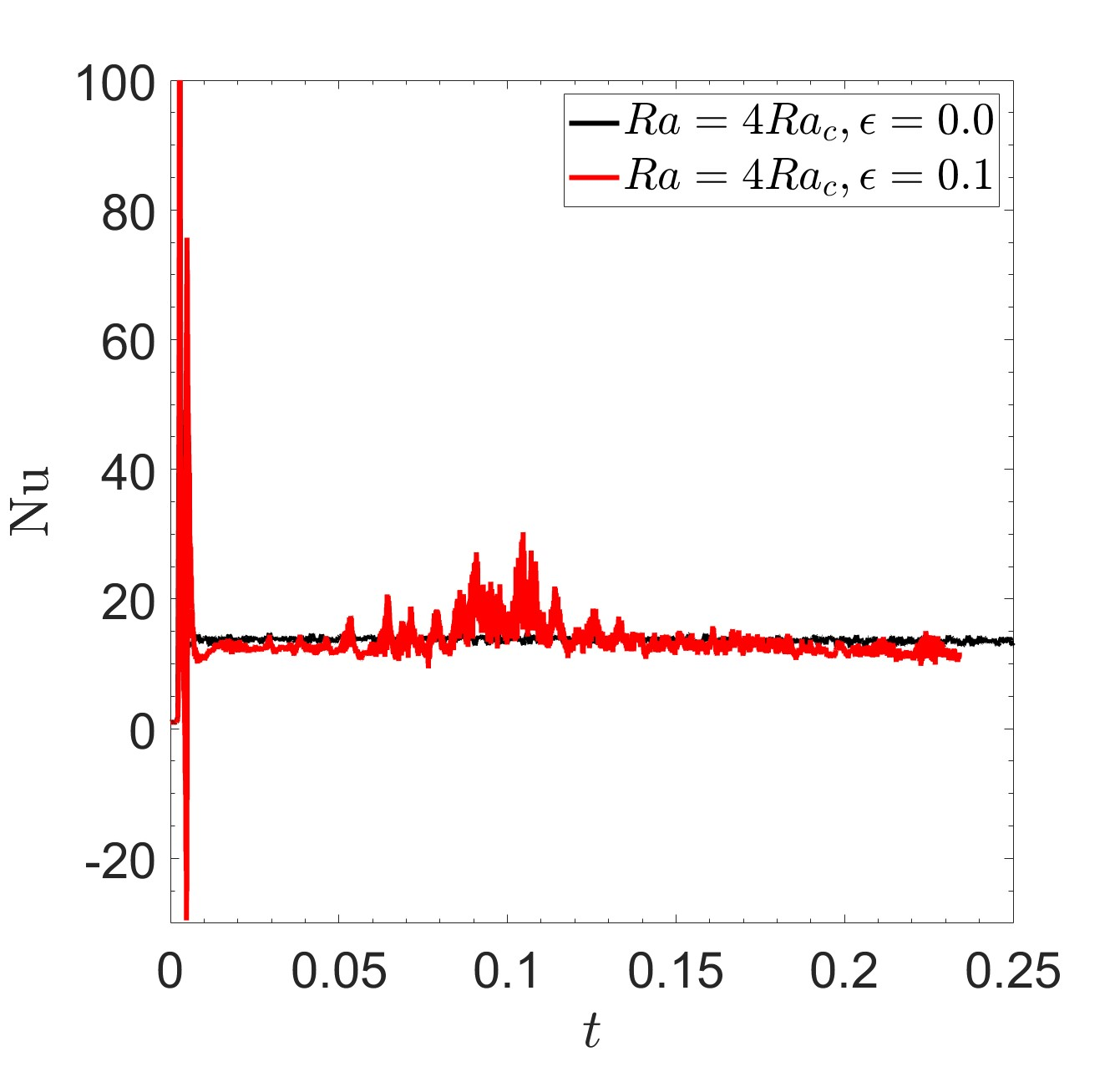}
    \label{fig:NusseltRa=4}
     \end{minipage}
     \hfill
     \begin{minipage}[b]{0.45\textwidth}
        \centering
    \includegraphics[width=0.9\linewidth]{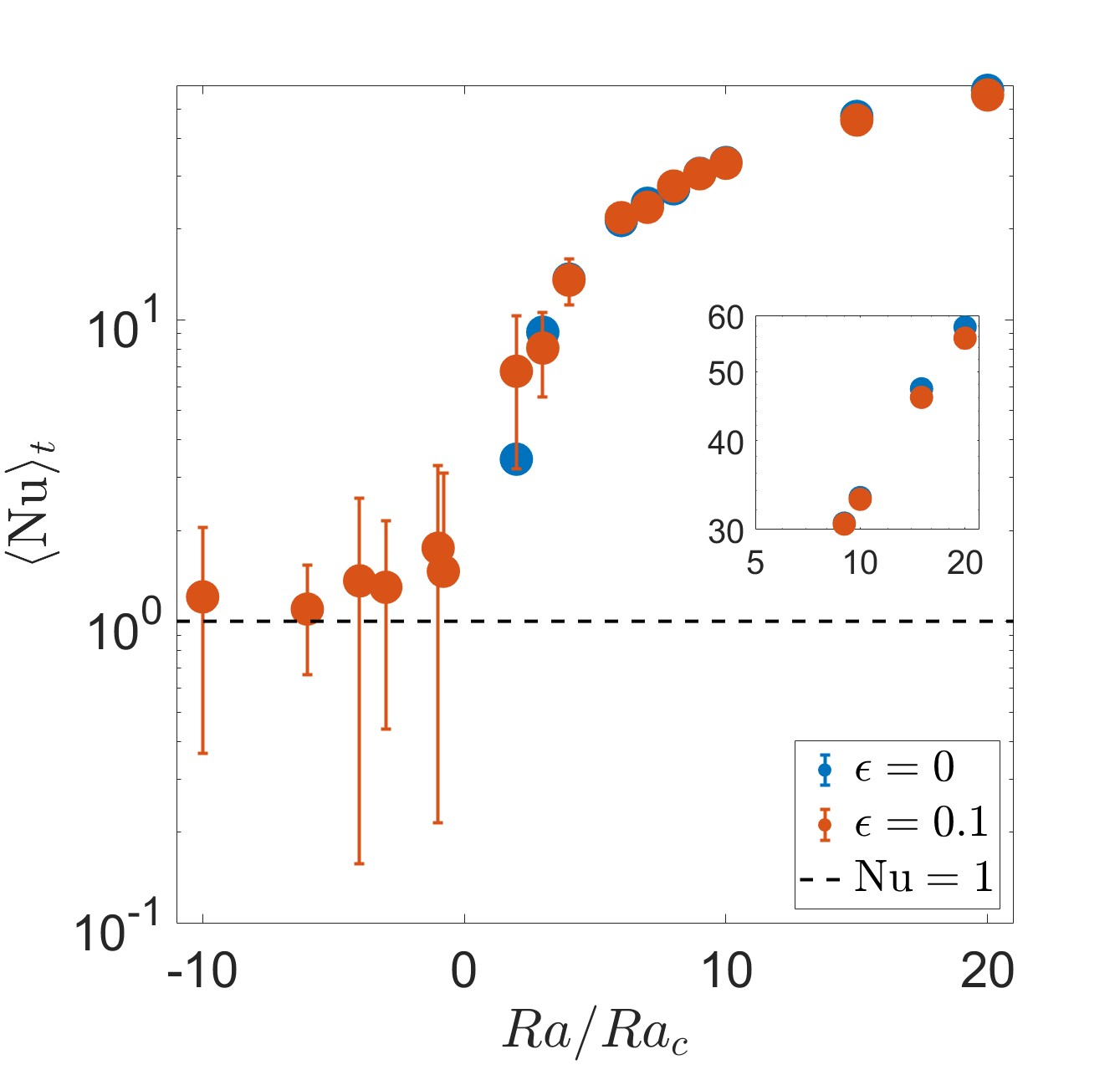}
    \label{fig:Nusseltimeavg}
     \end{minipage}
     \caption{Top: Nusselt number as a function of time at $\text{Ra}=4\text{Ra}_c$ with (red) and without (black) the elliptical instability with $\epsilon=0.1$, $\textrm{Ek}=5\cdot10^{-5.5}$. Bottom: Time average of the Nusselt number with (orange) and without (blue) the elliptical instability. The elliptical instability results in heat transport when it operates, even in the stably stratified regime ($\text{Ra}<0$). $\text{Nu}=1$ means there is no heat transport by fluid motions (black dashed line).}
    \label{fig:Nusselt}
\end{figure}

A further effect of the elliptical instability is to modify heat transport. The inertial waves excited by the elliptical instability are capable of transporting heat in the system \citep[][]{Lavorelexperimentalellip,Cebron2010}. The elliptical instability can occur and affect heat transport in both stably stratified and convectively unstable fluids (see Eq.~\ref{eq:kerswellgrowthconvrotat}). We observe that the heat transport is tied to the bursts of elliptical instability, and is similarly bursty in its operation. The Nusselt number represents the heat transport in the simulation and is shown as a function of time for $\epsilon=0.1$, $\textrm{Ra}=4\textrm{Ra}_c$ in the top panel of Fig.~\ref{fig:Nusselt}. This shows the two-sided effects of the elliptical instability in this simulation. The bursts increase heat transport, temporarily increasing the Nusselt number. Then, the enhanced cyclonic vortex as a result of the elliptical instability slightly decreases the heat transport, similar to the reduction observed in the presence of convective LSVs by \citep[][]{CelineLSV,FavierLSV}, which they proposed occurs because cyclonic vortices act to effectively increase the rotation, thus further constraining the vertical motions, and as a consequence the heat transport, according to the Taylor-Proudman theorem\citep[e.g.][]{CurrieconvRMLT}.

In stably stratified fluids ($\text{Ra}<0$), or in convectively stable but unstably stratified fluids ($\text{Ra}<\mathrm{Ra}_c$), in the absence of the elliptical instability, there are no sustained vortices or vertical motions. The heat transport in this regime would then be purely conductive (at long times, after decay of transients) with $\textrm{Nu}=1$. The vertical motions introduced by the elliptical instability in either regime may though transport heat during the bursts, even for convectively stable fluids.

The effects of the elliptical instability on heat transport have been quantified using time averages of the Nusselt number in the bottom panel of Fig.~\ref{fig:Nusselt}, for simulations performed with $\epsilon=0$ and $\epsilon=0.1$. Due to the absence of bursts of the elliptical instability, there is no observable enhancement to the heat transport when the convection is strong. In fact, since the ellipticity of the equilibrium flow slightly enhances the 2D energy of the vortex it is likely to result in a slightly diminished total heat transport, possibly supported by the reduced Nusselt number in the inset at $\text{Ra}=15\text{Ra}_c$ and $\text{Ra}=20\text{Ra}_c$. At low positive Rayleigh numbers, the elliptical instability plays a major role in enhancing or hindering the heat transport as a function of time. For very weak convection, in which there is no convectively-generated LSV, the elliptical instability enhances the net heat transport strongly, while at higher convection strengths it cancels or slightly decreases the heat transport. Finally, the elliptical instability does indeed produce heat transport in stably stratified regimes, although the additional heat transport is highly variable in time, only occurring during a burst in energy injection of the elliptical instability, and decreases as the stratification increases (i.e., $\mathrm-Ra$ increases). This is represented by the Nusselt number tending towards one on average as the fluid becomes more stably stratified, where we note that all cases plotted are linearly unstable.\\

\subsection{Growth rate of 2D convective vortices}
\label{growth2D}

\begin{figure}
    \centering
    \includegraphics[width=0.8\linewidth]{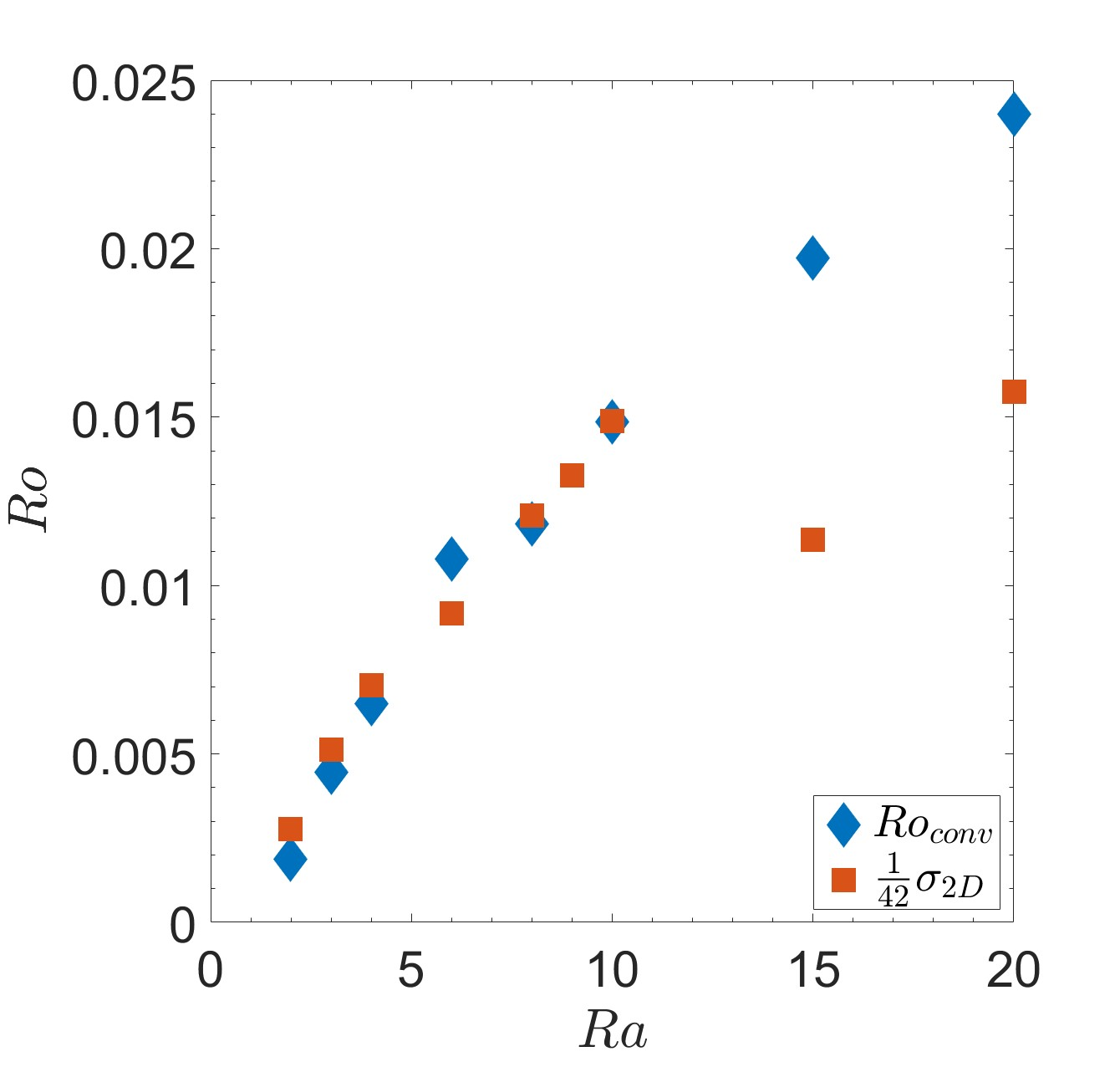}
    \caption{Convective Rossby number (blue) obtained from the vertical convective velocity, compared with the scaled growth rate of the vortex (orange). The growth rate is scaled by dividing it by a factor of $\frac{1}{42}$; the scaled growth rate and convective Rossby number agree up to $\text{Ra}=10\text{Ra}_c$.}
    \label{fig:sigma2d}
\end{figure}

In Fig.~\ref{fig:sigma2d} we examine the growth rate of the 2D energy $K_{2D}$ produced by convection during the initial burst as a function of the Rayleigh number (i.e.~we determine $\sigma_{2D}$ defined by $K_{2D}\propto \exp(\sigma_{2D}t)$ in the initial phases of the simulation). We examine here simulations with $\epsilon=0$ and ones with higher $\epsilon$, but all in the sustained energy injection regime. We find that non-zero ellipticities have little effect on this growth rate, modifying it slightly but not significantly. The growth rate $\sigma_{2D}$ is observed to increase as the Rayleigh number increases. Deviations in this growth behaviour are observed at higher Rayleigh numbers ($>10\text{Ra}_c$). A potential explanation could lie in the slow increase of the growth rate of the 2D component as the 3D component of the kinetic energy grows. Because the growth of the 3D component is rapid at these values of the Rayleigh number, the non-linear breakdown happens before the 2D component reaches its maximum growth rate. Comparing the scaled 2D growth rate with the convective velocity in each simulation (plotted as the blue diamonds), and by extension the convective Rossby number of our simulations ($\mathrm{Ro}_{conv}=\sqrt{\langle u^2_z\rangle}/(2\Omega d$)), we find good agreement. This implies that the growth of the 2D modes scales with the Rossby number in the initial regime, i.e. $\sigma_{2D}\propto \mathrm{Ro}_{conv}$. Note that this is not consistent with the predicted scaling due to weak nonlinear interactions between pairs of inertial waves (in the absence of convection) of e.g.\ \citet[][]{kerswell1999_intwavbreakdown}, which would predict $\sigma_{2D}\propto \mathrm{Ro}^2$. It however agrees with the prediction for generation of an LSV from interactions between inertial waves and geostrophic modes at sufficiently large Rossby number \cite{LeReun2020}, though a theory for convective LSVs has not yet been presented.

\section{Analysis of the sustained energy injection}
\label{sec:sustained}

\subsection{Origin and parameter regime for sustained tidal energy injection}
\label{sec:sustained parameters}

\begin{figure}
\centering
    \includegraphics[width=0.88\linewidth]{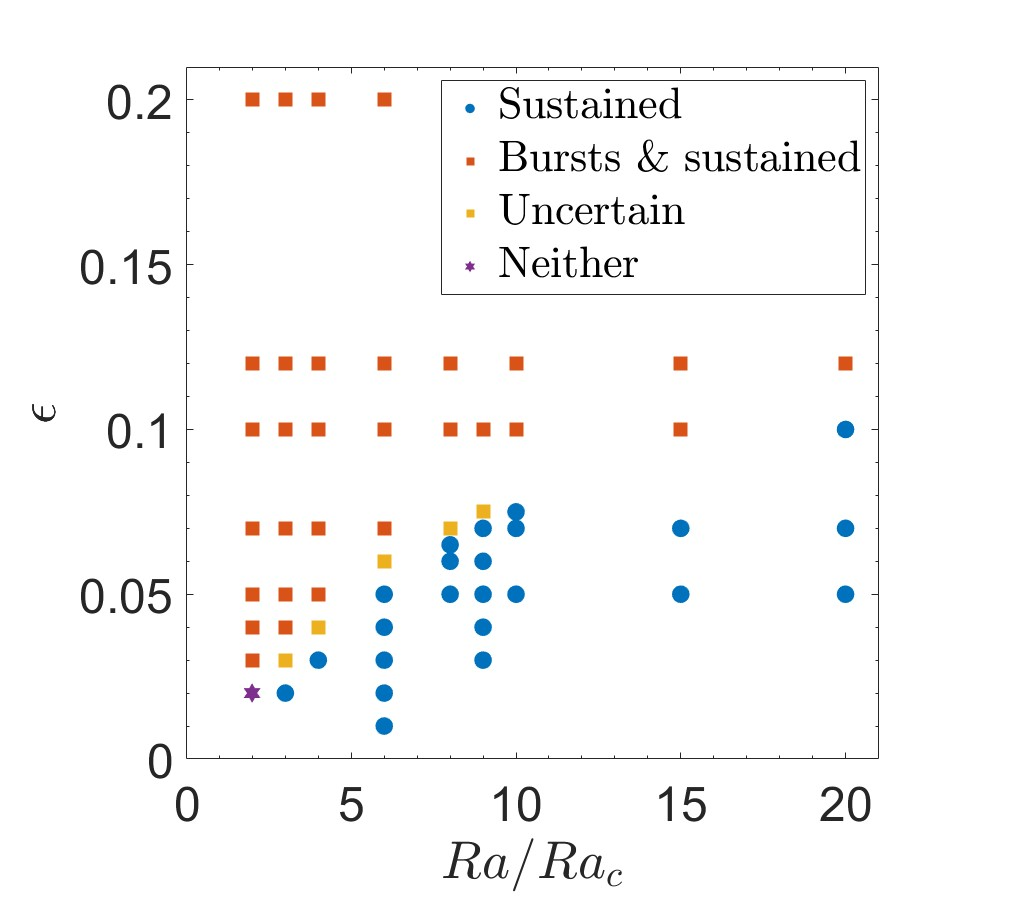}
    \caption{A `phase diagram' showing the observed behaviour in our simulations. Simulations where sustained behaviour but no bursts of elliptical instability are observed are marked in blue, those with clear bursts of the elliptical instability (and possible additional sustained behaviour) are marked in orange. The uncertain markers represent simulations close to where the regime transition is likely situated. The case $\text{Ra}=2\text{Ra}_c$, $\epsilon=0.02$ is marked in purple, as it shows neither sustained injection nor bursts of elliptical instability.}
    \label{fig:phasediagram}
\end{figure}

We have performed a range of simulations in which $\epsilon$ and Ra are varied for $\textrm{Ek}=5\cdot10^{-5.5}$ to determine when sustained energy injection (versus burstiness) is obtained. First, a `phase diagram' was created based on these simulations, plotted in  Fig.~\ref{fig:phasediagram}, which indicates in which simulations we observe bursts of the elliptical instability, and in which we observe sustained energy injection. Simulations containing any bursts of the elliptical instability, even just at the earliest times have been labelled bursty (orange markers), while simulations containing no such bursts have been labelled sustained (blue markers). Bursty simulations may still feature a sustained energy injection, however in the interest of determining where the bursts of the elliptical instability still exist we have classified them as bursty here. Some of the simulations are very difficult to tell by eye whether they are bursty or sustained, and have therefore been labelled as uncertain (yellow markers). The transition between the two `phases' is likely situated close to these markers. Finally, one point has been labelled `neither', namely the point corresponding to the simulation with $\text{Ra}=2\text{Ra}_c$, $\epsilon=0.02$. This simulation features no bursts as the elliptical instability is too weak to operate at this level of convection. However, it also doesn't display any sustained energy injection. Of further note is that at this supercriticality ($\text{Ra}/\text{Ra}_c$) the LSV is absent, which was also observed by \citet[][]{CelineLSV} to be independent of Ekman number.

\begin{figure*}
    \centering
     \hfill
     
     \subfloat[$t=0.05$, $\text{Ra}=0$, $\epsilon=0.1$.\label{fig:fluct3Dburst}]{\includegraphics[width= 0.45\textwidth]{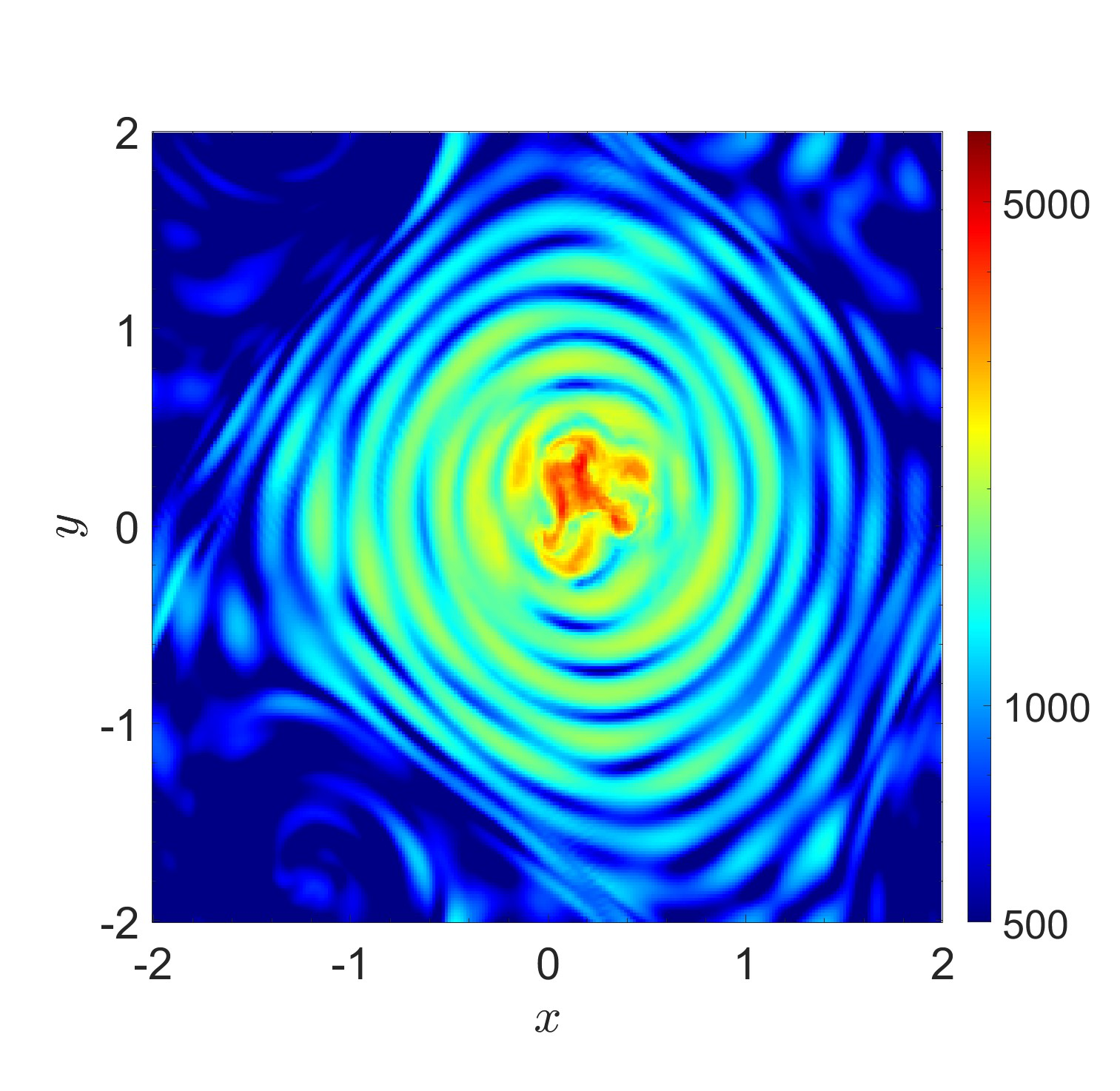}}
     \hfill
     \subfloat[$t=0.1$, $\text{Ra}=6$, $\epsilon=0.1$.\label{fig:Fluct3Dconvburst}] {\includegraphics[width=0.45\textwidth]{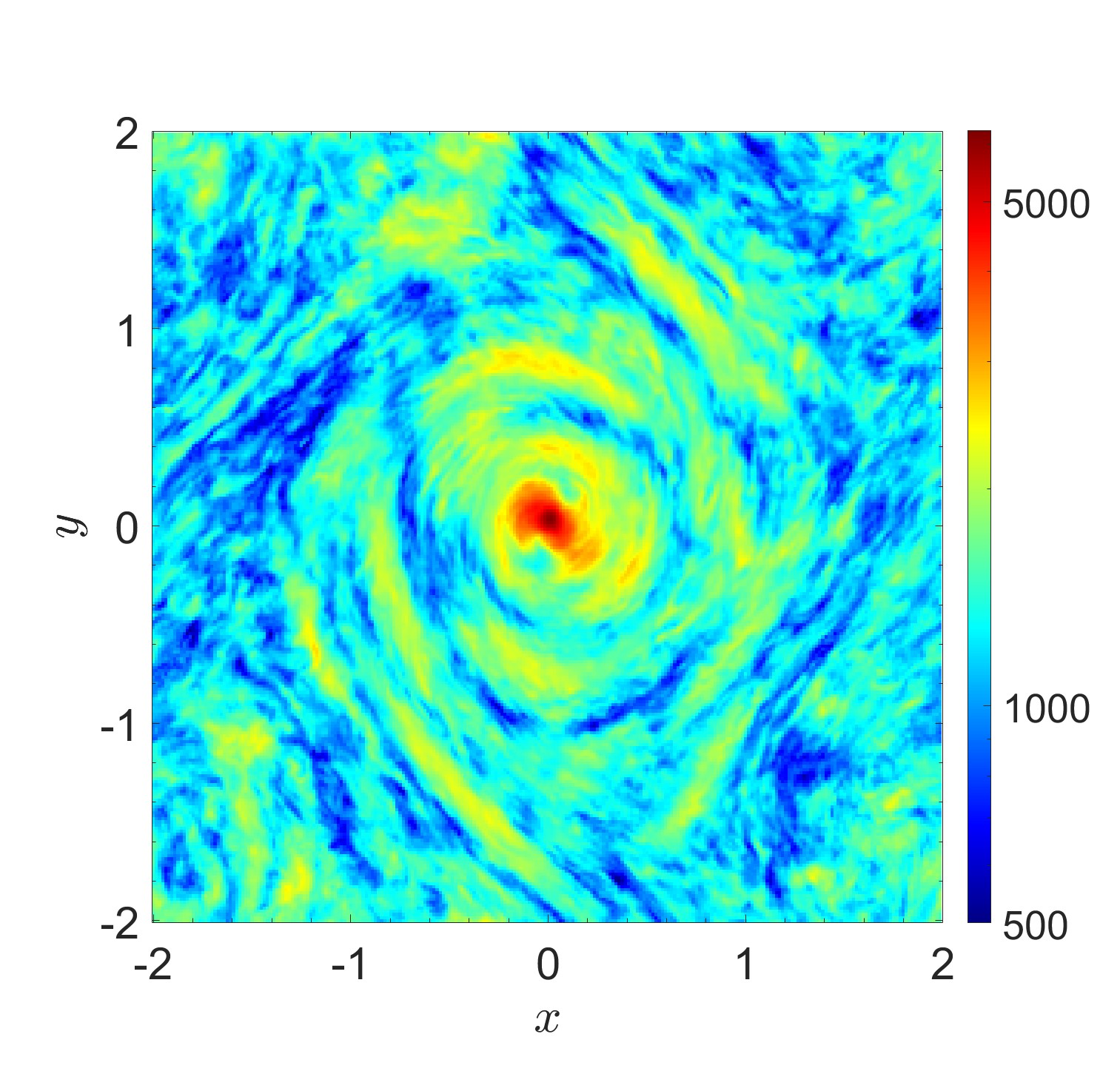}}
   
     \caption{Total velocity magnitude $u_{3D}$, in a burst of instability after the LSV has formed, showing its localisation within the vortices.  Left: elliptical instability in isolation with $\epsilon=0.1$. Right: elliptical instability and convection. The vortices are centred for clarity.}
     \label{fig:Fluct3Dall}
\end{figure*}

To explain the absence of bursts of elliptical instability we investigate the 3D modes in the flow in both real space and Fourier space. In real space we use a method akin to the one used in \citet[][]{Favier2019subcritLSV} to determine the 3D flow. The 3D velocity components are obtained by taking the difference between the total velocity and the $z$-averaged horizontal (or 2D) velocity components:
\begin{equation}
\begin{aligned}
u_{x,3D}&=u_x-\langle u_x\rangle_{z},\\
u_{y,3D}&=u_y-\langle u_y\rangle_{z},\\
u_{z,3D}&=u_z.
\end{aligned}
\end{equation}
Here, $\langle u_x\rangle_z$,$\langle u_y\rangle_z$ are the depth averaged $x$ and $y$ components of the velocity, respectively, i.e. the horizontal velocity components. From this, the magnitude of the 3D velocity is calculated. \citet[][]{Favier2019subcritLSV} showed that the convective LSV suppresses 3D motions, resulting in an area with lower 3D velocities inside the vortex. Since we observe similar LSVs, such a suppression of 3D modes might contribute to the suppression of the elliptical instability. 

First, we examine a case of the elliptical instability in isolation using this method in Fig.~\ref{fig:fluct3Dburst}, with parameters $\epsilon=0.1$, $\text{Ra}=0$. We show results for the total velocity magnitude $u_{3D}$ at $t=0.05$  at the mid-plane ($z=d/2$) during a burst of instability, after vortices have formed following initial saturation. In Fig.~\ref{fig:fluct3Dburst} we see that the power during a burst is concentrated inside the vortex, particularly in its centre. 

On the right panel, in Fig.~\ref{fig:Fluct3Dconvburst}, we show a similar result from a simulation with the elliptical instability and convection, using the parameters $\epsilon=0.1$, $\text{Ra}=6\text{Ra}_c$ during a burst of the elliptical instability at $t=0.1$. We again observe that the burst is concentrated in the centre of the vortex. Examining the same simulation in the absence of a burst of the elliptical instability as well as a simulation with $\text{Ra}=6\text{Ra}_c$, $\epsilon=0.05$ (not pictured) reveals that there is no such concentration of power in the centre. Instead the same picture of suppression of convective 3D motions is obtained as found by \citet[][]{Favier2019subcritLSV}. 

Bursts of the elliptical instability are thus primarily concentrated in the centre of LSVs according to these results. The LSV is therefore expected to have a strong effect on the growth of the elliptical instability, as the free inertial waves existing within these vortices will differ from those of the original flow, thus acting as a constraint to detune the elliptical instability. 

\subsection{Frequency-wavenumber spectrum analysis}
\label{sec:Freqspec}

\begin{figure*}
    \centering
        \subfloat[Linear growth phase of the simulation with $\text{Ra}=0$, $\epsilon=0.05$. \label{fig:omega-thetaeps=0.05lin}]{\includegraphics[width=0.45\textwidth]{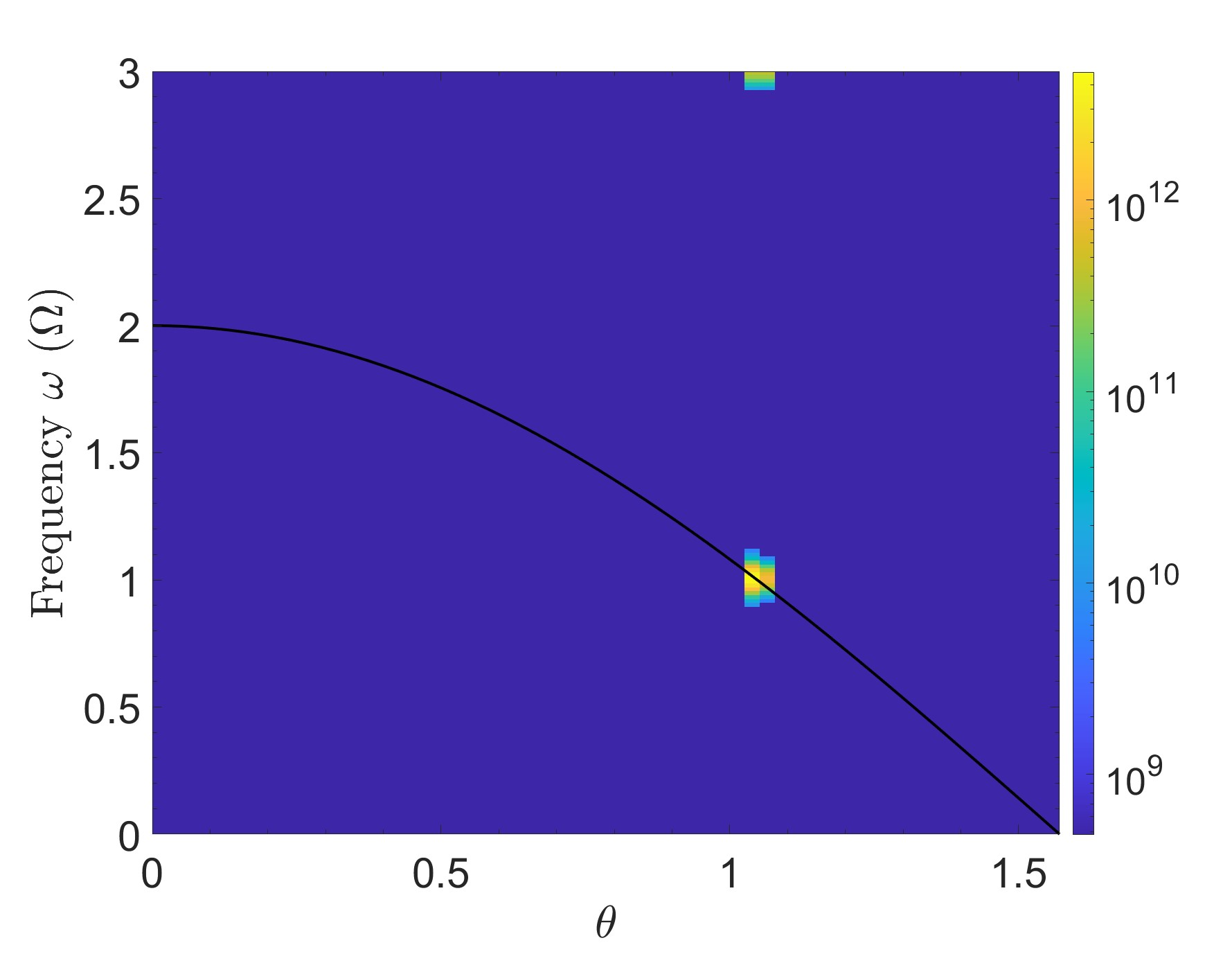}}
     \hfill
     \subfloat[Inertial wave breakdown of the simulation with $\text{Ra}=0$, $\epsilon=0.05$.\label{fig:omega-thetaeps=0.1lin+breakdown}]{\includegraphics[width=0.45\textwidth]{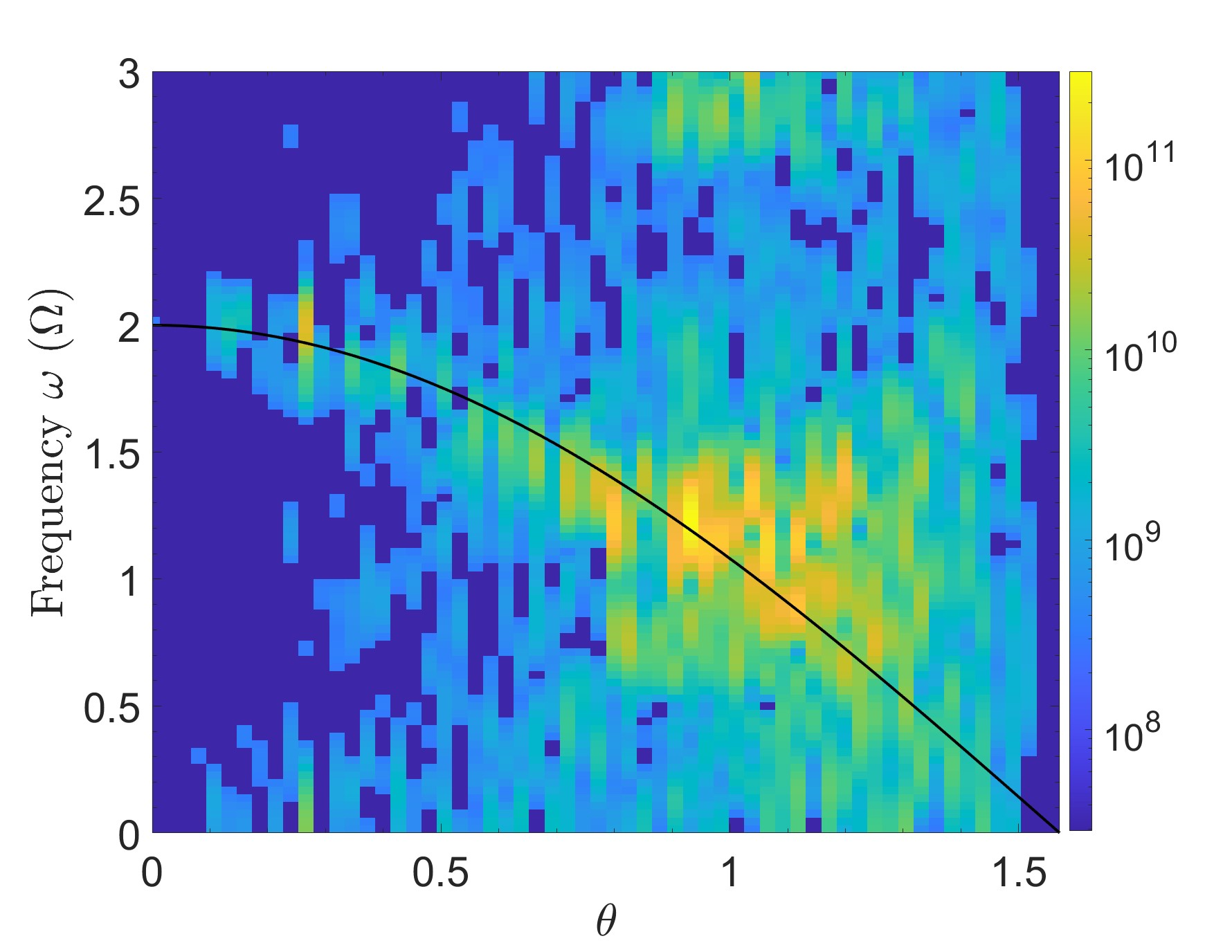}}\\
     
     \subfloat[$t=0.11-0.13$ of the simulation with $\text{Ra}=6\text{Ra}_c$, $\epsilon=0$. \label{fig:omega-thetaRa=6}]{\includegraphics[width=0.45\textwidth]{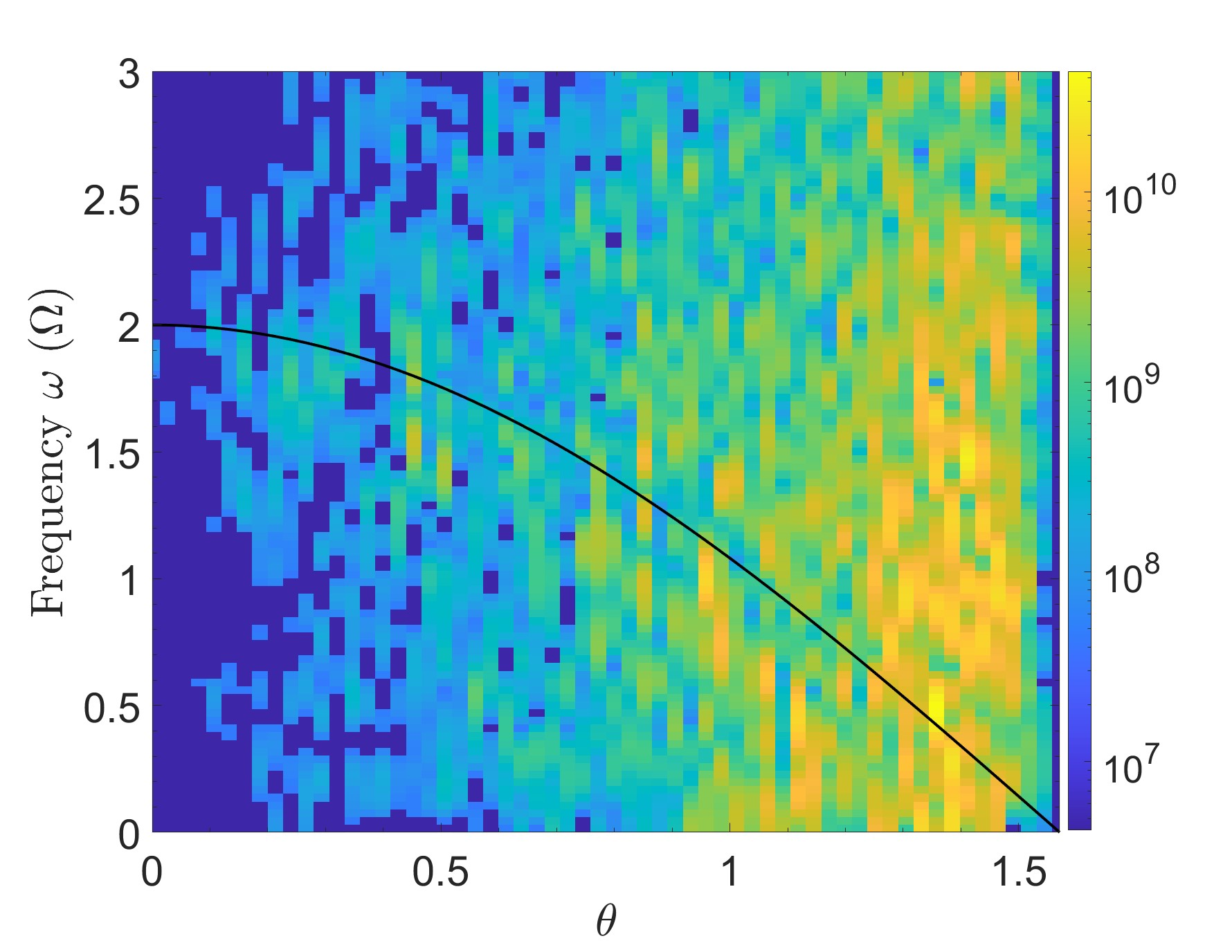}}
     \hfill
     \subfloat[$t=0.11-0.13$ of the simulation with $\text{Ra}=6\text{Ra}_c$, $\epsilon=0.05$.          \label{fig:omega-thetaRa6eps0.05}]{
         \includegraphics[width=0.45\textwidth]{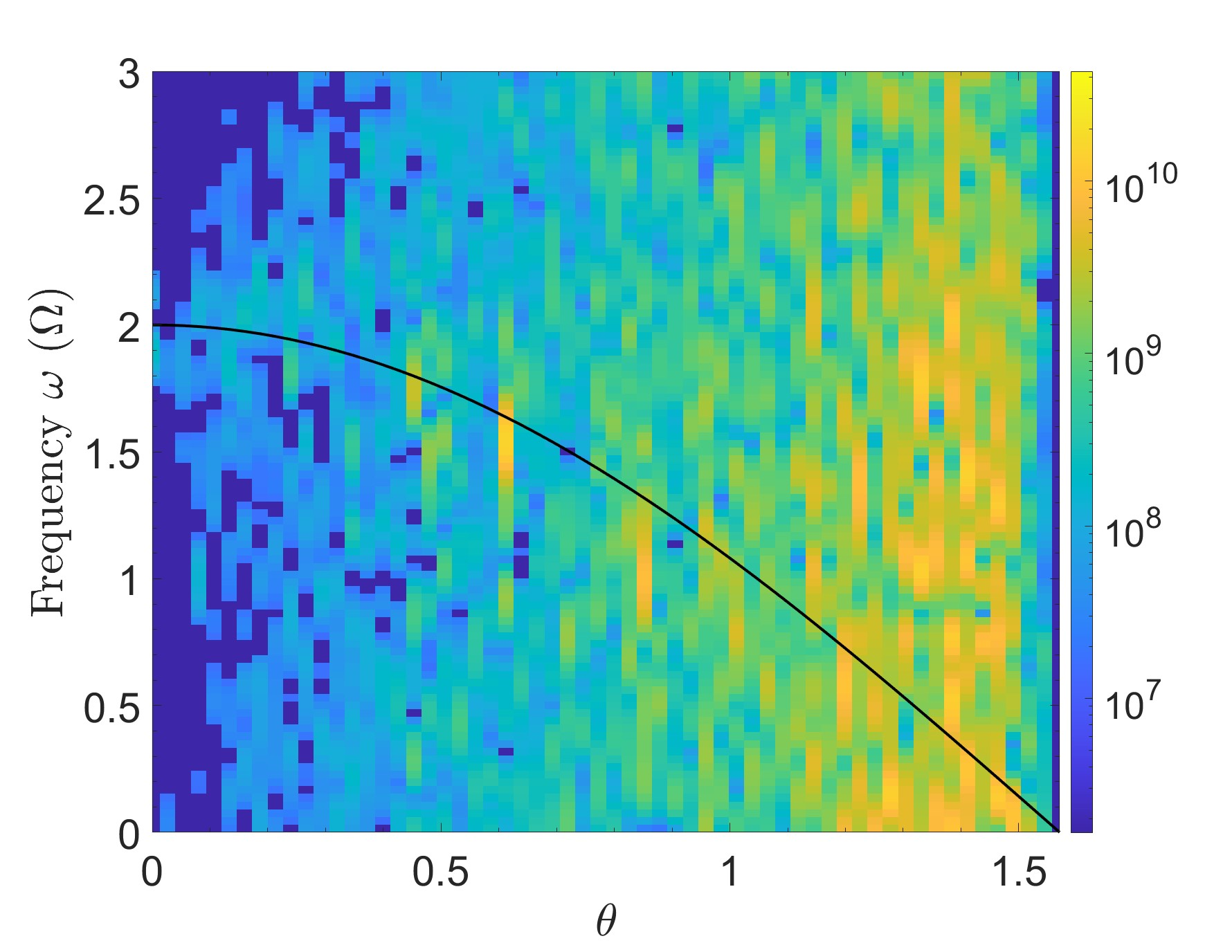}}
     \caption{Various $\theta-\omega$ energy spectra obtained by Fourier transforming the $t-\theta$ spectrum, where $\omega$ is given in units of $\Omega$. The interval of wavenumber bins used is: $k\in[2,50]$. The black solid line shows the dispersion relation for free inertial waves. For visibility the rightmost column containing the geostrophic modes which would otherwise dominate is set to zero.}
     \label{fig:omegathetaall}
\end{figure*}

We now present further analysis of our simulations using the approach devised by \citet[][]{LeReun2017} by computing the frequency-wavenumber spectrum of the flow to identify the inertial waves and the convection. This Fourier space analysis shown in Fig.~\ref{fig:omegathetaall} uses two properties of the elliptical instability to identify it in the spectrum: 1) it has a preferred direction (wavevector orientation) and 2) the dispersion relation of the inertial waves relate each direction to a particular frequency. 

The direction of the flow, i.e. angle the wavevector makes with the rotation axis $\theta$, is obtained from the ratio $k_z/k=\cos(\theta)$. Each velocity component, i.e. $u_x,u_y,u_z$, associated with a specific wavevector can be put into a bin corresponding to its angle and wavenumber (wavevector magnitude) at every timestep in a simulation. We use 60 bins for the angle $\theta$, with $\theta=[0,\pi/2]$; likewise we have chosen bins of size $\pi/2$ for the $k$-bins and enough of these to cover all values allowed by the spatial resolution of the simulation. To obtain a spectrum as a function of the frequency and angle we sum over the $k$-bins, resulting in the total velocity component in each wavevector angle bin, at each timestep. Equal timesteps of size $10^{-6}$ are used, such that the fastest inertial waves, with periods $\pi\cdot10^{-4.5}$ can be properly captured. The Fourier transform in time of all three velocity components gives the frequency spectrum of each velocity component. We then multiply the transformed velocities with their complex conjugates and add all three components to obtain the energy in each $\omega$ and $\theta$ bin. We consider the interval of wavenumber bins $k=[2,50]$ to avoid the contribution of small-scale motions which contain little energy. The geostrophic modes with $\theta=\pi/2$ strongly dominate the energy, so for clarity we set the rightmost column at $\theta=\pi/2$ to zero on these plots since we wish to analyse the waves. 

We plot the dispersion relation of inertial waves (in the absence of vortices and stratification) as a solid black line on all $\theta-\omega$ energy spectra. Thus the dispersion relation given in Eq.~\ref{eq:dispersionrelation} is plotted. This choice is suitable for the Rayleigh numbers plotted, because convection tends to reduce the magnitude of the buoyancy frequency in the bulk of the box, leading to an effective buoyancy frequency $N^2_{eff}>N^2$ (keeping in mind that $N^2$ is negative). At the plotted Rayleigh numbers of $\textrm{Ra}=4\textrm{Ra}_c$, $\textrm{Ra}=6\textrm{Ra}_c$, and $\textrm{Ra}=8\textrm{Ra}_c$ the respective effective buoyancy frequencies are: $N_{eff}^2\approx-1.5\textrm{Ra}_c$, $N_{eff}^2\approx-2.5\textrm{Ra}_c$, $N_{eff}^2\approx-3\textrm{Ra}_c$. Therefore implying $N_{eff}^2/\Omega^2\sim\mathcal{O}(10^{-2})$ at the Rayleigh numbers used in these simulations, thus the second term in Eq.~\ref{ref:dispersionrelation_Nsqr} is close to zero, only affecting the dispersion relation around $\theta\approx\pi/2$. The initial bursts of the elliptical instability are expected to be located at their preferred angle of $\theta=\arccos({1}/{2})=\pi/3$ (when $n=0$). Combined with their dispersion relation, the fastest growing mode of the elliptical instability is thus expected to be located at $\theta=\pi/3$, $\omega=\Omega$ on these figures. We can thus very easily identify the elliptical instability in such a Fourier spectrum. An example where we can clearly identify the elliptical instability is given in Fig.~\ref{fig:omega-thetaeps=0.05lin}, computed from the linear growth phase of the simulation with $\text{Ra}=0$, $\epsilon=0.05$. All modes with non-negligible energy during the linear growth phase are shown to be centred on this point, as well as at $\theta=\pi/3$, $\omega=3\Omega$, where the latter result from ``nonlinear" interactions between the background tidal flow with frequency $2\Omega$ (and wavenumber zero) and the dominant modes at $\omega=\Omega, \theta=\pi/3$.

Beyond the initial linear growth phase, inertial wave breakdown is observed, resulting in power concentrated around the initial fastest growing mode, but with a distribution primarily following the inertial wave dispersion relation. There is also energy in the geostrophic modes ($\theta\sim \pi/2$, not shown) as well as the ``mirrored dispersion relation" from secondary non-resonant interactions of the waves with the tidal flow \citep[][]{LeReun2017}. Fig.~\ref{fig:omega-thetaeps=0.1lin+breakdown} shows the same simulation as Fig.~\ref{fig:omega-thetaeps=0.05lin} but after the linear growth phase. Most of the power is concentrated around the initial fastest growing modes, however the energy is also spread throughout the figure, away from the dispersion relation, i.e. the resulting energy is no longer solely contained within the set of inertial waves.

Fig.~\ref{fig:omega-thetaRa=6} shows the spectra for a simulation of rotating convection without the elliptical instability, with $\text{Ra}=6\text{Ra}_c$, $\epsilon=0$ from $t=0.11$ to $t=0.13$. Convection is shown to introduce modes at high values of $\theta$. This can be understood from linear growth rate predictions, where we can show that convective instability of the conduction state requires $\theta\approx[1.4,\pi/2]$ for $n=1$ modes at this Rayleigh number. The dominant modes are indeed concentrated in convective modes at $\theta\approx[1.4,\pi/2]$ in this figure. The power away from these modes broadly follows the dispersion relation. This is particularly interesting because convective modes at onset are steady, so they should have $\omega\sim 0$ and be concentrated at the bottom of this figure. We might therefore speculate that the turbulence generated by convection is swept up by rotation into inertial waves, explaining the frequency of these modes. Inertial waves in rotating convection are expected to arise from oscillatory convective modes if $\text{Pr}<1$ (technically for Pr$<0.677$) \cite{Chandrsekharbook}, and have previously been observed in simulations \cite{Lin2021convinertialwaves} at $\text{Pr}<1$. Supporting our argument for inertial waves arising due to rotating convection at $\text{Pr}\geq1$ is the detection of inertial waves at $\text{Pr}>1$ in spherical shell simulations of convection \citep[][]{Bekki2022_nonlinsim}.

Spectra featuring both the elliptical instability and convection would be expected to look like a combination of these features, although it is likely to be difficult to distinguish the turbulent phases of the elliptical instability from the convective motions. However, the location of the elliptical instability bursts should shed some light on whether the sustained energy injection contains a weak (overshadowed) burst or whether the elliptical instability is absent entirely.
The simulation with $\textrm{Ra}=6\textrm{Ra}_c$, $\epsilon=0.05$ analysed from $t=0.11$ to $t=0.13$ is shown in Fig.~\ref{fig:omega-thetaRa6eps0.05}. This shows the expected convective modes, but no enhanced power at the expected location of the elliptical instability. Thus we conclude that the operation of the elliptical instability has been inhibited by convection at these parameters.

The convective motions are expected to be small-scale motions, based on the visible fluctuations in Fig.~\ref{fig:Fluct3Dconvburst}, \ref{fig:vorticity_flowpictures} and on the linear theoretical analysis of convection, which at these parameters predicts unstable modes in the range of wavenumber bins $k=[30,50]$. Meanwhile, the energetically dominant inertial waves are likely to be large-scale. This is a direct consequence of the elliptical instability being directional and thus choosing the mode with the smallest viscous effects (the largest possible modes, which also have the longest nonlinear cascade times). Therefore we have reproduced the plots in Fig.~\ref{fig:omegathetaall} with the limited wavenumber range of $k=[2,12]$ in Fig.~\ref{fig:app_omegathetaall} in \S \ref{sec:app_Fourierspectrum} of the appendix.

\begin{figure*}
    \centering\begin{minipage}[b]{0.45\textwidth}
         \centering
         \includegraphics[width=\textwidth]{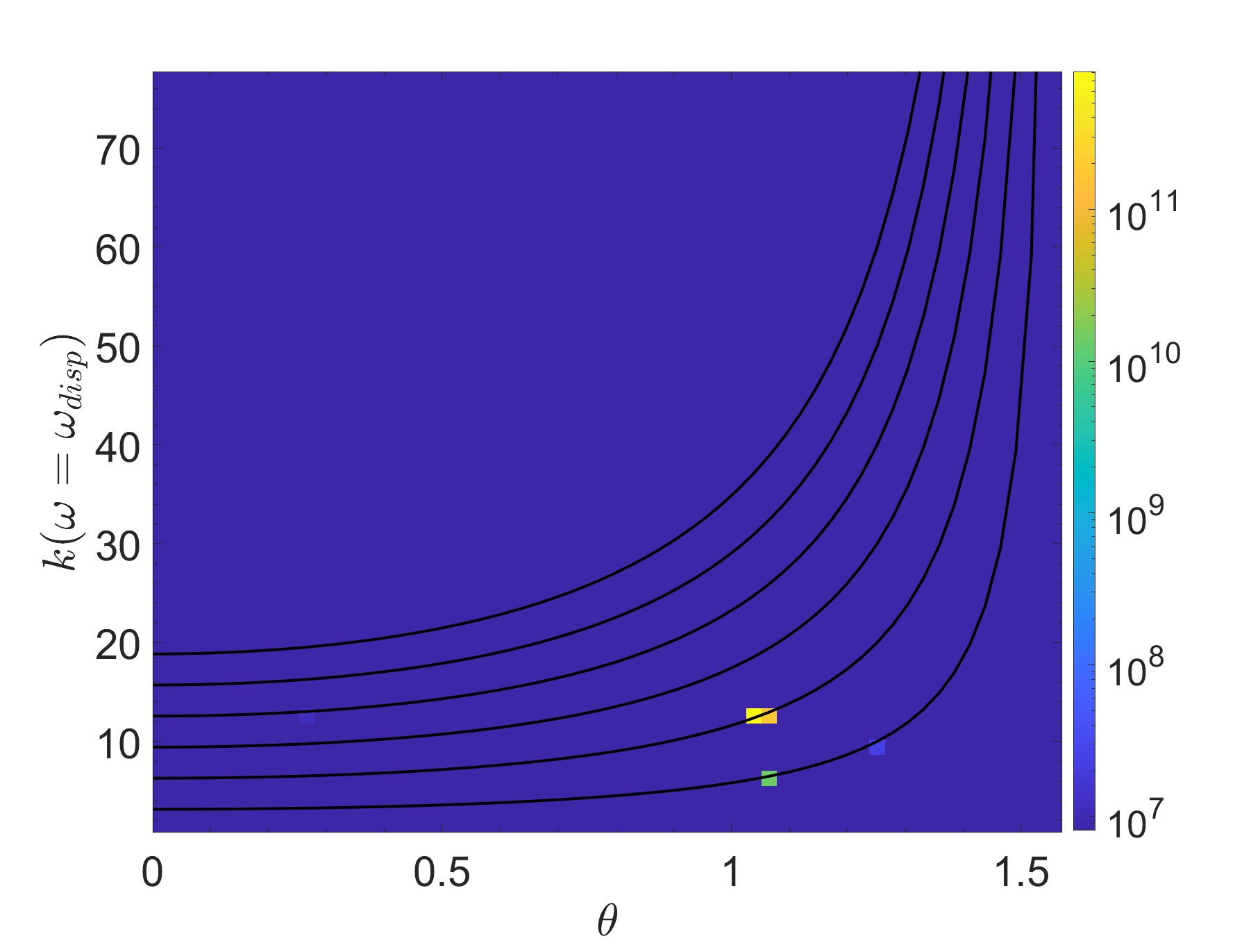}
         \label{fig:k-thetaeps=0.05lin}
     \end{minipage}
     \hfill
     \begin{minipage}[b]{0.45\textwidth}
         \centering
         \includegraphics[width=\textwidth]{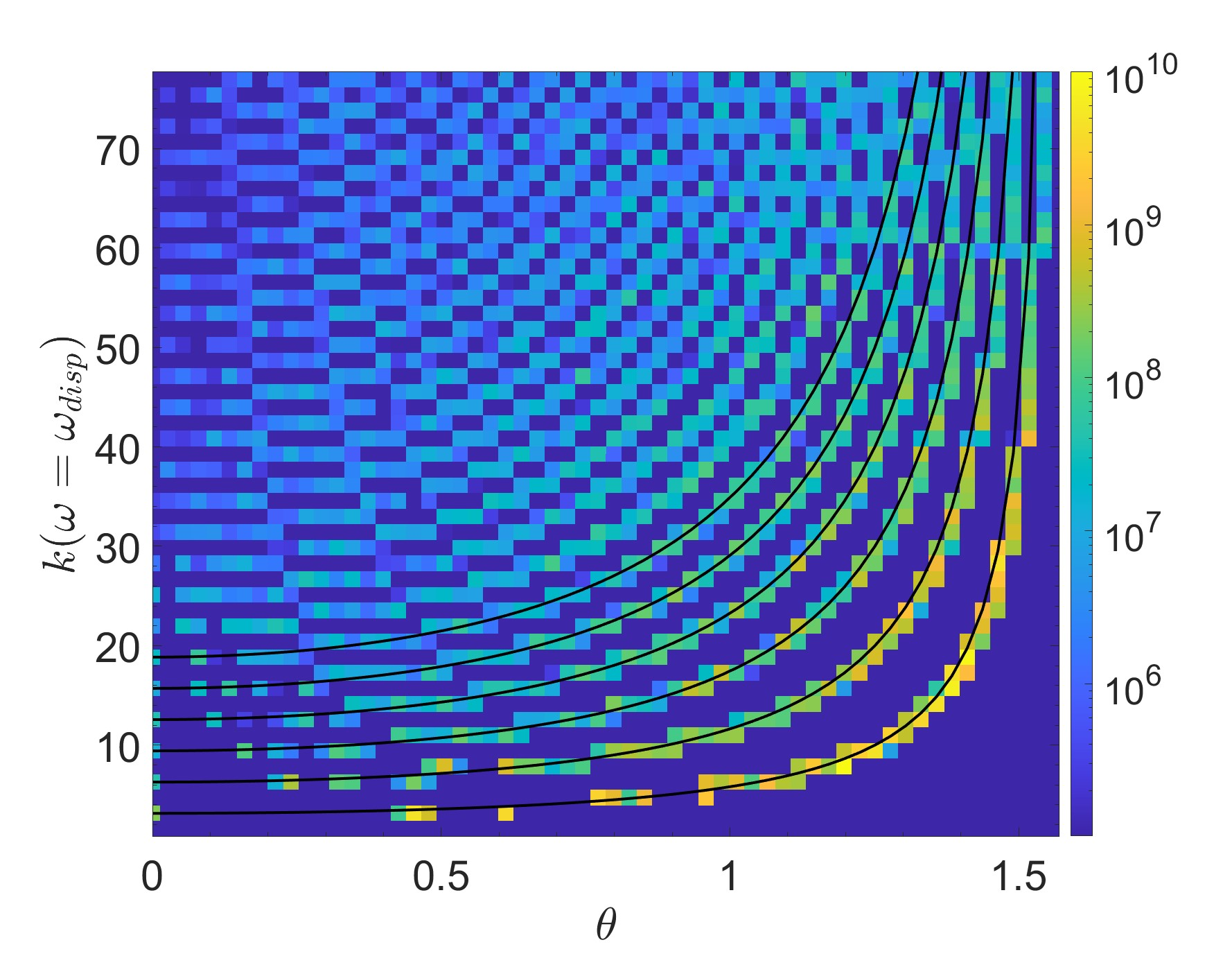}
         \label{fig:k-theta_Ra=6,eps=0.05}
     \end{minipage}
     \caption{Energy in each wavenumber as a function of $\theta$ on the dispersion relation, i.e.\ all wavenumbers have a frequency $\omega$ which satisfies $\omega=2\Omega\cos{\theta}$. The solid black curves are given by $k=n_z\pi/\cos(\theta)$, for $n_z=1,2,3,4,5,6$. The finite vertical resolution implies power must be along one of these curves. Left: during the linear growth phase for $\text{Ra}=0$, $\epsilon=0.05$. Right: $t=0.11-0.13$ for $\text{Ra}=6\text{Ra}_c$, $\epsilon=0.05$.}
     \label{fig:kthetaresults}
\end{figure*}

Using the Fourier spectrum we can determine the wavenumbers that contain the most energy in the simulation as a function of $\theta$ at a given $\omega$. To this end, we do not sum over all wavenumbers $k$, but instead construct a $\theta-k$ spectrum in Fig.~\ref{fig:kthetaresults}. We are interested in the wavenumber magnitudes that are active on the dispersion relation. Therefore we have done a Fourier transform on the wavenumbers, resulting in a $\omega-k-\theta$ matrix. Slices were then taken of the dispersion relation by taking the energy in all $k$-values at a combination of $\omega$ and $\theta$ that lies on the dispersion relation. The left panel of Fig.~\ref{fig:kthetaresults} shows the same simulation as Fig.~\ref{fig:omega-thetaeps=0.05lin}, i.e.~pure elliptical instability during its linear growth. The power here is concentrated along various black solid curves defined by the relation between $\theta$, the vertical wavenumber, $k_z=n_z\pi$, and total wavenumber, $k$: $\theta=\arccos({k_z}/{k})$. Each curve has a different integer vertical wavenumber $n_z$, with $n_z=1$ the lowest curve, $n_z=2$ the one above, etc. The curves with $n_z=1,2,3,4,5,6$ are plotted. In this simulation the energy of the elliptical instability burst is concentrated in modes with $n_z=2$, $k=4\pi$. The mode corresponding to this with horizontal wavenumber integers $n_x=n_y$ is the (5,5,2) mode. This method of analysis is powerful as it clearly shows which inertial modes are growing in the simulation. In the right panel of Fig.~\ref{fig:kthetaresults} this method is applied to the same simulation as in Fig.~\ref{fig:omega-thetaRa6eps0.05} of both convection and the elliptical instability in the sustained regime. This panel shows that there is less power on the (5,5,2) mode. Instead, we see power concentrated on the $n_z=1$ curve, and concentrated towards higher wavenumbers and higher angles, as we would expect of convective motions. So indeed, we again observe no clear sign of the elliptical instability during this simulation.

\begin{figure*}
    \centering\begin{minipage}[b]{0.45\textwidth}
         \centering
         \includegraphics[width=\textwidth]{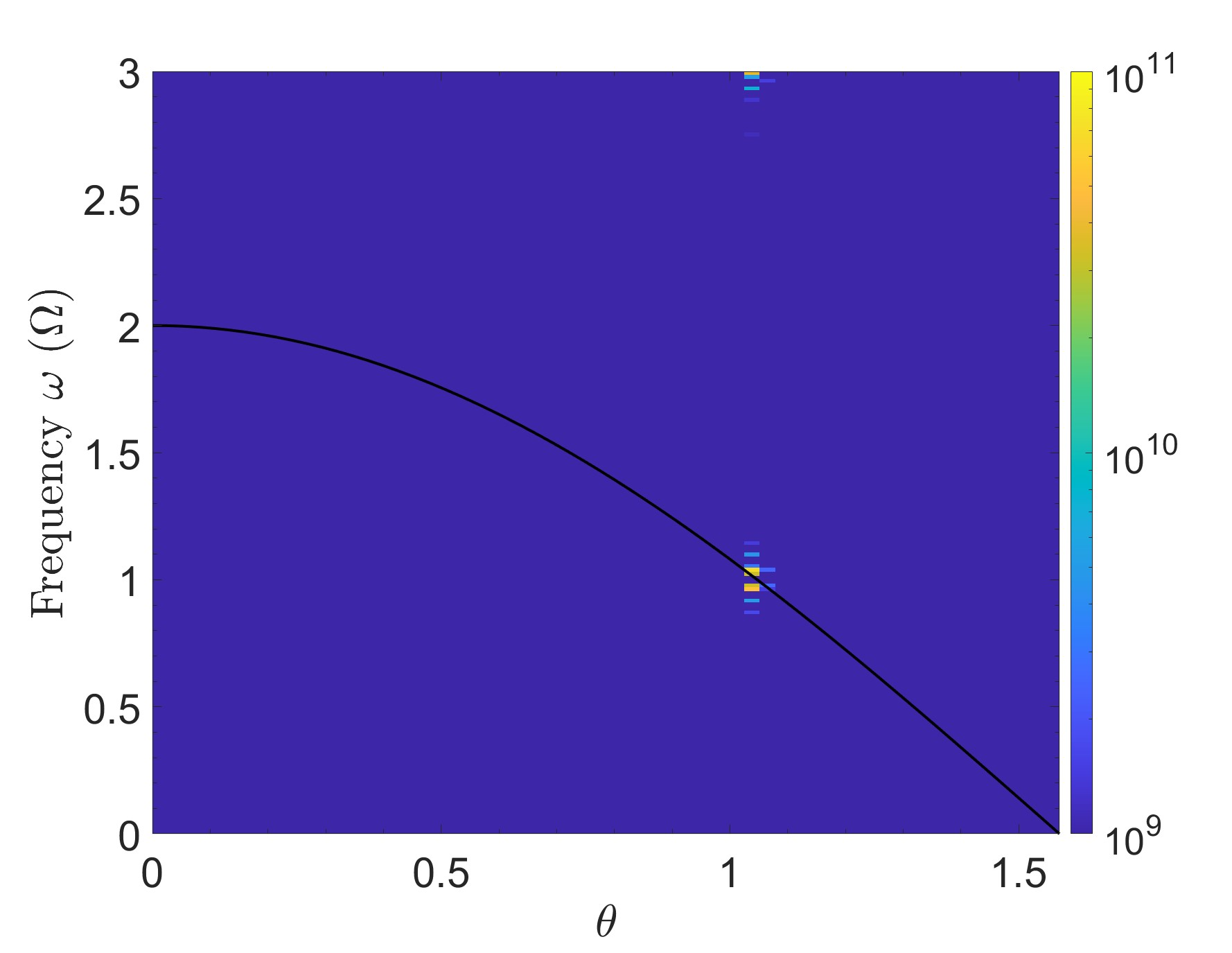}
         \label{fig:uAuomega-thetaeps=0.05lin}
     \end{minipage}
     \hfill
     \begin{minipage}[b]{0.45\textwidth}
         \centering
         \includegraphics[width=\textwidth]{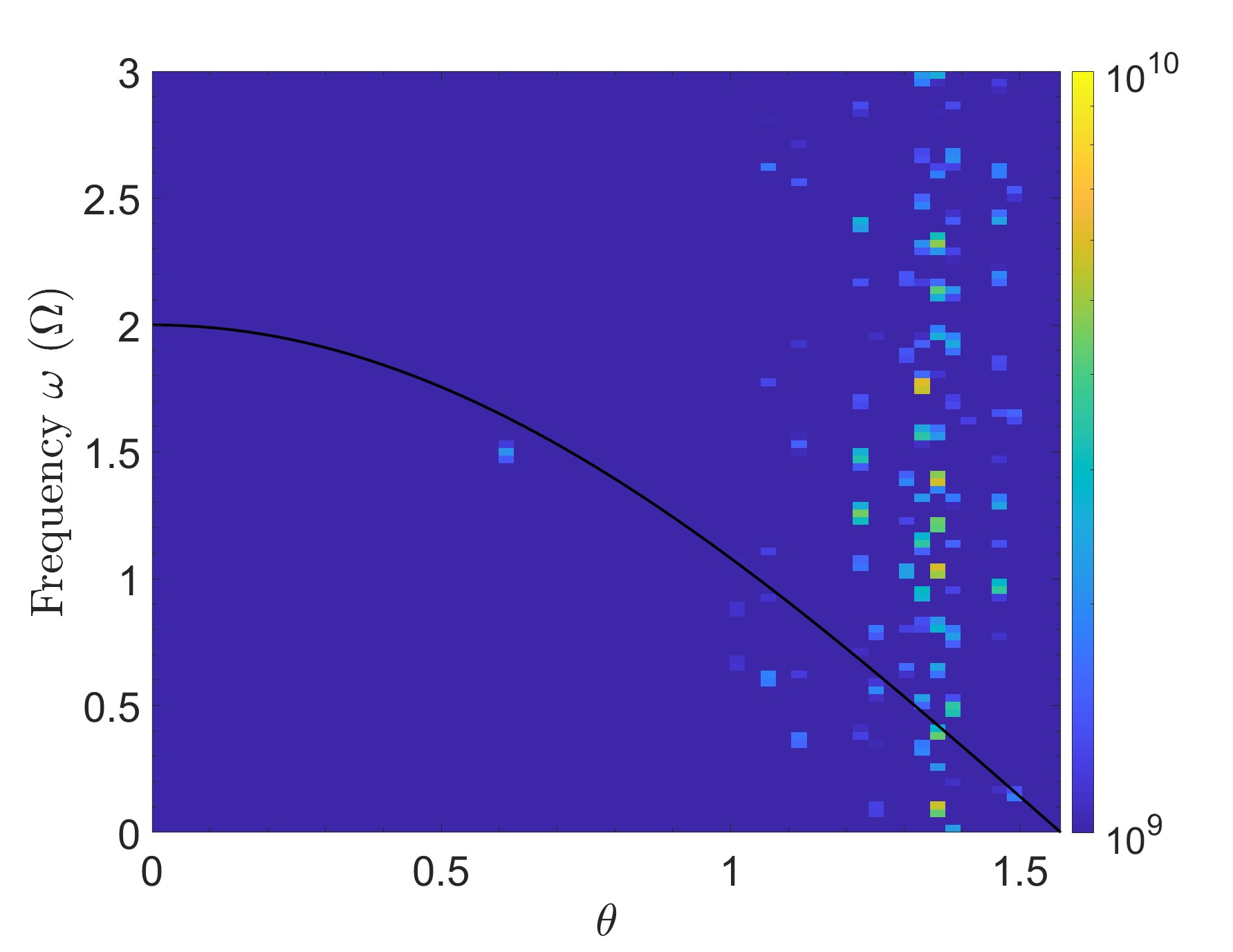}
         \label{fig:uAuomega-thetaeps=0.1lin+breakdown}
     \end{minipage}
     \caption{The $\theta-\omega$ $I_{3D}$ spectrum obtained by calculating $I_{3D}$ and taking the real part of the Fourier transform of the $t-\theta$ spectrum; $\omega$ is given in units of $\Omega$. The interval of wavenumber bins is $k=[2,50]$. The black solid line is the dispersion relation. For visibility the rightmost column containing the geostrophic modes which dominate the flow is set to zero. Left: linear growth phase of $\text{Ra}=0$, $\epsilon=0.05$. Right: $t=0.11-0.13$ of $\text{Ra}=6\text{Ra}_c$, $\epsilon=0.05$.}
     \label{fig:uAuomegathetaresults}
\end{figure*}

Finally, we are interested in analysing further the energy injection term $I_{3D}$. The 3D component is any component that is not in the rightmost column of these $\theta-\omega$ spectra (since that column has $\theta=\pi$, implying that $n_z=0$). Therefore to study $I_{3D}$ we just set the rightmost column to zero, which was already done for visibility in the above plots. $I_{3D}$ is calculated from the Reynolds stress components $u_xu_y$, $u_x^2$ and $u_y^2$. After calculation, this quantity is Fourier transformed, and its real part is then plotted here. The colourbar minimum has been increased compared to previous figures to reduce the impact of convective noise on the figure. On the left of Fig.~\ref{fig:uAuomegathetaresults} we show just the elliptical instability, for the same simulation as Fig.~\ref{fig:omega-thetaeps=0.05lin}. We see that the energy injection is into the resonant inertial waves during this initial burst. It seems two frequency bins in particular contain the majority of the energy injection, one on the dispersion relation and one on the mirrored dispersion relation. To compare, $I_{3D}$ of the same simulation as Fig.~\ref{fig:omega-thetaRa6eps0.05} is plotted on the right panel of Fig.~\ref{fig:uAuomegathetaresults}. Energy injection is present predominantly on the right-hand side of the figure and is no longer concentrated along the dispersion relation, suggesting the energy injection is primarily into convective motions, instead of the inertial waves of the elliptical instability. This would be consistent with treating the energy transfer between tidal and convective flows as being due to a turbulent effective viscosity from the convective motions.

\subsection{Linear growth of elliptical instability on a convective background with an LSV}
\label{sec:LingrowthEIonLSV}

\begin{figure*}
    \centering
    \includegraphics[width=\linewidth]{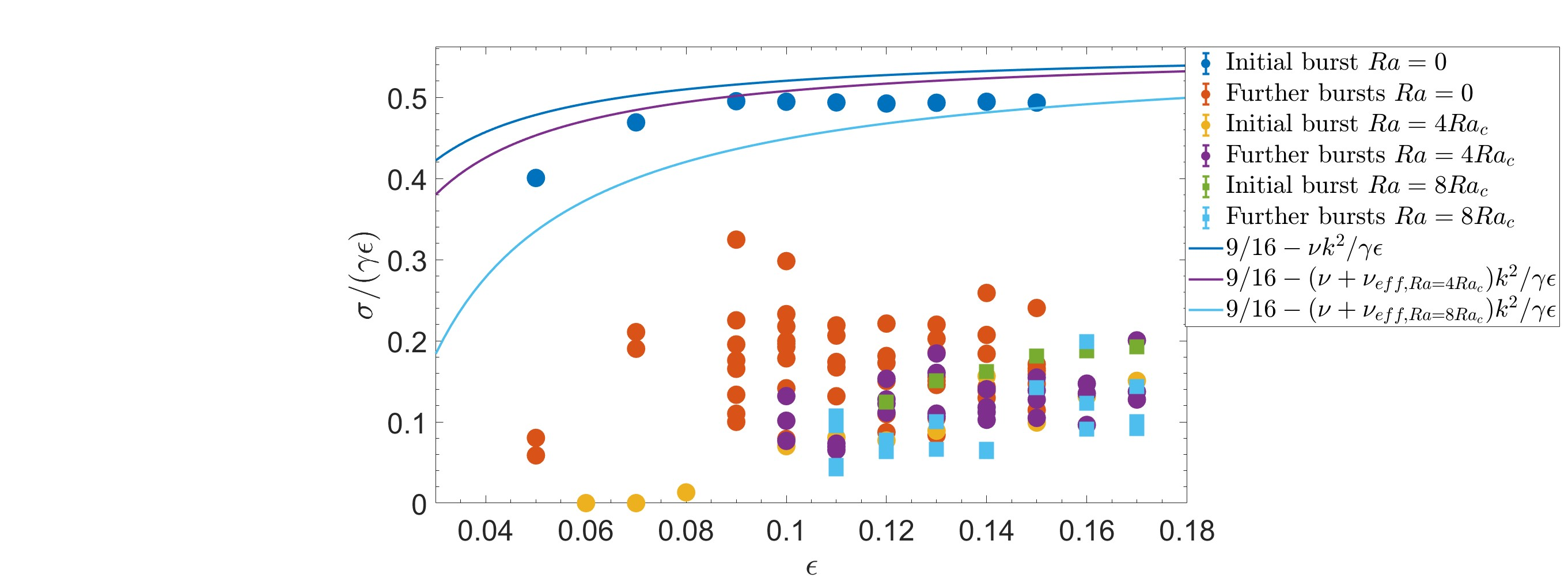}
    \caption{Growth rate of the initial burst and further bursts of the elliptical instability measured in simulations without convection $\text{Ra}=0$ (blue and orange), and simulations restarted from a turbulent purely convective state (with an LSV) at $\text{Ra}=4\text{Ra}_c$ (yellow and purple) and $\text{Ra}=8\text{Ra}_c$ (green and cyan). The theoretical prediction for the inviscid linear growth rate without convection is plotted as a blue line, while predictions assuming an effective viscosity (with damping rate $-\nu_{eff}k^2$) from convection at $\text{Ra}=4\text{Ra}_c$ and $\text{Ra}=8\text{Ra}_c$ are plotted as a purple and cyan line, respectively. This figure implies that the reduction in the growth rate originates from something other than a simple effective viscosity, as the prefactor of the growth rate is changed, indicated by the growth rates tending to a value lower than 9/16 as $\epsilon$ is increased.}
    \label{fig:initconvsigma}
\end{figure*}

Based on the real space analysis in \S~\ref{sec:sustained parameters} we concluded that the convective flow and its resulting LSV modifies subsequent growth of the elliptical instability, similar to the modification of the LSV resulting from the elliptical instability. Furthermore, based on the Fourier space results in \S~\ref{sec:Freqspec} we found that the elliptical instability is largely inhibited by the convective flow (and its LSV) for lower values of $\epsilon$.

To further examine the effects of convection on the elliptical instability, we analyse the growth rate of bursts of the elliptical instability on a convective background. First we measured the growth rates in simulations without convection ($\textrm{Ra}=0$) as a reference. We split the results of each simulation into the initial burst and any further bursts. The initial burst should then be close to the linear prediction, while any subsequent bursts are affected by non-linear effects, such as the LSV. The growth rate of the bursts of the elliptical instability is obtained from (half) the growth rate of $K_{3D}$, and our results normalised by $\sigma \epsilon$ are shown in Fig.~\ref{fig:initconvsigma}. The growth rates of the initial burst without convection (blue diamonds) lie close to the linear theoretical prediction, $(9/16)\epsilon\gamma$, plotted as the solid black line. Further bursts of these simulations (orange diamonds), however, substantially deviate from this prediction. A large reduction of the growth rate is found, likely due primarily to the LSV created by the initial burst. Note that these growth rates reduce further as the simulation continues and the energy in the LSV grows.

To compare these with similar results in the presence of convection, we ran new simulations which have been initialised with the flow and temperature fields from a purely convective simulation long after saturation of initial instability. We started several simulations with various ellipticities $\epsilon>0.05$ from our simulation with $\text{Ra}=4\text{Ra}_c$ ( $\epsilon=0$), and several with $\epsilon>0.1$ from our simulation with $\text{Ra}=8\text{Ra}_c$ ($\epsilon=0$). These results are shown in Fig.~\ref{fig:initconvsigma} using yellow circles for the initial burst and purple circles for further bursts at $\text{Ra}=4\text{Ra}_c$ and green and cyan squares at $\text{Ra}=8\text{Ra}_c$, respectively. Focusing first on $\text{Ra}=4\text{Ra}_c$, we see that the suppression of the initial burst of the elliptical instability occurs for $\epsilon\lesssim 0.07$. Our previous results and phase diagram (Fig.~\ref{fig:phasediagram}) indicated bursts of instability for $\epsilon\gtrsim 0.05$. These results also show that the growth rate is strongly affected by the convection, as the markers are substantially below the theoretical growth rate. The further bursts show a wide spread of growth rates. The highest measured growth rates overlap with those of the simulations of the pure elliptical instability. This implies that the convective LSV inhibits the elliptical instability in the same way as the LSV generated by the elliptical instability itself, but can inhibit it more strongly.

Our results for the case of more turbulent convection with $\text{Ra}=8\text{Ra}_c$ show higher growth rates than at $\text{Ra}=4\text{Ra}_c$, particularly at $\epsilon>0.14$. This is possibly indicative of a reduced suppression of the elliptical instability or an enhancement of the growth rate as the convection becomes stronger. A possible explanation for the enhanced growth rate could lie in Eq.~\ref{eq:kerswellgrowthconvrotat}. Increasing the Rayleigh number increases $-N^2$ for the conduction state. However, at $\text{Ra}=8\text{Ra}_c$ the growth rate is only increased by a factor of $\approx1.13$ compared to $\text{Ra}=0$, and even at $\text{Ra}=20\text{Ra}_c$ the increase is only a factor of $\approx1.3$ compared to $\text{Ra}=0$. Furthermore, this factor is likely to be less important than this would predict, as convection acts to reduce $-N^2$, and the efficiency of rotating convection increases with\citep{CurrieconvRMLT} Ra.

We compare these results with some theoretical arguments in the figure. We use the energy injection $I_{3D}$ of each simulation as a function of the Rayleigh number to obtain an effective viscosity $\nu_{eff}$ that corresponds with this sustained rate of injection (strictly acting on the tidal flow rather than the waves). We add this effective viscosity to the fluid viscosity $\nu$ to obtain the ``total viscosity", which is used to compute a viscous damping rate $-(\nu+\nu_{eff})k^2$. Predictions for the growth rate after introduction of an effective viscous damping rate corresponding with $\text{Ra}=4\text{Ra}_c$ and $\text{Ra}=8\text{Ra}_c$ are plotted in purple and cyan lines, respectively, assuming the dominant wavenumbers and resonance conditions are unchanged. Incorporating the microscopic viscosity and/or effective viscosity decay rates are both inconsistent with the numerically-obtained growth rates. Indeed, results from simulations with just the elliptical instability imply that the suppression by an LSV is much stronger than would be predicted by such an effective viscosity. Furthermore, the slope of the growth rate as a function of $\epsilon$ for simulations initialised on a convective background deviates from the 9/16 prediction in a similar manner with and without the convective background. To modify this 9/16 value there must be some change in the resonance conditions and the dominant wavenumbers, due to either detuning as previously remarked upon in this work, or the phases of the inertial waves, to explain the burst behaviour.
\begin{figure*}
    \centering\begin{minipage}[b]{0.45\textwidth}
         \centering
         \includegraphics[width=\textwidth]{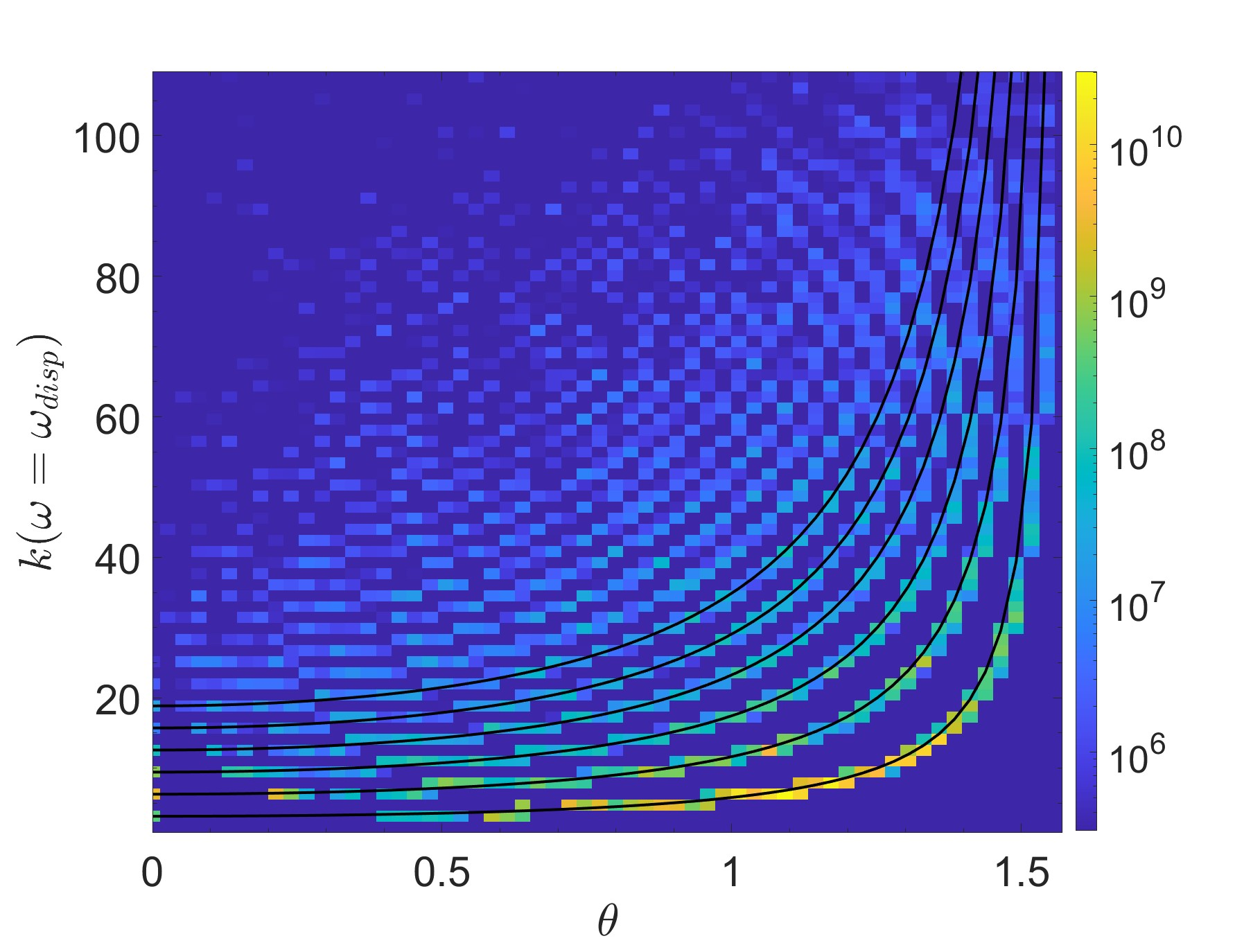}
         \label{fig:initconvomega-thetaeps=0.05lin}
     \end{minipage}
     \hfill
     \begin{minipage}[b]{0.45\textwidth}
         \centering
         \includegraphics[width=\textwidth]{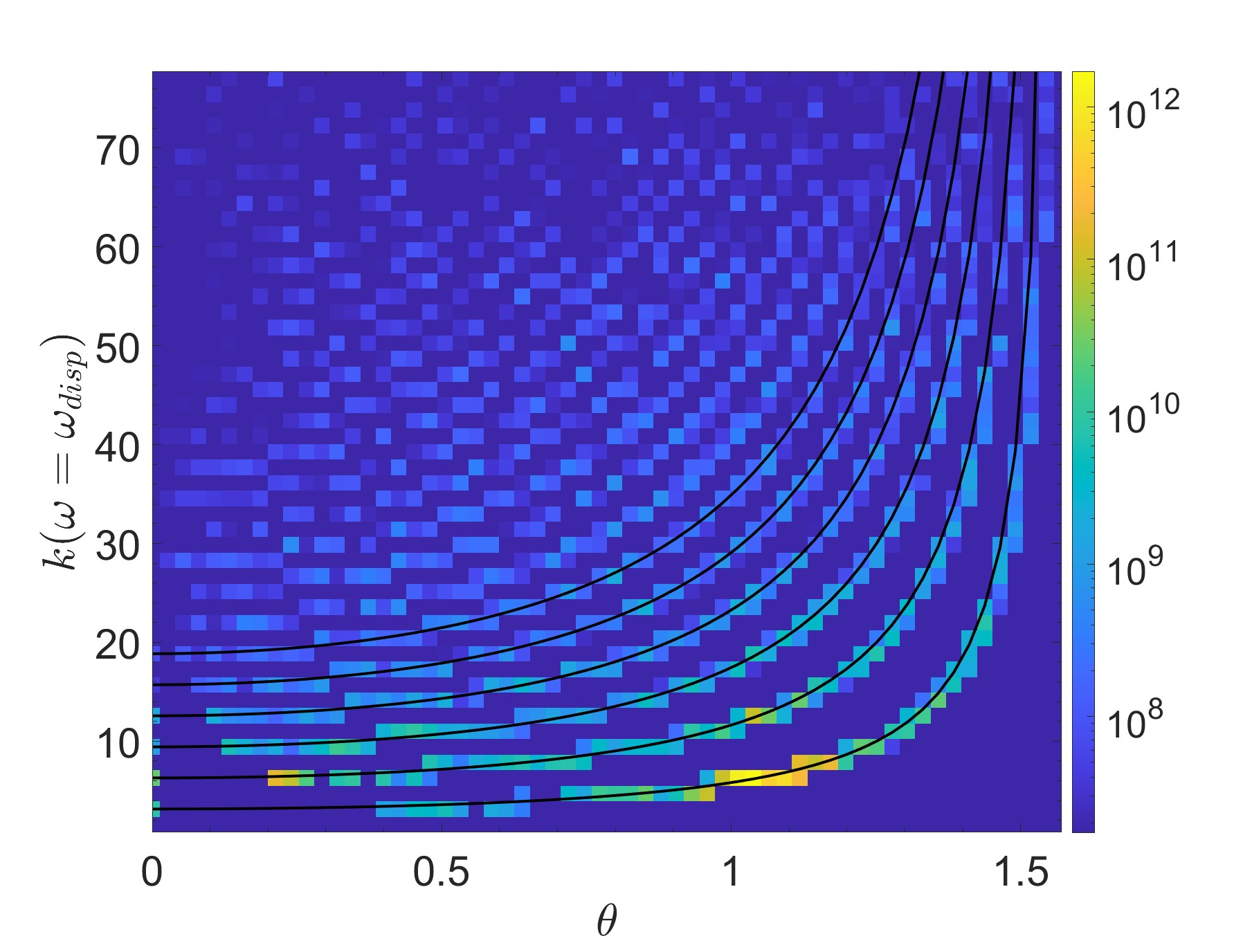}
         \label{fig:initconcomega-thetaeps=0.1lin+breakdown}
     \end{minipage}
     \caption{Same as Fig.~\ref{fig:kthetaresults} for simulations initialised from a convective simulation. Left: initial burst of $\text{Ra}=4\text{Ra}_c$, $\epsilon=0.15$, initiated from $\text{Ra}=4\text{Ra}_c$, $\epsilon=0$. Right: initial burst of $\text{Ra}=8\text{Ra}_c$, $\epsilon=0.15$, initiated from $\text{Ra}=8\text{Ra}_c$, $\epsilon=0$.}
     \label{fig:initconv_kthetaresults}.
\end{figure*}

We investigate the dominant wavenumber in each simulation, to examine if weakening of the elliptical instability occurs because the LSV changes this dominant wavenumber, using the approach described in \S~\ref{sec:Freqspec}. The $\theta-k$ spectrum for inertial modes following the dispersion relation is shown with $\epsilon=0.15$ for two $\text{Ra}$ values in Fig.~\ref{fig:initconv_kthetaresults}.
This shows that there is indeed a modification of the dominant wavenumber of the initial burst when initialising on a convective background. The $\theta-k$ spectrum in the left panel shows the first elliptical instability burst in the simulation with $\text{Ra}=4\text{Ra}_c$, $\epsilon=0.15$. The energetically dominant wavenumber is no longer the $(5,5,2)$ mode satisfying the ideal resonance condition without convection, and instead the power is concentrated at $n_z=1$ with $\theta$ close to but larger than the ideal value $\theta=\pi/3$. In the right panel of Fig.~\ref{fig:initconv_kthetaresults} we show the same for the first elliptical instability burst, but with $\text{Ra}=8\text{Ra}_c$, $\epsilon=0.15$. At both values of the Rayleigh number the subsequent inertial wave breakdown results in power in larger wavenumbers, including the (5,5,2) mode, which is viscously dissipated but otherwise maintained until the next burst. The availability of the (5,5,2) mode in the subsequent bursts results in higher growth rates compared to the initial burst, but still far below the ideal linear prediction. However, we observe power at the expected $k_z/k={1}/{2}$ and $\omega=\gamma$, therefore apparently arguing against the hypothesis that detuning the dominant resonance is responsible for most of the reduction in growth rate. This leaves the perturbed phase argument originally proposed by \citet[][]{Barker2013} to potentially explain the observed change of the growth rate prefactor.

\section{Scaling laws of the energy injection}
\label{sec:scalinglaws}

\begin{figure*}
    \centering
    \includegraphics[width=0.8\linewidth]{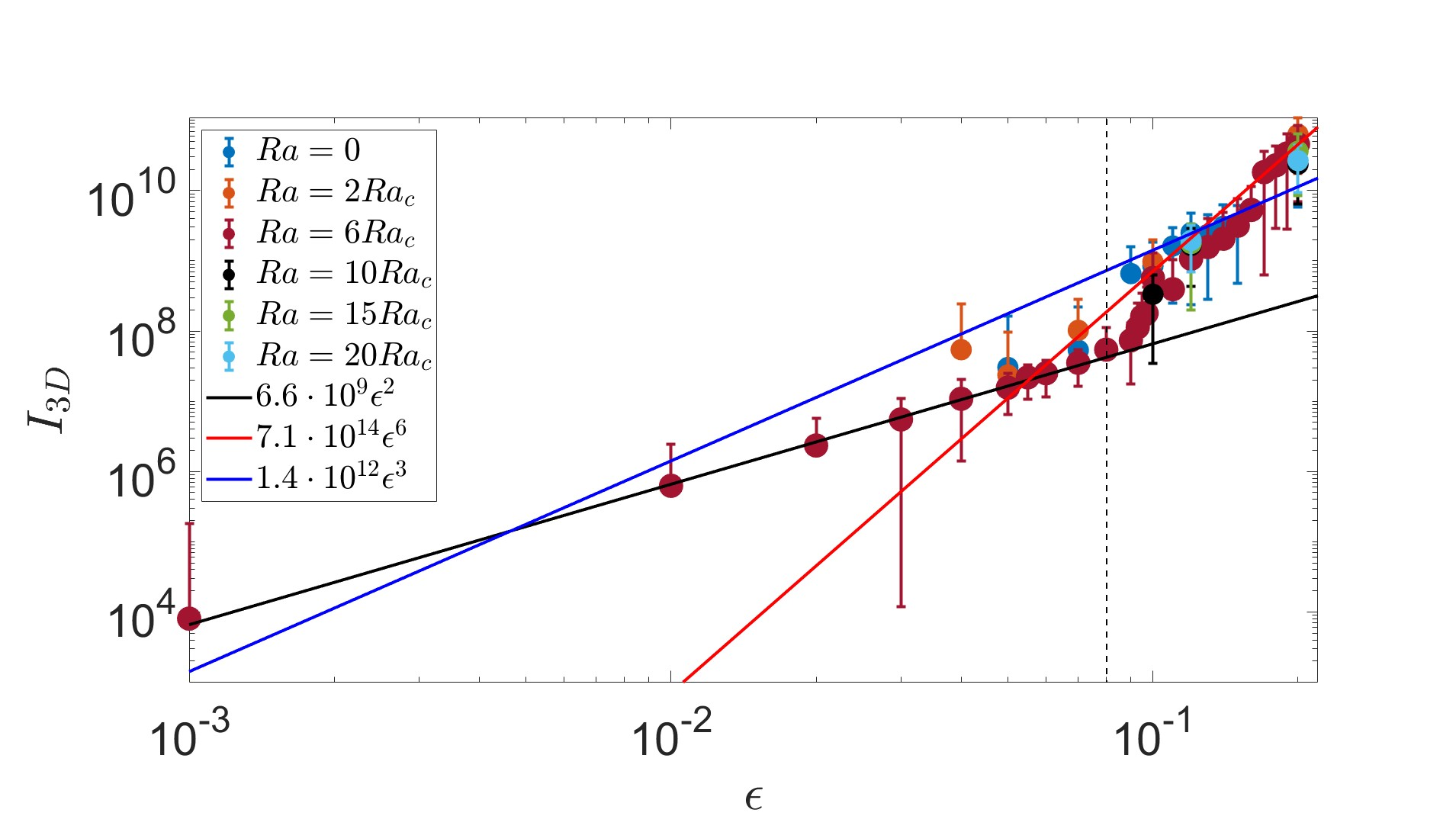}
    \caption{Energy injection rate (into 3D modes) $I_{3D}$ as a function of $\epsilon$ for various Rayleigh numbers. The vertical dashed line at $\epsilon=0.08$ marks the transition between sustained behaviour on the left, and bursts on top of sustained behaviour on the right. Three lines are fitted to the data at $\text{Ra}=6\text{Ra}_c$. The sustained behaviour is consistent with an $\epsilon^2$ scaling, represented by the black line. Bursts of the elliptical instability contribute on top of this sustained energy injection, resulting in a much larger energy injection. The sustained+bursts energy injection is fitted using an $\epsilon^3$ fit in blue, and an $\epsilon^6$ fit in red.}
    \label{fig:I3Dfitepsilonsqr}
\end{figure*}

The energy injection rate ($I_{3D}$) due to the elliptical instability on its own scales consistently with $\epsilon^3$ when the flow is sufficiently turbulent \citep[][]{Barker2013,Barker2014}. However, the sustained energy injection in our simulations isn't the result of the elliptical instability in isolation. We plot the energy injection $I_{3D}$ as a function of $\epsilon$ at various values of Ra at fixed $\textrm{Ek}=5\cdot10^{-5.5}$ in the top panel of Fig.~\ref{fig:I3Dfitepsilonsqr}, which we divide into two regimes by a vertical dashed line located at $\epsilon=0.08$, in accordance with our discussion of the simulations initialised on a convective background. 

To the left of this line the simulations show sustained energy injection without obvious bursts for $\textrm{Ra}\gtrsim2\textrm{Ra}_c$, whereas to the right of the line the simulations show clear bursts. We fit both sides separately using the $\textrm{Ra}=6\textrm{Ra}_c$ data. The black line is the fit to the sustained energy injection using points on the left side, which scales like $\epsilon^2$. This is predicted if the convection acts like an effective viscosity in Eq.~\ref{eq: I_and_nueff}. 

The bursts of elliptical instability on the right side of the figure contribute on top of this sustained energy injection. We fit using both the (naive) theoretically-predicted $\epsilon^3$ scaling and one like $\epsilon^6$ as previously observed  \citep[][]{Barker2013}. Both are consistent with the data on the right hand side, and are inconsistent with data on the left. Furthermore, the fits are consistent with data from simulations at all values of the Rayleigh number, indicating that this scaling may be independent of Ra. In this regime the elliptical instability is much more efficient than the effective viscosity of convection, and would only be surpassed by the latter when $\epsilon\approx0.01$ if we extrapolate the former with an $\epsilon^3$ scaling.

\section{Discussion and conclusion}
\label{sec:discussion}
\subsection{Comparison with previous work}

As mentioned previously in \S \ref{sec:governeq}, in the linear study of the elliptical instability in an unbounded strained vortex convection enhances the growth rate of the elliptical instability  \citep[][]{Ellipticalinstability}. In our nonlinear simulations however we observe the opposite, as the suppression of the elliptical instability increases with increasing Rayleigh number. Linear analysis of the elliptical instability in a heated cylindrical annulus on the other hand finds the growth rate can decrease with increasing Rayleigh number \citep[][]{LeBars2006_linellipcylinder}. Expanding on this linear analysis are the experiments of the elliptical instability with convection \citep[][]{Lavorelexperimentalellip}. These experiments also measured the growth rate of the elliptical instability with convection and obtained the same result. In addition, they found that smaller Ekman number leads to a faster growth rate of the instability. This result is also supported by previous numerical simulations in ellipsoids \citep[][]{Cebron2010}. Although none of these experiments or simulations have the same geometry as our simulations or clearly feature an LSV, a suppression of the elliptical instability due to convection is also observed. We conclude that this result is likely to be universal, that the elliptical instability is weakened by convection. We find similar heat transport as a result of the elliptical instability as \citet[][]{Lavorelexperimentalellip, Cebron2010} in the stably stratified regime, as well as the enhanced heat transport in simulations with Rayleigh number below or just above the critical Rayleigh number. However, we do not observe a constant Nusselt number but rather observe a weakening of this effect as the stratification increases. We also find decreased heat transport at $\text{Ra}\gg\text{Ra}_c$ compared to the purely convective simulations, likely due to the stronger vortices formed.

\subsection{Future work}
In our current setup the local box is appropriate to model the poles of a planet, with the gravity and rotation axis both pointing in the $z$-direction. The location of the box, and thus accounting for misaligned gravity and rotation could have important effects on the interactions of these instabilities. Rotating convection has been studied with misaligned gravity and rotation by \citet{CurrieconvRMLT}. They found that convective plumes (for rapid rotation) align with the rotation axis, and that strong zonal flows tend to predominantly form rather than LSVs at non-polar latitudes. These zonal shear flows have an important effect on heat transport, but could also modify the excitation and saturation of inertial waves due to the elliptical instability that should be explored in further work. Global simulations that are sufficiently turbulent and rapidly rotating to capture regimes similar to those we have explored would also be worthwhile, somewhat along the lines of the previous laminar simulations presented in e.g.~\citet[][]{Cebron2010}. Following \citet[][]{Barker2016} one could also study the interaction of the bursty non-linear dynamics of the elliptical instability with convection. One advantage of global simulations (in full ellipsoids) is that the linearly-excited inertial waves are no longer constrained by the (artificial) aspect ratio of the box.

It would also be of interest to further explore the parameter regime in our simulations, particularly by varying the Prandtl number. In particular, low Prandtl number ($\text{Pr}<0.67$) rotating convection itself excites inertial waves \citep[][]{Lin2021convinertialwaves}. These convectively-created inertial waves might also be unstable to the elliptical instability and could, due to their constant generation by the oscillatory convection, result in another source of potentially continuous tidal dissipation.
Finally, Hot Jupiters, like Jupiter itself, tend to have strong magnetic fields \citep[][]{Cauley_HJ_strongB}. Therefore the inclusion of MHD is likely to be important, and can have significant effects on tidal dissipation. In these simulations the LSVs of convection and the elliptical instability are likely to be suppressed, as magnetic fields inhibit the formation of large-scale structures. This should allow for a continuous operation of the elliptical instability\cite{Barker2014}. It is however unclear what the influence of convection will be in this interaction.

\subsection{Conclusion}
We have investigated the interactions of the elliptical instability and rotating Rayleigh-B\'{e}nard convection in a Cartesian model using psuedo-spectral hydrodynamical numerical simulations involving horizontal shearing waves. First, we simulated the elliptical instability without convection in wide boxes (with stress-free impenetrable boundaries in the vertical) for the first time, and found the nonlinear evolution of the instability to produce geostrophic vortices that dominate the flow to an even greater extent than in cubical boxes. The introduction of convection leads to a suppression of the elliptical instability that we argue is primarily due to the convectively-generated LSV. It also gives rise to a sustained energy injection into the flow (i.e. transfer from the elliptical/tidal flow) that scales as $\epsilon^2$, which can be interpreted such that the convection operates as an effective viscosity (independent of $\epsilon$) in damping the tidal flow.

The suppression of the elliptical instability by convection was investigated in detail using numerous approaches. Measuring the 3D motions, which are weakened by the LSV, we showed that during a burst of the elliptical instability the power is concentrated in the centre of the vortex. We also presented a detailed analysis of the frequency and wavenumber Fourier spectra of the energy in our simulations to clearly identify inertial modes and convective flows.  We observed that the elliptical instability (and energy injection into inertial modes more generally) is indeed inhibited by convective flows. Rotating convection also weakly excites inertial modes, which are identified as power in modes along the dispersion relation, in the absence of the elliptical instability.

When initialising simulations of the elliptical instability from a convective turbulent state including an LSV it was found that this LSV reduces the growth rate of the elliptical instability compared with the inviscid or viscous growth rate prediction. It is also reduced compared with the prediction modified by crudely adopting the aforementioned effective turbulent viscosity. The reduction of the growth rate by the LSV indicates that the dominant resonances are de-tuned by it or that there are significant perturbations in the phases of the waves by the LSV. Our Fourier space analysis showed that the fastest growing mode with an LSV is the same as the one found in the absence of an LSV for all bursts of elliptical instability. This indicates that the latter argument may be more applicable.

We also found that the inertial waves excited by the elliptical instability can transport heat; when the elliptical instability is weak relative to convection or suppressed this has little effect on the Nusselt number, but when the elliptical instability is comparable in strength to the convection, it can significantly enhance transport. The elliptical instability can also result in heat transport in stably stratified regimes, but this weakens as the stratification becomes stronger (within the linearly unstable regime).

The elliptical instability leads to an energy transfer rate from the tidal/elliptical flow (and hence dissipation rate) that is approximately proportional to $\epsilon^3$, as previously found in the absence of convection \cite{Barker2013,Barker2016}. This scaling is similar to results obtained for related instabilities like the precessional instability \cite{Barker2016precession,Pizzi2022}. Indeed, when the elliptical instability operates, the energy transfer rates are quantitatively similar to those found in prior work \cite[e.g.][]{Barker2014}. This implies that when it is not suppressed by convection, the astrophysical energy transfer rates (e.g. in Hot Jupiters) from the elliptical instability are negligibly affected by convection. However, we should point out that in the narrow range of simulations where the $\epsilon^3$ scaling is obtained without being suppressed by convection, the data is also consistent with a stronger $\epsilon^6$ scaling observed previously. Further work exploring more turbulent regimes at higher Ra and smaller Ek and Pr would be beneficial to further explore the scaling laws to allow robust extrapolation to stars and planets.

\begin{acknowledgments}
We thank Thomas Le Reun for kindly supplying his routines in Snoopy that were modified and used to produce the Fourier spectra in \S \ref{sec:sustained}. We also thank the 3 referees for their positive and constructive comments that helped us to improve the article. NBV was supported by EPSRC studentship 2528559. AJB and RH were supported by STFC grants ST/S000275/1 and ST/W000873/1. RH would like to thank the Isaac Newton Institute for Mathematical Sciences, Cambridge, for support and hospitality during the programme ``Frontiers in dynamo theory: from the Earth to the stars'' where work on this paper was undertaken. This work was supported by EPSRC grant no.\ EP/R014604/1. RH's visit to the Newton Institute was supported by a grant from the Heilbronn Institute. This work was undertaken on ARC4, part of the High Performance Computing facilities at the University of Leeds, and using the DiRAC Data Intensive service at Leicester, operated by the University of Leicester IT Services, which forms part of the STFC DiRAC HPC Facility (\href{www.dirac.ac.uk}{www.dirac.ac.uk}). The equipment was funded by BEIS capital funding via STFC capital grants ST/K000373/1 and ST/R002363/1 and STFC DiRAC Operations grant ST/R001014/1. DiRAC is part of the National e-Infrastructure.
\end{acknowledgments}

\section*{Author declarations}
\subsection*{Conflict of Interest}
The authors have no conflicts to disclose.

\subsection*{Author Contributions}

\textbf{NBV}: Conceptualization (equal); Formal analysis (lead); Software (equal); Writing - original draft (lead); Writing - review and editing (equal) \textbf{AJB}: Conceptualization (equal); Software (equal); Supervision (equal); Writing - review and editing (equal) \textbf{RH}: Conceptualization (equal); Supervision (equal); Writing - review and editing (equal)

\section*{Data Availability}

The data that support the findings of this study are available from the corresponding author upon reasonable request.


\appendix

\section{Resolution}
Multiple tests were performed to ensure our simulations were properly resolved. We ensure that heat transport is well resolved by testing the Nusselt number. If upon increasing the vertical (or horizontal) resolutions ($n_x$, $n_y$, $n_z$ are the numbers of grid points in each direction) the Nusselt number was negligibly altered it meant that the previous resolution was suitable for resolving the convection.
\begin{figure}
    \centering
    \includegraphics[width=\linewidth]{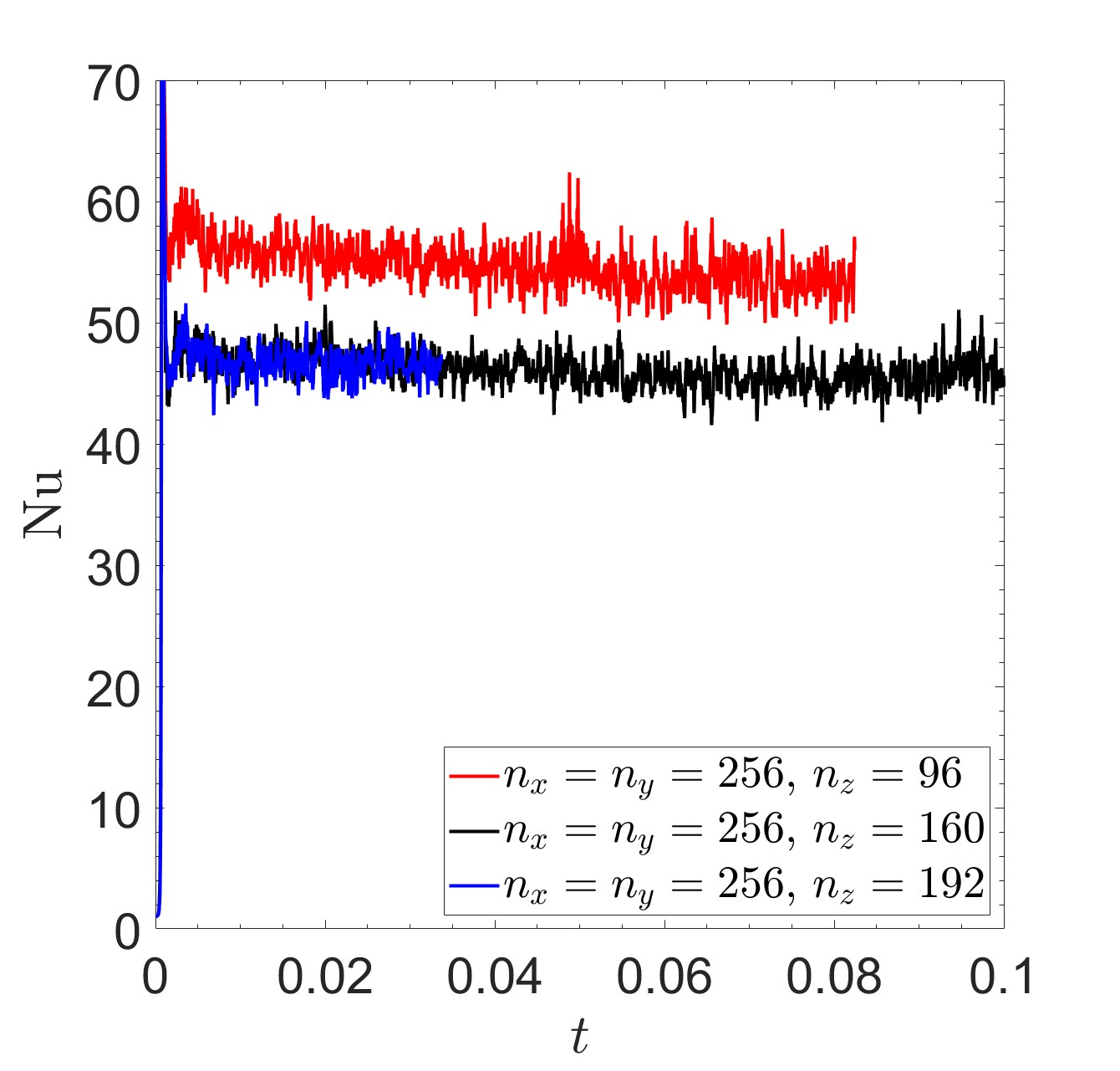}
    \caption{ Nusselt number versus time for different vertical resolutions for $\text{Ra}=15\text{Ra}_c$, $\epsilon=0.1$ and $\textrm{Ek}=5\cdot10^{-5.5}$. The convection is well resolved here for $n_z=160$, as increasing $n_z$ further doesn't affect Nu.}
    \label{fig:Nusresolutioncompare}
\end{figure}
In Fig.~\ref{fig:Nusresolutioncompare} the Nusselt number for different vertical resolutions for one of our most demanding simulations with $\text{Ra}=15\text{Ra}_c$ and $\epsilon=0.1$ ($\textrm{Ek}=5\cdot10^{-5.5}$) is plotted. It is clear that too small a vertical resolution influences Nu, and that we get convergence numerically in this case when $n_z\geq 160$. Similar test simulations were done for all Ra to ensure a good vertical resolution. 

\begin{figure}
    \centering
    \includegraphics[width=\linewidth]{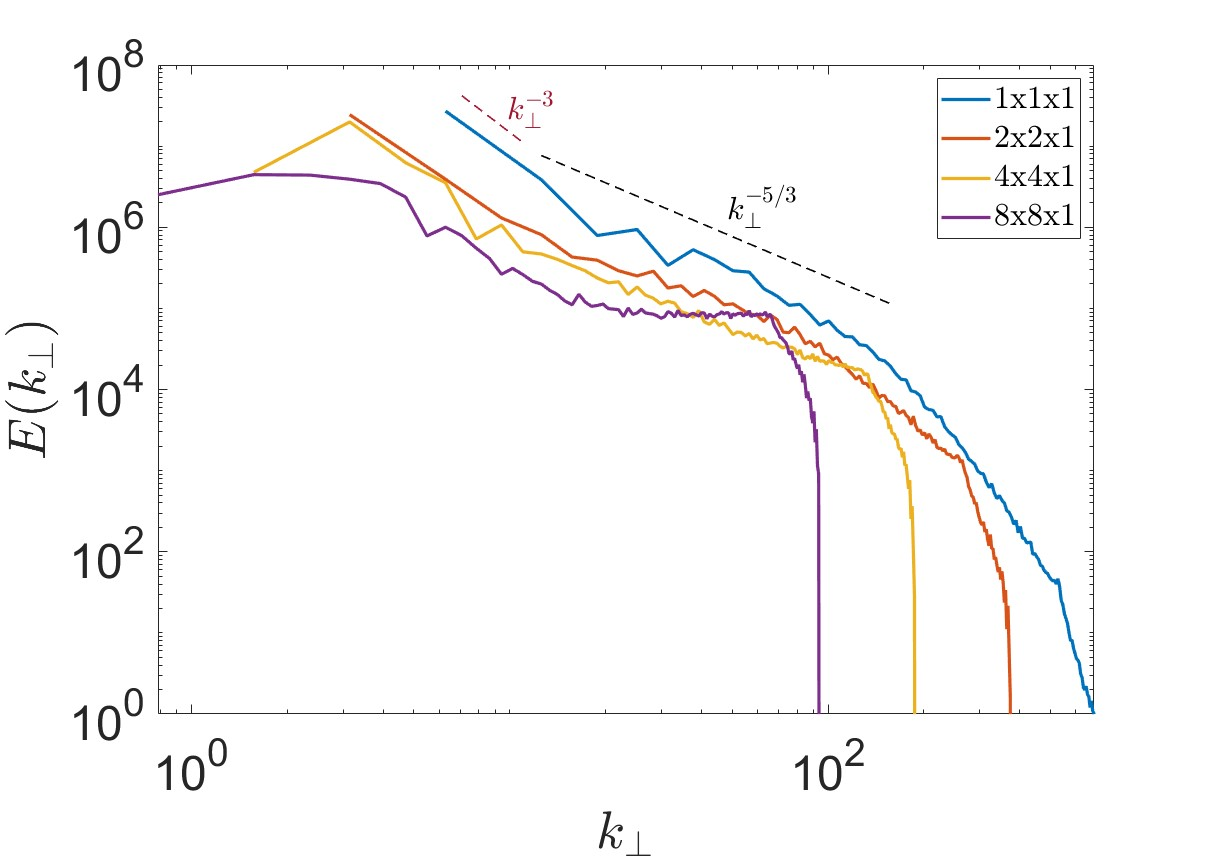}
    \caption{Horizontal power spectrum for simulations with $\textrm{Ra}=20\textrm{Ra}_c$, $\epsilon=0.2$, $\textrm{Ek}=5\cdot10^{-5.5}$, with resolution $n_x=n_y=256$, $n_z=224$, for various box sizes. Anti-aliasing scale of the simulations are: 536 (blue), 268 (orange), 134 (yellow), 67 (purple). All spectra are at $t=0.005$, i.e.\ during the initial burst of elliptical instability, to ensure these bursts are well resolved. Indicated is the Kolmogorov power law in dashed-black with slope of $k_\perp^{-5/3}$ and the wave turbulence power law in dashed-red with expected slope of $k_\perp^{-3}$.}
    \label{fig:horzpowerspec}
\end{figure}

A horizontal wavenumber power spectrum of the kinetic energy showing simulations with various horizontal box sizes is shown in Fig.~\ref{fig:horzpowerspec} for a demanding case with $\text{Ra}=20\text{Ra}_c$ and $\epsilon=0.2$ ($\textrm{Ek}=5\cdot10^{-5.5}$). The LSV caused by convection leads to the smallest wavenumber modes becoming dominant. Choosing a larger box only serves to let the vortex grow larger over time. However, during the initial burst phase at the start of the simulation, around $t=0.005$, the LSV is still forming and power is not yet contained in the largest scales. Therefore we can check these early phases to see whether the box size is appropriate to accommodate the convection and elliptical instability. Additionally, we can examine the power at the anti-aliasing scale, as this will reveal whether the flow is well resolved. We desire that the power here is at least a factor of $10^3$ lower than that in the peak to consider a simulation to be ``well resolved". Based on the horizontal power spectra in Fig.~\ref{fig:horzpowerspec} we conclude that a horizontal resolution of $256\times256$ with a box size of $4\times4$ is suitable. In addition, two power laws are plotted, in dashed-black the $k_\perp^{-5/3}$ associated with the Kolmogorov spectrum of turbulence, which matches the middle or inertial subrange of all spectra quite well, except for 8x8x1. Furthermore, the dashed-red $k_\perp^{-3}$ power law associated with wave turbulence is plotted, possibly matching the smaller wavenumbers.

\begin{figure}
    \centering
    \includegraphics[width=\linewidth]{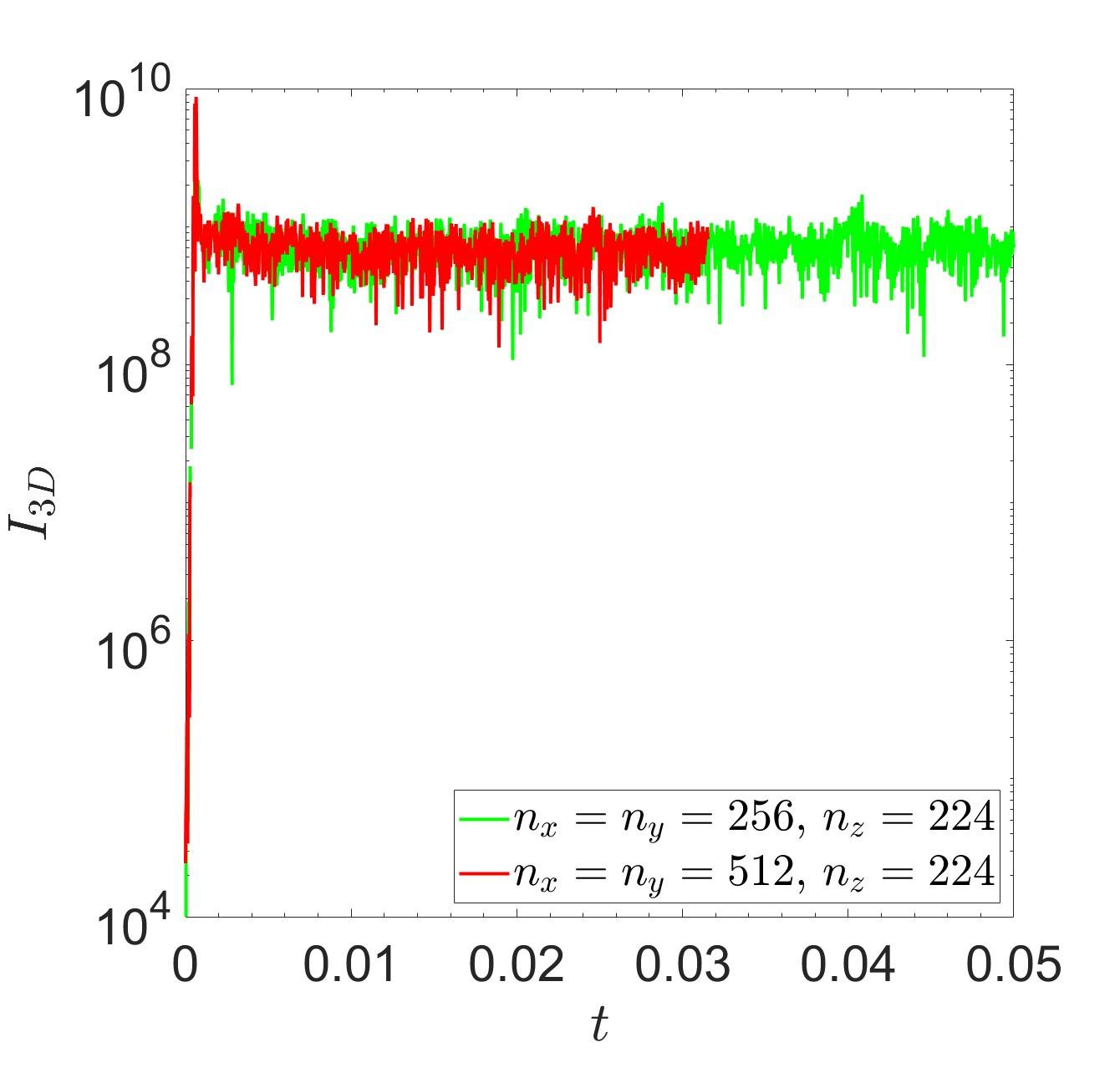}
    \caption{Comparison of $I_{3D}$ for two simulations with $\text{Ra}=20\text{Ra}_c$, $\epsilon=0.1$, and resolutions $256\times256\times224$ and $512\times512\times224$. Aside from small fluctuations the two simulations are in agreement.}
    \label{fig:resolutioncomparisonuau}
\end{figure}

Increasing Ra leads to more turbulent simulations, requiring higher resolutions to accurately capture small-scale effects, which becomes computationally expensive. To minimise computational expense we desire to minimise resolution subject to the simulation being well resolved. One further check is that the most important quantity we study, the energy injection term $I$, is numerically converged. To this end an additional test was performed with a simulation with $\text{Ra}=20\text{Ra}_c$, $\epsilon=0.1$ and a resolution of $512\times512\times224$. The energy injection term $I_{3D}$ of this simulation is compared with one with resolution $256\times256\times224$ in Fig.~\ref{fig:resolutioncomparisonuau}. Aside from small fluctuations the two simulations are in agreement, indicating that horizontal resolutions of $256\times256$ are appropriate to study the energy injection accurately. The resolutions used for all Ra at fixed Ek are given in Table~\ref{tab:resolutiontable}.

\begin{table}
\caption{Table of resolutions used in simulations at different Ra.}
\centering
\begin{tabular}{|l|l|l|}
\hline
$\textrm{Ek}=5\cdot10^{-5.5}$                                 & $n_x\times n_y$   & $n_z$  \\ \hline
$\text{Ra}/\text{Ra}_c=-6,-4,-3,-1,-0.8,0.3,0.8,1.99,3,4,6$ & 256x256 & 96  \\ \hline
$\text{Ra}/\text{Ra}_c=7,8$                        & 256x256 & 128 \\ \hline
$\text{Ra}/\text{Ra}_c=-10,9,10,15$                    & 256x256 & 160 \\ \hline
$\text{Ra}/\text{Ra}_c=20$                       & 256x256 & 224 \\ \hline

\end{tabular}
\label{tab:resolutiontable}
\end{table}

\section{Different box sizes}

\begin{figure}
    \centering
    \includegraphics[width=\linewidth]{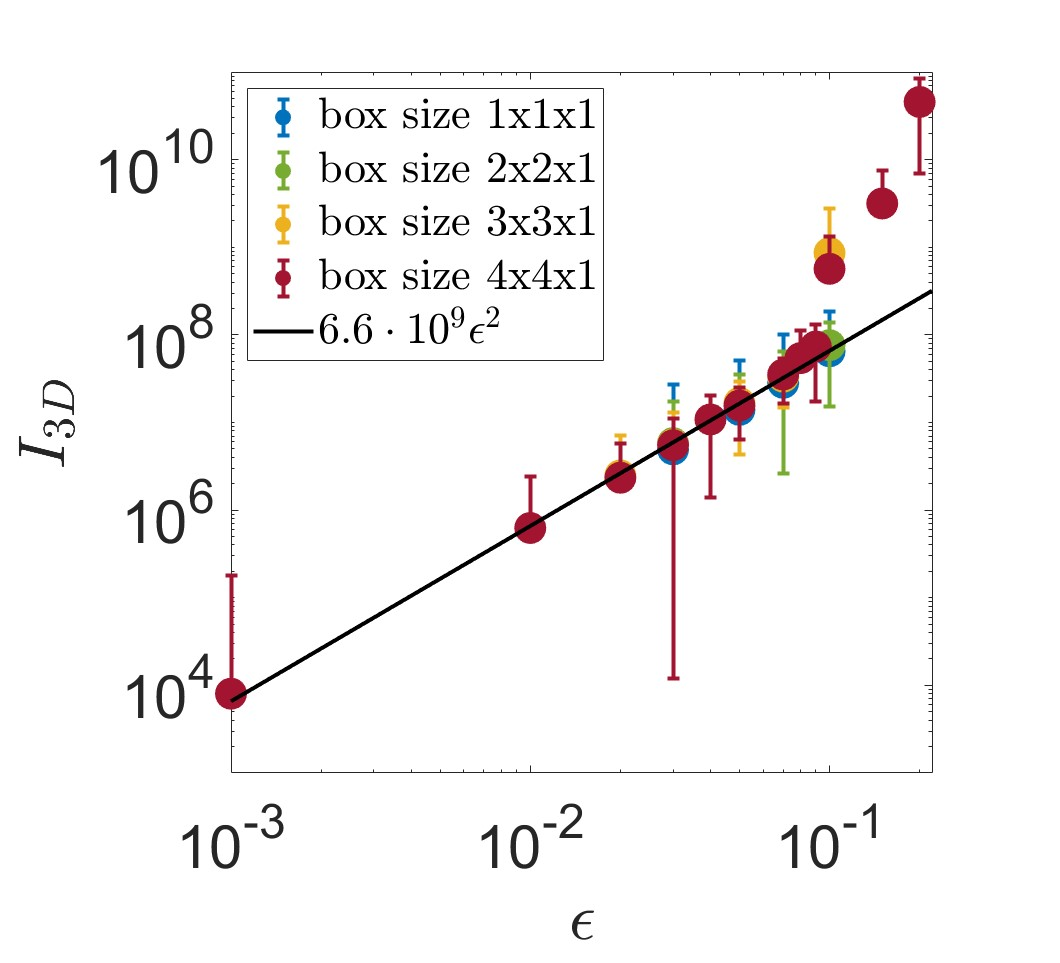}
    \caption{Same as Fig.~\ref{fig:I3Dfitepsilonsqr} showing $I_{3D}$ but for different box sizes. Changing the box size has no impact on the sustained energy injection, but smaller boxes result in the elliptical instability being suppressed for higher $\epsilon$.}
    \label{fig:app_boxsizecomp}
\end{figure}

To test the effect of different box sizes on energy transfers and Fourier spectra we ran multiple simulations with $\textrm{Ra}=6\textrm{Ra}_c$ and different $\epsilon$ ($\textrm{Ek}=5\cdot10^{-5.5}$). We have plotted $I_{3D}$ as a function of $\epsilon$ with different markers denoting different box sizes in Fig.~\ref{fig:app_boxsizecomp}. The blue markers represent results at 1x1x1, the green markers are 2x2x1, the yellow markers 3x3x1 and the burgundy markers 4x4x1. The mean energy injection in the sustained regime is shown to be independent of box size, and all markers follow the same $\epsilon^2$ scaling. The values on the right in the bursty regime do change, as the 4x4x1 results becomes bursty for $\epsilon\geq 0.05$; whereas simulations with smaller horizontal box sizes do not. Results at 3x3x1 only indicate burstiness at $\epsilon\geq 0.1$, while 2x2x1 and 1x1x1 remain in the sustained regime beyond $\epsilon=0.1$.

\begin{figure}
    \centering
    \includegraphics[width=\linewidth]{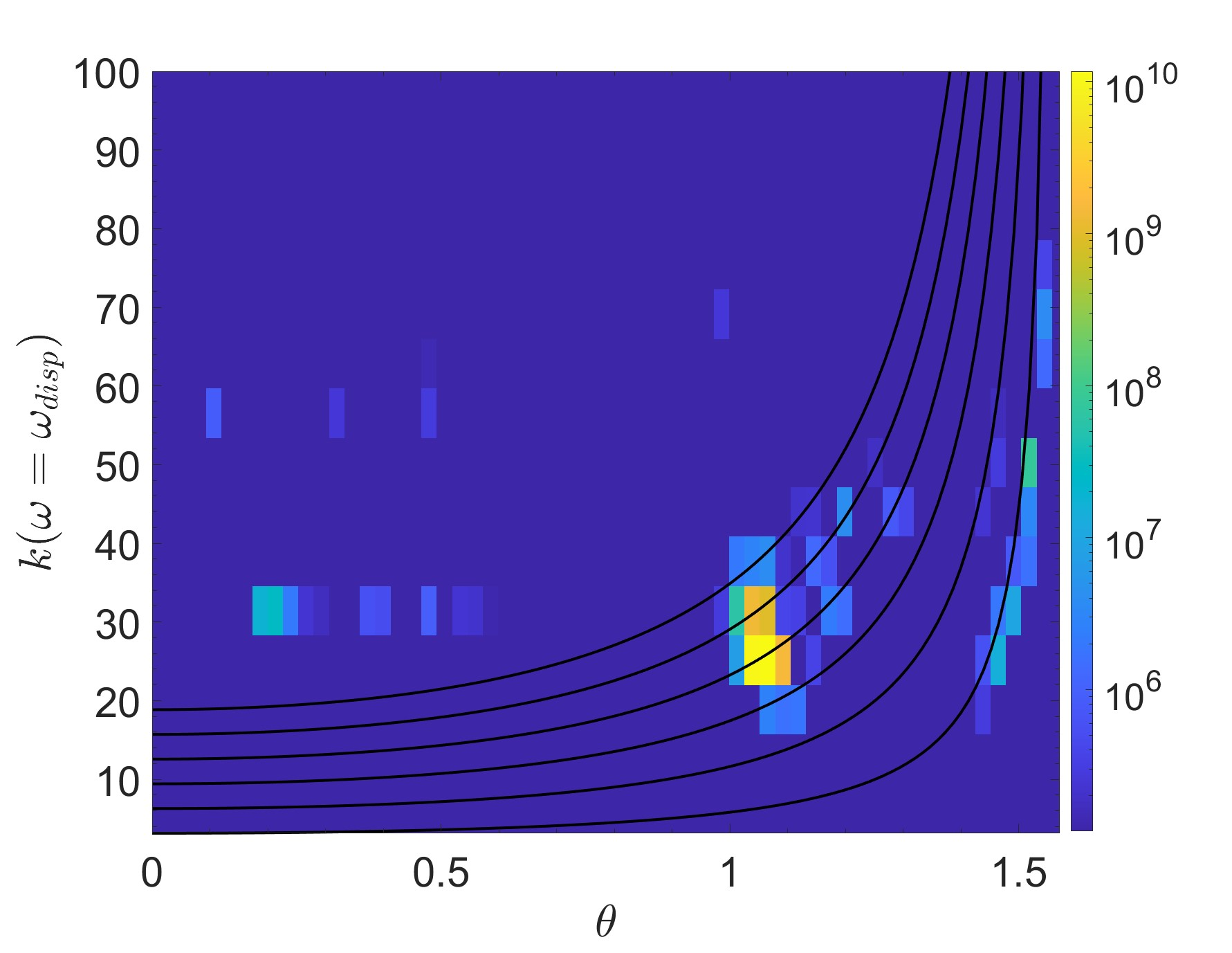}
    \caption{Same as the left panel of Fig.~\ref{fig:kthetaresults} showing the linear growth phase for $\text{Ra}=0, \epsilon=0.1$ in a 1x1x1 box.}
    \label{fig:app_ktheta1x1x1}
\end{figure}

The explanation for this behaviour lies in the allowed values of $k_\perp$ as the box size is varied. For smaller box sizes, the values of $k_\perp$ and hence $k$ increase (for the same $k_z$), so for a resonance with $k_z/k=\pm {1}/{2}$, $k_z$ (and hence $k$) must also be larger. Fig.~\ref{fig:app_ktheta1x1x1} shows the $\theta-k$ spectrum on the dispersion relation for 1x1x1, $\textrm{Ek}=5\cdot10^{-5.5}$, $\text{Ra}=0$, $\epsilon=0.1$, during the initial burst of elliptical instability; these are the linearly most unstable modes. The dominant $k$ modes lie on lines of $n=4$ and $n=5$.
These larger values of $k$ imply larger decay rates $-\nu k^2$, and therefore a decreased growth rate due to viscosity. This suppresses the elliptical instability for larger $\epsilon$ when the box is smaller. The suppression of the elliptical instability is thus artificially enhanced (reduced) by the choice of a smaller (larger) box. Upon extrapolating this effect to a full planet it is expected that the viscosity suppression of the elliptical instability is weak due to the large scales available to the system.

\section{}
\label{sec:app_Fourierspectrum}
\begin{figure*}
    \centering
         \subfloat[Linear growth phase of the simulation with $\text{Ra}=0$, $\epsilon=0.05$.
         \label{fig:app_omega-thetaeps=0.05lin}]{\includegraphics[width=0.45\textwidth]{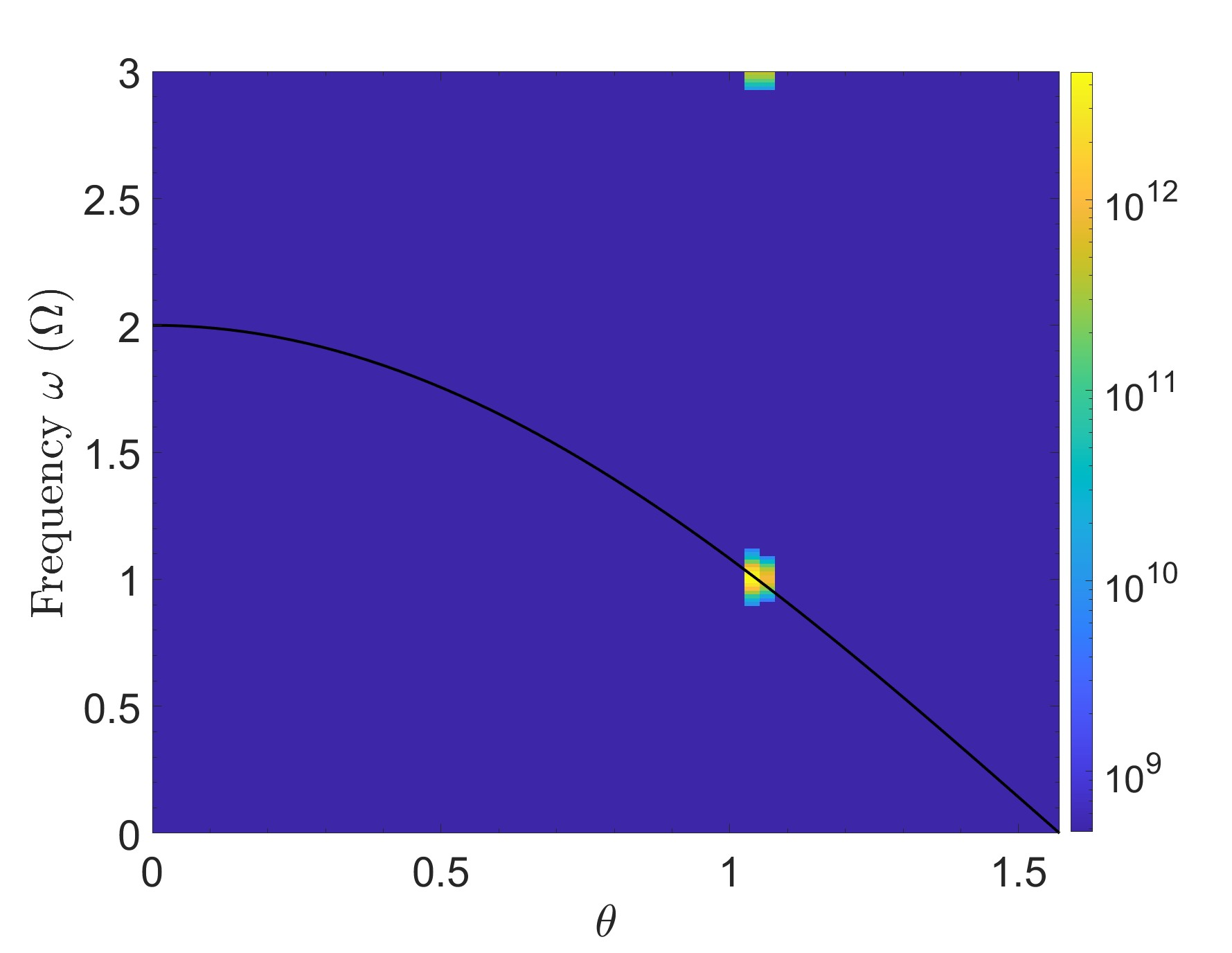}}
     \hfill
         \subfloat[Inertial wave breakdown of the simulation with $\text{Ra}=0$, $\epsilon=0.05$.
         \label{fig:app_omega-thetaeps=0.1lin+breakdown}]{\includegraphics[width=0.45\textwidth]{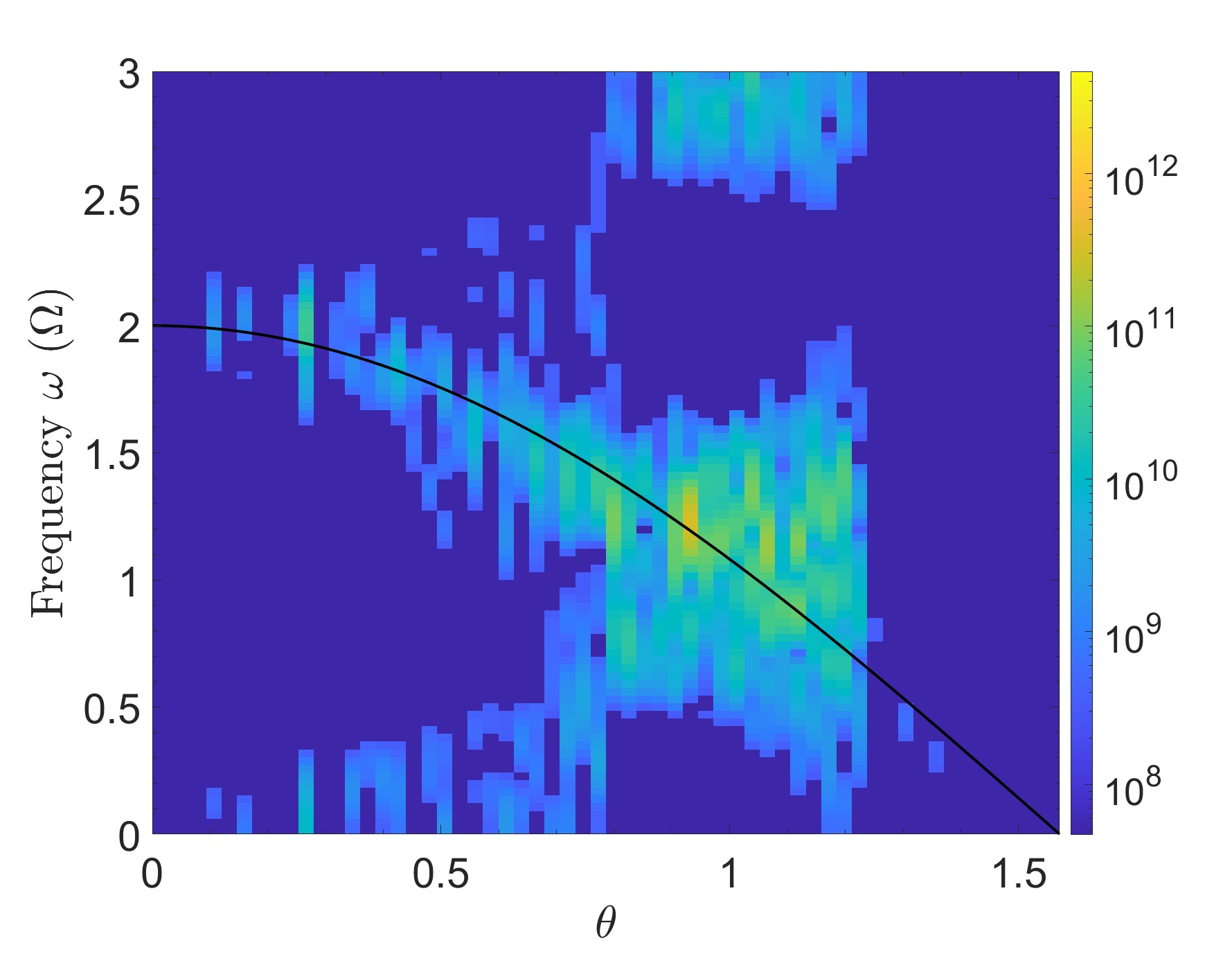}}\\
         \subfloat[$t=0.11-0.13$ of the simulation with $\text{Ra}=6\text{Ra}_c$, $\epsilon=0$.
         \label{fig:app_omega-thetaRa=6}]{\includegraphics[width=0.45\textwidth]{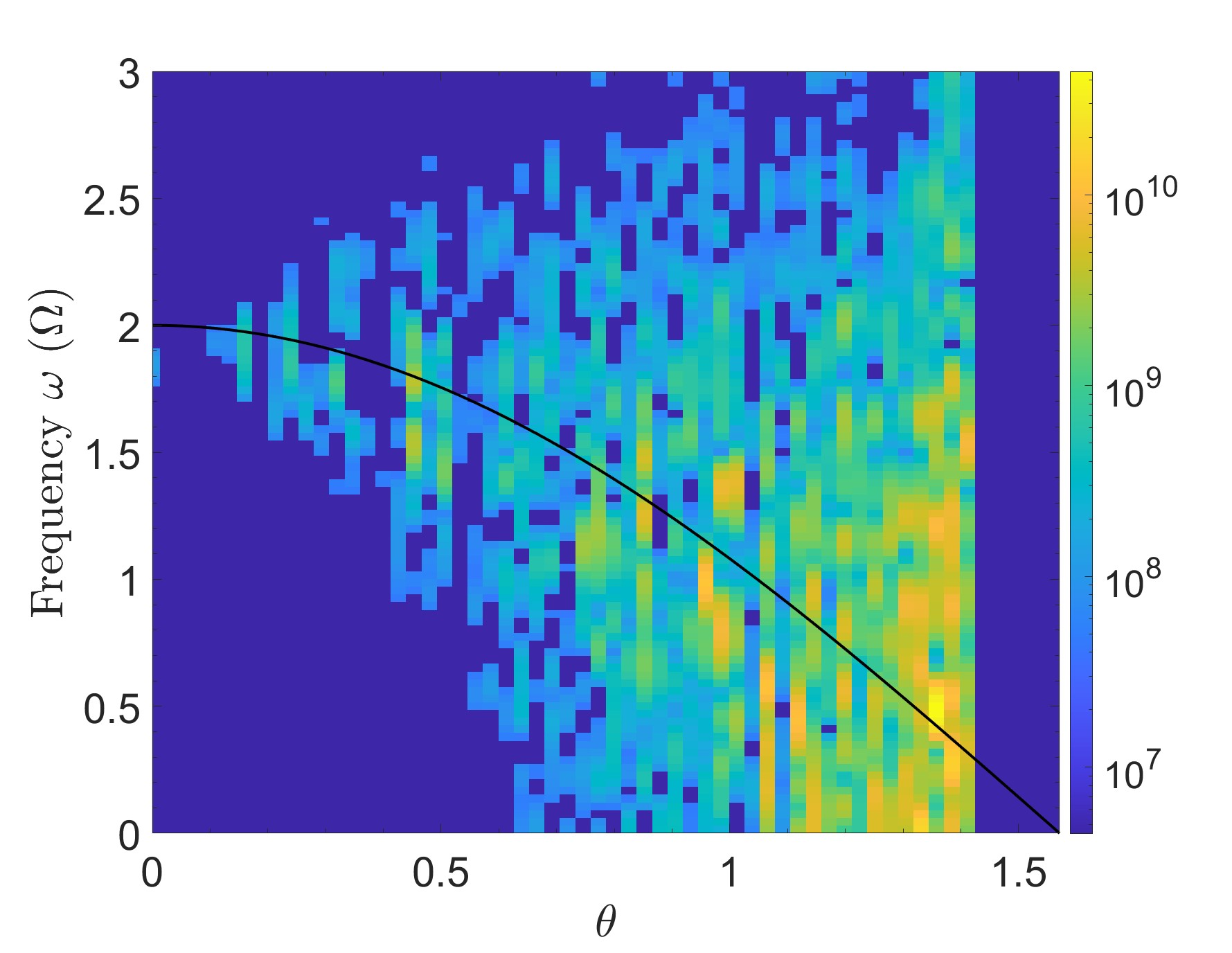}}
     \hfill
         \subfloat[$t=0.11-0.13$ of the simulation with $\text{Ra}=6\text{Ra}_c$, $\epsilon=0.05$.
         \label{fig:app_omega-thetaRa6eps0.05}]{\includegraphics[width=0.45\textwidth]{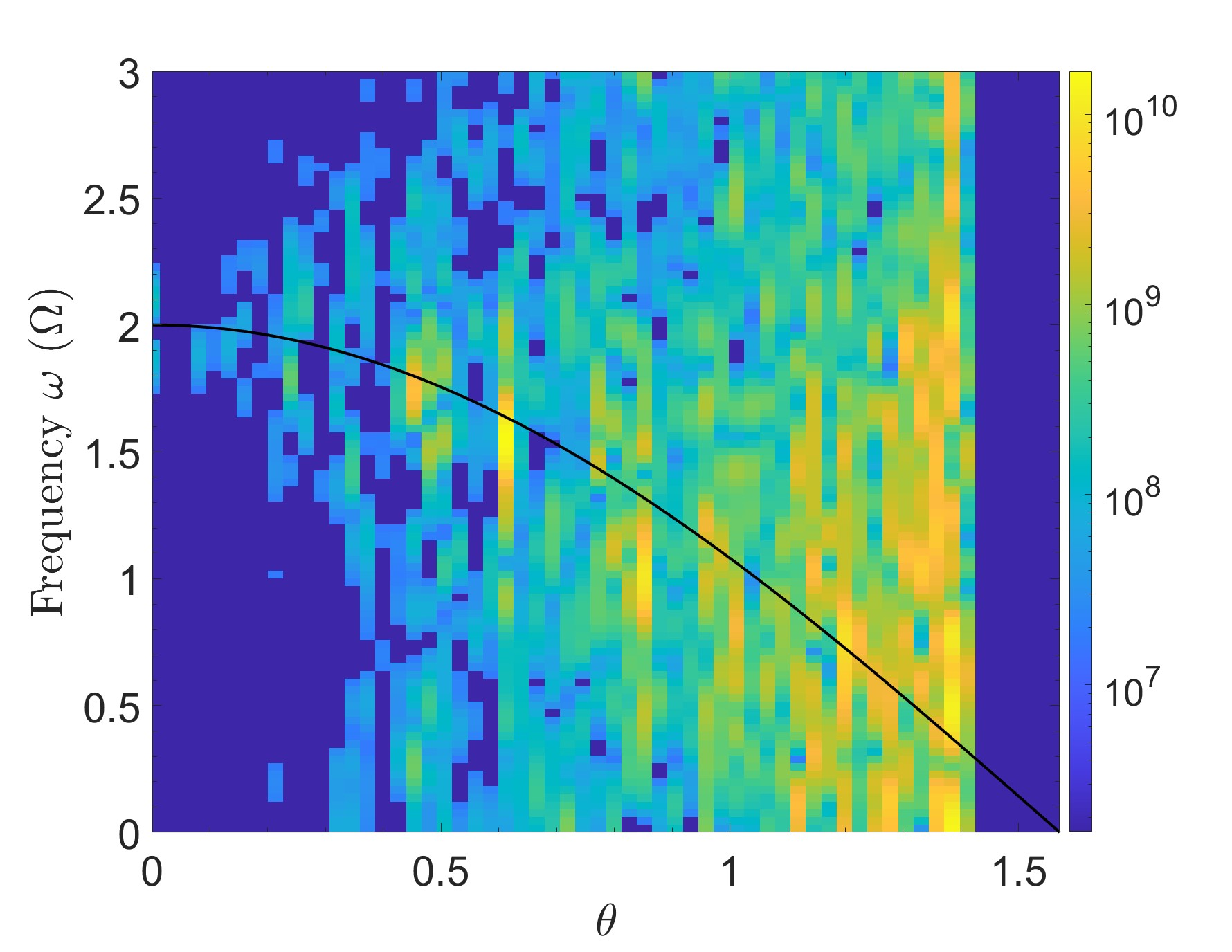}}
     \caption{Same as Fig.~\ref{fig:omegathetaall}, except the interval of wavenumber bins is reduced to $k\in [2,12]$. This removes most convective modes from the right hand side of the figure, allowing for clearer visibility of the inertial waves. The black solid line is the dispersion relation. The geostrophic modes which dominate the flow are set to zero for visibility of other modes.}
     \label{fig:app_omegathetaall}
\end{figure*}

We present the same plots as in Fig.~\ref{fig:omegathetaall} with a limited wavenumber range of $k=[2,12]$. One effect of the limited wavenumber range, combined with our finite grid, is that a number of columns on the right will contain no energy. Only higher wavenumbers can have these angles. The limited $k$ range doesn't affect the linear growth spectrum in Fig.~\ref{fig:app_omega-thetaeps=0.05lin} as the power is concentrated in wavenumbers within our adopted range. During the inertial wave breakdown in Fig.~\ref{fig:app_omega-thetaeps=0.1lin+breakdown} we can clearly see the power concentrated along the inertial wave dispersion relation, as well as the ``mirrored dispersion relation" representing the secondary non-resonant interactions between the waves and the background tidal flow \citep[][]{LeReun2017}. In the convective simulations in the bottom two panels there is indeed power along the dispersion relation where inertial waves are expected, providing another tentative hint for inertial waves in rotating convection.

\nocite{*}
\bibliography{aipsamp}

\end{document}